\documentclass[aps, prd, 10pt, notitlepage, superscriptaddress, nofootinbib,numbers,showpacs,showkeys,twocolumn]{revtex4-1}
\usepackage[utf8]{inputenc}
\usepackage[T1]{fontenc}
\usepackage{anyfontsize}
\usepackage{amsmath}
\usepackage{amssymb}
\usepackage{amsfonts}
\usepackage{mathrsfs}
\usepackage{bm} 
\usepackage{subfigure}
\usepackage{microtype}

\usepackage{graphicx}
\usepackage{epsfig}

\usepackage[dvipsnames]{xcolor}
\usepackage{hyperref}
\hypersetup{
    colorlinks=true,
    citecolor=Purple,
    linkcolor=Purple,
    urlcolor=Purple,
    linktocpage=true,
    breaklinks=true
}
\usepackage{slashed}

\usepackage{hyperref}
\hypersetup{colorlinks,breaklinks,
			citecolor=[rgb]{0,0.0,1.0},
            urlcolor=[rgb]{0.0,0.0,1.0},
            linkcolor=[rgb]{0,0.5,0.9}}

\def\lal{&&{}}
\def\eq{Eq.\,}

\def\beq{\begin{equation} }
\def\eeq{\end{equation}}
\def\bear{\begin{eqnarray}}
\def\bearr{\begin{eqnarray} \lal}
\def\ear{\end{eqnarray}}
\def\earn{\nonumber \end{eqnarray}}
\def\nn{\nonumber\\ {}}

\def\nnn{\nonumber\\ \lal }

\def\E{{\mathbb E}}
\def\R{{\mathbb R}}

\def\mN{_\mu^\nu}

\def\const{{\rm const}}
\def\rf{\eqref}

\def\ssph{static, spherically symmetric}
\def\asflat{asymptotically flat}
\def\wh{wormhole}
\def\whs{wormholes}
\def\bh{black hole}
\def\bhs{black holes}

\def\grav{gravitational}
\def\Scw{Schwarz\-schild}

\def\RN{Reiss\-ner-Nord\-str\"om}

\usepackage{color}

\begin{document}

\title{From massless wormholes to massive black bounces: Shadows and multiple light rings}
\author{G. Alencar}
\email{geova@fisica.ufc.br}
\affiliation{Departamento de F\'isica, Universidade Federal do Cear\'a, Caixa Postal 6030,\\ Campus do Pici, 60455-760 Fortaleza, Cear\'a, Brazil.}
\author{Kirill A. Bronnikov}
\email{kb20@yandex.ru}
\affiliation{Center of Gravitation and Fundamental Metrology, Rostest, Ozyornaya ulitsa 46, Moscow 119361, Russia}
\affiliation{Peoples' Friendship University of Russia, ulitsa Miklukho-Maklaya 6, Moscow, 117198, Russia}
\affiliation{National Research Nuclear University ''MEPhI'', Kashirskoe shosse 31, Moscow 115409, Russia}
\author{T. M. Crispim}
\email{tiago.crispim@fisica.ufc.br}
\affiliation{Departamento de F\'isica, Universidade Federal do Cear\'a, Caixa Postal 6030,\\ Campus do Pici, 60455-760 Fortaleza, Cear\'a, Brazil.}
\author{Diego S\'aez-Chill\'on G\'omez}
\email{diego.saez@uva.es} 
\affiliation{Department of Theoretical Physics, Atomic and Optics, and Laboratory for Disruptive Interdisciplinary Science (LaDIS), Campus Miguel Delibes, \\ University of Valladolid UVA, Paseo Bel\'en, 7,
47011 - Valladolid, Spain}
\affiliation{Departamento de F\'isica, Universidade Federal do Cear\'a, Caixa Postal 6030,\\ Campus do Pici, 60455-760 Fortaleza, Cear\'a, Brazil.}

\author{Marcos V. de S. Silva
}
\email{marcos.sousa@uva.es}
\affiliation{Departamento de F\'isica, Universidade Federal do Cear\'a, Caixa Postal 6030,\\ Campus do Pici, 60455-760 Fortaleza, Cear\'a, Brazil.}
\affiliation{Department of Theoretical Physics, Atomic and Optics, and Laboratory for Disruptive Interdisciplinary Science (LaDIS), Campus Miguel Delibes, \\ University of Valladolid UVA, Paseo Bel\'en, 7,
47011 - Valladolid, Spain}

\date{\today}

\begin{abstract}
We develop a general framework for constructing massive black bounce geometries from massless wormhole seeds by introducing a position-dependent mass function while preserving the underlying areal-radius profile. We apply this procedure to the generalized Ellis–Bronnikov wormhole and obtain a new family of generalized Bardeen-like black bounce space-times, which continuously interpolates between traversable wormholes and regular black holes. We investigate the motion of massive particles and photons and show that, in contrast with the simpler orbital structure of the massless generalized Ellis–Bronnikov geometry, the generalized Bardeen-like space-time can exhibit multiple circular orbits and a rich light ring structure, including configurations with two unstable circular photon orbits separated by a stable one. We determine the corresponding photon sphere and shadow radii and compare our predictions with Event Horizon Telescope observations of Sgr~A*, deriving observational constraints on the parameter space of the model. We further investigate the optical appearance produced by a geometrically and optically thin accretion disk through ray tracing. Multiple light rings generate characteristic nested structures in the high-resolution intensity profiles, whose relative brightness is strongly affected by the mass deformation and gravitational redshift. However, after modeling the finite angular resolution of the EHT with a Gaussian beam convolution, these fine structures are largely washed out, revealing a strong observational degeneracy between these exotic compact objects and standard black hole geometries at current EHT resolution.
\end{abstract}


\maketitle


\section{Introduction}\label{sec:intro}

It is well known that, within the context of General Relativity (GR), the presence of space-time singularities represents a fundamental theoretical limitation. When considering matter sources with physically reasonable properties undergoing gravitational collapse, such as in the classic Oppenheimer-Snyder model \cite{Oppenheimer:1939ue}, the formation of singularities is generically unavoidable under the conditions established by the singularity theorems of Hawking and Penrose \cite{Penrose1965,Hawking:1967ju,Hawking:1970zqf}. In an attempt to overcome such limitations and obtain physically acceptable objects across all regions of the manifold, several frameworks have been proposed to construct regular metrics. In this context, the first proposal for a regular black hole was introduced by Bardeen in 1968 \cite{Bardeen:1968pbd}. However, a well-defined field model for this metric was only obtained later by Ayón-Beato and García in \cite{Ayon-Beato:2000mjt}, considering the coupling between gravity and a nonlinear electromagnetic theory.

However, it is a remarkable fact that the concept of a wormhole, initially conceived in the work of Einstein and Rosen in the form of so-called topological ``bridges'' \cite{Einstein:1935tc}, is historically and conceptually older than the modern description of black holes itself \cite{Herdeiro:2018ldf}. Like black holes, static and spherically symmetric wormholes are solutions to Einstein’s field equations; however, in their standard traversable realization, they differ from black holes by being free of event horizons and central singularities and can connect two asymptotically flat regions. The modern formulation of traversable wormholes was developed independently by Ellis and Bronnikov \cite{Bronnikov:1973fh,Ellis:1973yv} and later consolidated by Morris and Thorne \cite{Morris1988WormholesIS}. Recently, interest in the properties of these structures has grown substantially due to the realization that certain wormhole solutions are capable of mimicking several observational features of black holes (\textit{black hole mimickers}) \cite{Bambi:2008jg,Ortiz:2015rma,Bambi:2013nla,Shaikh:2018kfv,Bronnikov:2012ch,Crispim:2025cql,Cardoso:2016rao,Konoplya:2016hmd}. This has motivated an intense search for optical, gravitational, and lensing signatures that can distinguish them in the current era of precision astrophysics \cite{Guerrero:2021ues,Guerrero:2022qkh,Alencar:2026qeb,Novikov:2025elo,Bugaev:2024dte,Bugaev:2023mlc,Silva:2026mlo,Pereira:2026ffn,Pereira:2025fvg}.

To bridge the gap between regular black holes and traversable wormholes, a promising new class of regular geometries has recently emerged: black bounce space-times. Originally proposed by Simpson and Visser \cite{Simpson:2018tsi}, and subsequently extensively studied in various contexts \cite{Bronnikov:2006fu,Muniz:2024wiv,Bronnikov:2021uta,Bronnikov:2019sbx,Alencar:2025jvl,Bronnikov:2021liv,Crispim:2024yjz,Bolokhov:2012kn,Bronnikov:2022bud,Bronnikov:2024izh,Alencar:2026ypy,Alencar:2026quc,Alencar:2024yvh,deSousaSilva:2026kvp,Silva:2025fqj}, these solutions modify the areal-radius function of the manifold, $r(x)$, replacing the singular center at $r=0$ with a non-zero minimum radius $r_0 > 0$ (a throat or \textit{bounce}). Depending on the relative magnitude of the mass and deformation parameters, the \textit{bounce} can be located within an event horizon (configuring a regular black hole with a Kantowski-Sachs-type interior) or in a horizonless static region (characterizing a traversable wormhole). As extended by Lobo \textit{et al.} \cite{Lobo:2020ffi}, the concept of a black bounce has come to designate, in a broad sense, an entire family of regular space-times with two asymptotic regions that lack a singular center at $r=0$, encompassing everything from regular black holes with an internal throat to inter-universe wormholes.

A natural question in this context is how a gravitational mass scale can be introduced into a massless wormhole seed while preserving its underlying areal-radius profile. The simplest case of a wormhole geometry occurs when the temporal metric function is constant, $f(x) \equiv 1$, corresponding to the absence of gravitational redshift and gravitational acceleration for static observers. In this massless limit, the ADM mass vanishes ($m=0$), although the curvature of space-time still guarantees non-zero effective matter density and pressure. To introduce a nonzero asymptotic mass scale and enrich the strong-field phenomenology of these configurations, we can redefine the mass function via a deformation of the form $M(x) = m \mu(x)$, where $m$ represents the ADM mass and $\mu(x)$ dictates the spatial distribution profile. Under this reparametrization, the non-vanishing components of the Einstein tensor can be decomposed exactly as the sum of the seed geometry tensor $G_\mu^\nu [\text{seed}]$ and the term proportional to $m$ that encodes the modifications imposed by the presence of mass, $m\Delta G_\mu^\nu$.

This framework, aligned with recent developments in linear deformations of seed geometries \cite{Ovalle2017,Ovalle2019,Misyura:2024fho}, allows the conversion of a massless wormhole into a massive black bounce or a regular black hole without altering the intrinsic geometric structure of the radius $r(x)$. The parameter $m$ establishes a characteristic scale for fundamental observables, such as the photon sphere, shadow radius, and gravitational lensing profile. For suitable choices of the function $\mu(x)$, as the mass parameter $m$ increases, the space-time smoothly departs from the massless limit and forms a continuous family of geometries comprising, for comparatively small $m$, massive wormholes and, for large $m$, regular black hole space-times. Depending on the choice of the function $f(x)$, the minimum of the areal radius $r(x=x_m)$ is not necessarily located inside a non-static Kantowski-Sachs region. If horizons are present while this minimum remains in a static region ($f(x_m)>0$), it acts as a wormhole throat within the regular black hole. There is also the intermediate case where the minimum of $r(x)$ coincides exactly with an event horizon ($f(x_m)=0$). In these cases, the resulting regular black holes lack a center (since $r(x)>0$ throughout the manifold) but possess two asymptotic regions inherited from the seed geometry. 

In this work, we present a general framework to construct such massive black bounce geometries starting from massless seed wormholes and apply this methodology to a class of wormholes that has recently gained prominence in the literature \cite{Kar:1995jz,Crispim:2024dgd,deSSilva:2025ncg}: the Generalized Ellis-Bronnikov (GEB) wormhole, characterized by the areal radius profile $r(x) = (x^n + a^n)^{1/n}$, where $a$ denotes the throat scale parameter and $n\geq 2$ is an even integer. By promoting the GEB model via the inclusion of the mass parameter $m$, we obtain what we denote as the Generalized Bardeen-like (GBL) black bounce. For $n=2$, this metric reduces to the original ``Bardeen-like black bounce'' of Ref.~\cite{Lobo:2020ffi}. In the limit $a = 0$, the throat collapses and the space-time reduces to the standard Schwarzschild solution, whereas setting $m = 0$ recovers the massless GEB wormhole seed. Finally, at large spatial distances ($\vert{}x\vert{} \to \infty$), the geometry is approximately Schwarzschild, naturally guaranteeing asymptotic flatness.

This paper is organized as follows. In Sec.~\ref{sec:general}, we present the general framework for constructing massive black bounces from massless seed wormholes, detailing their regularity properties, asymptotic flatness, and the decomposition of the Einstein tensor. In Sec.~\ref{spacetimes}, we apply this methodology to the GEB wormhole to obtain the new GBL black bounce geometry, while in Sec.~\ref{sec:massive} we analyze the dynamics and orbital structure of massive particles. Sec.~\ref{massless} is devoted to the study of null geodesics, the occurrence of multiple light rings, and the determination of the shadow radius, comparing our theoretical results with the observational data from the EHT for Sgr~A*. In Sec.~\ref{sec:OA}, we simulate the optical appearance of the geometry illuminated by a thin accretion disk via ray tracing. Finally, in Sec.~\ref{sec:con}, we evaluate the effects of the spatial resolution limit of the EHT by applying a Gaussian filter convolution and present our concluding remarks. Throughout this work, we adopt units where $c=8\pi G = 1$ and use the metric signature $(-,+,+,+)$.
\section{From massless wormholes to massive black bounces}\label{sec:general}
\subsection{Basic notions}

  The most general static, spherically symmetric line element may be brought to the form
\beq    \label{bbgeneral}
    		ds^2 = -f(x)\,dt^2 + \frac{dx^2}{f(x)}+ r^2(x)\, d\Omega_2^2,
\eeq    
  where $d\Omega_2^2 = d\theta^{2} + \sin^{2}\theta\, d\varphi^{2}$ is the line element 
  on a 2-sphere of unit radius. Here, the radial coordinate $x$, often called a Buchdahl
  or quasiglobal coordinate (see, e.g., \cite{Bronnikov:2001tv, Bronnikov:2012ws}), is 
  chosen according to the coordinate condition $g_{00} g_{11} = -1$ which is convenient 
  for describing the geometry of both \bhs\ and wormholes, in particular, because it allows regions on both sides of a horizon to be described within a single coordinate system; horizons then appear as regular zeros of the function $f(x)$. The function $r(x)$ determines 
  the radius of coordinate 2-spheres $x = \const$, the 2-sphere area being equal to
  $4\pi r(x)^2$.
  
  Concerning \wh\ geometies, the basic notion is that of a {\it throat\/}, defined as the 
  location where the area of the 2-sphere has a local minimum, that is,
\beq    
    r(x_0) = \min r(x),
\eeq    
  which implies for regular functions $r(x)$
\beq 		\label{min-r}
		r'(x_0) =0, \qquad  r''(x_0) \geq 0.
\eeq  
  where the prime denotes $d/dx$.
  A (traversable) {\it \wh\/} can be defined as a space-time with the geometry \rf{bbgeneral}
  containing at least one throat $x = x_{\rm th}$ and such that $r(x) \gg r(x_{\rm th})$
  on both sides from the throat(s). One often adds the requirement $f(x) > 0$ in the whole
  range of $x$, meaning the absence of horizons; however, one cannot exclude a \wh\
  geometry with a horizon somewhere far from the throat (if, for example, a \wh\ 
  resides in an asymptotically de Sitter space-time, see, e.g., \cite{Bronnikov:2017kvq}),
  therefore, the requirement $f(x) > 0$ is not universal. 

  With the line element \rf{bbgeneral}, many authors have investigated different aspects 
  of wormhole physics, including their geometric properties, stability, lensing features, 
  and potential astrophysical signatures, see, e.g.,  
  \cite{Bronnikov:2021liv, Ellis:1973yv,Crispim:2024dgd, Crispim:2024lzf,Visser:1995cc,Damour:2007ap,Konoplya:2025mvj,Bronnikov:2013coa,Alencar:2026qeb,Lobo:2005us,Lobo:2005yv}.

  Descriptons of more general \wh\ geometries, not restricted to spherical symmetry 
  and static, can be found, for instance, in the reviews   
  \cite{Visser:1995cc, Hochberg:1998ha, KordZangeneh:2025qel, Bambi:2021qfo}, see 
  also references therein.
  
  Let us remark here that following the famous paper \cite{Morris1988WormholesIS},
  many authors discuss \whs\ using the quantity $r$ as a radial coordinate, by analogy 
  with \Scw\ and other well-known \bh\ space-times, and write the \wh\ metric as 
\beq                   \label{morris}
		ds^2 = - e^{2\Phi(r)}dt^2 + \frac{dr^2}{1- b(r)/r} + r^2 d\Omega^2,		
\eeq  
  where $\Phi(r)$ is called the redshift function and $b(r)$ the shape function of a \wh.
  Identifying \rf{morris} with \rf{bbgeneral} term by term, we obtain the relationships
\beq               \label{x-r}
		f(x) = e^{2\Phi(r)}, \quad  dx = \pm \frac {e^{\Phi(r)}dr}{\sqrt{1- b(r)/r}}.
\eeq  
  Using $r$ as a coordinate is very common and often useful, since it simplifies the field equations. In the case of wormholes, however, this choice introduces a coordinate singularity, 
  where $g_{rr} \to \infty$, at the most important sphere, the throat (where $b(r) =r$), 
  while in any other (admissible) coordinate system this sphere is regular. It also makes it impossible to cover the whole space with a single chart. In the metric \rf{morris},
  the transparent condition\footnote 
  		{At a higher-order minimum of $r(x)$ there is $r''(x_{\rm th})=0$, but then
  		we have inevitably $r'' >0$ in its close neighborhood.}
  $r'' > 0$ of a minimum of $r$ transforms to the rather 
  vague, so-called ``flaring-out condition'' $b(r) > r db/dr$ near the throat. 
  So, in what follows we adhere to the metric \rf{bbgeneral}.
  
  Furthermore, a possible space-time region where $f(x) < 0$ that generally occurs in 
  \bh\ interiors comprises, in fact, the geometry of a Kantowski-Sachs homogeneous
  cosmology since the coordinate $x$ has there a temporal nature; in this case, 
  the quantities $\sqrt{-f(x)}$ and $r(x)$ are cosmological scale factors describing 
  expansion or contraction of space in the directions of $t$ (which is now a spatial 
  coordinate) and the angles $\theta, \varphi$, respectively. Therefore, a possible 
  minimum of $r(x)$ in such a region is not a throat but a bounce of one of these 
  scale factors, and it has acquired the name of a {\it black bounce\/} 
  \cite{Simpson:2018tsi} due to its relation to \bh\ physics. In what follows we will 
  discuss space-times with black bounces in addition to \whs.
  
\subsection{Regularity and asymptotic flatness}  

  The independent nonvanishing Riemann tensor components for the metric
  \rf{bbgeneral} are given by
\begin{eqnarray}         \label{Riem} 
		R^{tx}{}_{tx} &=& - \frac{f''}{2}, 
\qquad
		R^{t\theta}{}_{t\theta} = R^{t\varphi}{}_{ t\varphi} = -\frac{f'r'}{2r},
\nn
		R^{x\theta}{}_{ x\theta} &=& R^{x\varphi}{}_{ x\varphi} = -\frac{f'r' + 2fr''}{2r},
\nn
 		R^{\theta\varphi}{}_{\theta\varphi} &=& \frac{1-fr'^{2}}{r^{2}}.
\end{eqnarray}
  All these expressions are finite and regular if $r >0$ and the functions $r(x)$ and $f(x)$ 
  are regular (at least $\sl C^2$). Since all algebraic curvature invariants can be written as
  polynomials of the components \rf{Riem}, regularity of the functions $r(x) >0$ and 
  $f(x)$ guarantees regularity of all these invariants. For example, the Kretschmann scalar is
\beq    
	\begin{split}
	K &= R^{\mu\nu\alpha\beta}R_{\mu\nu\alpha\beta}
\\
	&= 4\left(R^{tx}{}_{tx}\right)^{2}
		+ 8\left(R^{t\theta}{}_{ t\theta}\right)^{2}
		+ 8\left(R^{x\theta}{}_{ x\theta}\right)^{2}
		+ 4\left(R^{\theta\varphi}{}_{\theta\varphi}\right)^{2}.
	\end{split}
\eeq    
  As long as $K$ is a sum of squares, its finiteness is not only necessary but also a 
  sufficient condition for regularity of the metric. Let us also stress that at $r(x) >0$ 
  the metric \rf{bbgeneral} is singularity-free both with $f(x) > 0$ (\ssph\ geometry) 
  and $f(x) < 0$ (Kantowski-Sachs geometry). Moreover, it is regular at horizons defined by
  $f(x) =0$, irrespective of whether these horizons are generic (first-order, with $f' \ne 0$),
  extremal ($f'=0,\ f''\ne 0$), or superextremal, with higher-order zeros of $f(x)$.   
  In what follows we will deal with wormhole and black bounce space-times with 
  everywhere regular functions $f(x)$ and $r(x) >0$, thus all of them are globally regular. 
  
  For island-like objects with weak gravity far from their strongly curved regions, we 
  should require asymptotic flatness. For a general metric of the form \rf{bbgeneral}, 
  the corresponding conditions read \cite{Bronnikov:2012ws}
\beq
		r(x) \to \infty, \ \ 
		f(x) = f_0 + O\Big(\frac 1r\Big), \ \
		f r'{}^2 = 1+ O\Big(\frac 1r\Big), 
\eeq    
  where $f_0 = \const >0$, and the last condition means that the circumference to radius 
  ratio for large circles $x=\const$ should tend to the correct Euclidean value $2\pi$.
  While considering \whs\ and black bounce objects with two large-$r$ asymptotic
  regions, it is natural to expect that they are \asflat\ on both ends, $x \to \pm \infty$, 
  where, in general, the clock rates can be different, leading to different $f_0$ values.
  However, in the present paper we deal with solutions symmetric with respect to the 
  change $x \to -x$, so without loss of generality we can put 
\beq               \label{asflat}
		f(x) \to 1, \quad\  r'(x) \to \pm 1 \quad {\rm as}\ \ \ x \to \pm \infty. 
\eeq  
  Furthermore, assuming that in such far regions the \grav\ field of our objects should be 
  approximately Newtonian, we can require a \Scw-like behavior of $f(x)$, that is, 
  $f(x) \approx 1 - 2 m/r(x)$, where $m = \const$ is the \Scw\ (or ADM) mass. In our 
  symmetric configuration, the mass $m$ is the same on both ends $x\to \pm \infty$. 
  
  The condition $r' \to 1$ as $x \to \infty$ implies $r(x) \approx x + x_0$, where 
  $x_0 = \const$, and redefining $x + x_0 \to x$ without loss of generality, we achieve 
  $r(x) \approx x$ as $x \to \infty$. Furthermore, under our assumption of symmetry,
   we have $r(x) \approx -x$ at large negative $x$, or, in other words, $r(x) \approx |x|$
   as $x \to \pm\infty$.
    
\subsection{Introducing a mass scale}

  The effective matter distribution in configurations of interest can be characterized in 
  different ways. Thus, the Einstein equations have the following nontrivial components:
\bearr              \label{EE00}
		G^t_t = \frac 1{r^2}[-1 + f(2 r r'' + r'^2) + f' rr'] = - T^t_t, 
\\  \lal             \label{EE11}
		G^x_x = \frac 1{r^2}[-1 + f r'^2 + f' rr'] = -T^x_x, 
\\  \lal             \label{EE22}
		G^\theta_\theta = G_\varphi^\varphi=
		   \frac 1{r}\big[ fr'' + \tfrac 12 r f'' + f' r'\big] = -T^\theta_\theta,
\ear   
  where $T\mN$ is the total stress-energy tensor of matter.\footnote
  		{We are using the units in which $8\pi G = c =1$.}
  In particular, the difference of \rf{EE00} and \rf{EE11} gives 
\beq                   \label{01}
        T^t_t - T^x_x = -2 f(x) r''/r.		 
\eeq  
  In static space-time, where $f >0$, $T^t_t = \rho$ is the energy density, and 
  $-T^x_x = p_x$ the radial pressure. Since $r'' >0$ at a \wh\ throat, \eq \rf{01} leads 
  to the widely known result \cite{Morris1988WormholesIS} that at such a throat 
  $\rho + p_r < 0$, which expresses a violation of the Null Energy Condition (NEC). 
  A less well-known result \cite{Bronnikov:2021uta} follows from \rf{01} in the case $f < 0$, 
  where the condition $r'' >0$ is valid at and near a black bounce time $x_{\rm bb}$. 
  In such regions, the density $\rho$ coincides with $T^x_x$ while $-T^t_t = p_t$ is 
  pressure in the spatial $t$-direction. Thus with $f < 0$ and $r'' >0$, \rf{01} leads 
  to $\rho+p_t <0$, again expressing a NEC violation. We conclude that
  {\it in the framework of GR, the NEC is inevitably violated both at \wh\ throats and 
  at black bounce times.} 
  
  Another useful characteristic of matter distributions is the Hernandez-Misner-Sharp 
  (HMS) \cite{Misner:1964je,Hernandez:1966zf} quasi-local mass function $M_{\rm HMS}$, initially defined while using the spherical 
  radius $r$ as a coordinate: specifically, it is assigned
\beq                \label{M-HMS}
		g_{rr} = \bigg(1 - \frac {2M_{\rm HMS}(r)} {r}\bigg)^{-1}.
\eeq  
  The quantity $M_{\rm HMS}(r)$ is well known as an expression for the total gravitating 
  mass of matter inside a given radius $r$ for stellar models with a regular center.
  Comparing \rf{M-HMS} with \eq \rf{morris}, we see that $2M_{\rm HMS} = b(r)$
  (which, in our opinion, makes it redundant to introduce the separate term 
  ``shape function'' for $b(r)$, whereas the \wh\ shape is much better characterized by 
  the function $r(x)$). On the other hand, comparing \rf{M-HMS} with \rf{bbgeneral}, 
  it is easy to obtain $M_{\rm HMS}$ in terms of $x$:
\beq                \label{M-HMS(x)}
	 M_{\rm HMS}(x) = \frac{r(x)}{2} \left[ 1 - f(x)\,r'^2(x) \right].
\eeq
  By definition, $M_{\rm HMS} \to m$, that is, tends to the ADM mass at both asymptotics
  $x \to \pm \infty$. It is also clear from \rf{M-HMS(x)} that $M_{\rm HMS} = r/2$ 
  at a \wh\ throat or black bounce (where $r'=0$) and at a horizon (where $f=0$).   
  
  Another convenient definition of a mass function $M(x)$ can be introduced
  by analogy between $g_{tt} = f(x)$ and its \Scw\ expression: we put   
\beq    \label{eq:f_def_intro}
    	f(x)= 1 - \frac{2 M(x)}{r(x)}  \ \ \Longleftrightarrow\ \ M(x) = \frac{r(x)}{2}[1 - f(x)].
\eeq    
   Again, by definition, $M \to m$ as $x \to \pm \infty$, and also $M = r/2$ at horizons,
   if any. However, at throats and black bounces the expression \rf{eq:f_def_intro}
   is different from $r/2$: it is smaller than that at throats but larger at black bounces. 
   
   It is also easy to obtain a relationship between the two mass functions:
\beq    		\label{eq:m_relation_smooth}
	  M_{\rm HMS}(x) = \frac{r}{2}\bigl(1-r'^2\bigr) + M(x)\,r'^2.
\eeq  
  Evidently, they coincide at large $|x|$ (assuming asymptotic flatness with $|r'|\to 1$) 
  and at horizons ($f =0,\ M = r/2$).
  
  The simplest and important special case of \wh\ geometries is obtained assuming 
  $f(x) = \const$, or, choosing the proper time scale, without loss of generality, 
  $f(x)\equiv 1$. It corresponds to zero \grav\ attraction and zero tidal forces acting 
  on moving bodies. In this case, the ADM mass $m =0$ and $M(x) \equiv 0$, but since 
  space is curved, the effective matter density and pressure are in general nonzero, 
  and according to \eq \rf{M-HMS(x)}, $M_{\rm HMS}(x) = r(1 - r'^2)$ is also in general
  nonzero.  
  
  We can redefine $M(x)$ as $M(x)=m\,\mu(x)$, then the nonvanishing components 
  of the Einstein tensor are given by
\bear
       G_t^t &=& \frac{r'^2 -1}{r^2} + \frac{2r''}{r} 
       		-2m\frac{\mu'r'}{r^2} -4m\frac{\mu r''}{r^2},
\\
    G_x^x &=& \frac{r'^2 -1}{r^2} -2m\frac{\mu'r'}{r^2},
\\
    G_\theta^\theta &=& G_\varphi^\varphi 
    		= \frac{r''}{r} - m\frac{\mu''}{r} - m\frac{\mu r''}{r^2}.
\ear
  If we look carefully at the equations above, we notice that they can be written as
\beq         \label{seed}
    	G_\mu^\nu = G_\mu^\nu [\text{seed}] + m\Delta G_\mu^\nu,
\eeq    
  where the term $G^{\mu}{}_{\nu}[\text{seed}]$ is the Einstein tensor of a massless 
  wormhole with the same profile $r(x)$, while introduction of the mass $m$ through 
  the function $f(x)\not\equiv 1$ allows us to reinterpret the full solution 
  as a deformation of this seed geometry. The contribution proportional to $m$
  in \eq \rf{seed} encodes modifications introduced by the mass. Thus the 
  massless wormhole acts as the underlying background, and the function $\mu(x)$
  shows how the geometry is smoothly deformed away from this seed configuration. 
  It is worth noting that linear deformations of a given seed geometry have been 
  extensively explored in the recent years in several contexts
  \cite{Ovalle2017,Ovalle2019,Misyura:2024fho}.

  In this way, with a sufficiently large $m$, a massless wormhole metric can be converted 
  to a massive black hole geometry without altering the intrinsic areal structure $r(x)$. 
  The parameter $m$ now sets a characteristic scale for observable features such as the 
  photon sphere, the shadow radius, and the lensing profile. With growing mass 
  parameter $m$, the space-time smoothly departs from the massless limit and 
  forms a family of geometries comprising, at comparatively small $m$, massive \whs,
  and at large $m$ black-hole space-times. 
  
In this case, depending on the choice of $f(x)$, the minimum of $r(x)$ ($x=x_m$) does 
  not necessarily get into a nonstatic Kantowski-Sachs region of the \bh, thus becoming 
  a black bounce. If this minimum is still located in a static region, that is, $f(x_m) >0$, it 
  is, as before, a \wh\ throat, and we can assert that the obtained regular \bh\ has a \wh\ 
  region in its interior. In what follows we will deal with an example of such a configuration. 
  There can be an intermediate case where the minimum of $r(x)$ coincides with a horizon, 
  that is, $f(x_m) =0$. In all such cases we are dealing with regular \bhs\ which have no 
  center (since everywhere $r(x) >0$) but instead possess two asymptotic regions like 
  the ``seed'' \wh\ (unlike the conventional \bh\ space-times containing $r=0$, be it a regular 
  center or a singularity like \Scw's). Following \cite{Lobo:2020ffi}, we will call such \bhs\
  {\it black bounce space-times\/} in an extended sense.

\subsection{3D visualization}

  To obtain a 3D visualization of \wh\ geometries, it is useful to employ embedding 
  diagrams that provide an effective way to illustrate how a 2D spatial slice of the 
  geometry is curved when embedded into a higher-dimensional Euclidean space. 
  More precisely, we consider an equatorial slice of space-time ($\theta = \pi/2$) at 
  fixed time $t = \const$. The 2D metric of this slice can be interpreted as the geometry 
  of a surface isometrically embedded in $\E^3$.

  The spatial metric written for a fixed time at $\theta = \pi/2$ reads
\beq    \label{bb2d}
    	ds^2 = \frac{dx^2}{f(x)} + r(x)^2d\varphi^2, \qquad f(x) >0.
\eeq    
  On the other hand, the metric of 3D Euclidean space $\E^3$ in cylindrical coordinates is
\beq    \label{ds_3E}
	    d\sigma_{3D}^2 = d\rho^2 + dz^2+ \rho^2 d\varphi^2 .    
\eeq    
  Our aim is to consider the metric \rf{bb2d} as that of a surface of rotation in $\E^3$.
  To do that, we identify the angle $\varphi$ and the radius $\rho \equiv r(x)$ in the 
  two metrics and consider the coordinate $z$ as a function of $x$, so that 
\beq        \label{ds_comp}
	  d\sigma_{3D}^2 = (r'^2 + z'^2) dx^2 +  r(x)^2d\varphi^2,
\eeq     
  where, as before, the prime denotes $d/dx$.  Comparing \rf{bb2d} and \rf{ds_comp}, 
  we get an equation for the forming curve of our surface of rotation
\beq           \label{z_x}
		\frac{dz(x)}{dx} = \pm \sqrt{B}, \quad B :=\frac {1- f(x)r'(x)^2}{f(x)}. 
\eeq    
   Thus for a well-defined wormhole geometry, a necessary feature of the embedding 
   condition is
\beq
    	r'^2f \leq 1, \ \ \forall\, x \in (-\infty, +\infty).
\eeq    	
       
   To analyze the shape of the embedded surface, let us calculate the derivatives
   $dr/dz$ and $d^2 r/dz^2$. We obtain:    
\bearr                \label{r_z}
		  \frac {dr}{dz} = \frac {r'(x)}{z'(x)} = \pm \frac{r'}{\sqrt{B}},
\\ \lal		  \label {r_zz}
		  \frac {d^2r}{dz^2} = \frac 1{z'} \frac{d}{dx}\!\left( \frac{r'}{\sqrt{B}} \right)
		  = \frac{2\,r'' B - r' B'}{2 B^2} .
\ear       
   At the throat $x = x_{\rm th}$ we have $r'=0$, so $B(x_{\rm th})= 1/f(x_{\rm th})$, 
   and \eq \rf{r_zz} becomes
\beq             \label{r_zz-throat}
	\frac{d^2 r}{dz^2} \Big|_{x_{\rm th}} = r''(x_{\rm th})\, f(x_{\rm th}) .
\eeq    
   Thus, as expected, the throat, defined as a minimum of $r(x)$, is also a minimum of 
   $r(z)$, forming the narrowest section of the obtained surface of rotation. 

   At large $|x|$, where $f \to 1$ and $r'^2 \to 1$, we have $B \to 0$, hence 
   $dz/dx \to 0$, so that the slope of the curve $z(x)$ tends to zero while $r\to\infty$.
   
   All that concerned \whs, while black bounce space-times, being a particular version
   of \bhs, do not admit such visualization; one can note that the best 
   representation of their causal structure is provided by Carter-Penrose diagrams.

\section{From GEB wormhole to GBL black bounce}\label{spacetimes}
\subsection{Revisiting GEB}

  The simplest massless wormhole model is described by the Ellis-Bronnikov (EB) 
  solution, whose line element can be written as Eq.\,\eqref{bbgeneral} with $f\equiv 1$ 
  and $r(x) = \sqrt{x^2 + a^2}$ \cite{Bronnikov:1973fh,Ellis:1973yv}:
\beq    
         ds^2=-dt^2+dx^2+(x^2+a^2)d\Omega_2^2,            \label{metric_EBx}
\eeq    
  where $a$ is the radius of the wormhole throat. In this coordinate system, $x$ represents 
  the proper radial distance, taking values in $\R$, and the throat is located at $x=0$.

   A generalization of this model was introduced in Ref. \cite{Kar:1995jz}, in which the areal-radius function is given by $r(x) = (x^n + a^n)^{1/n}$, and the line element is written as
\beq        \label{metric_EBGx}
        ds^2=-dt^2+dx^2+(x^n+a^n)^{2/n}d\Omega_2^2,
\eeq    
   where $n \geq 2$ is taken to be an even integer. At $n = 2$, we recover the EB case. In order for the angular term 
   to be invariant under the transformation $x \to -x$, we will assume that the 
   parameter $n$ can only take even values.

\begin{figure*}[!htb]
    \centering
    \includegraphics[width=1\linewidth]{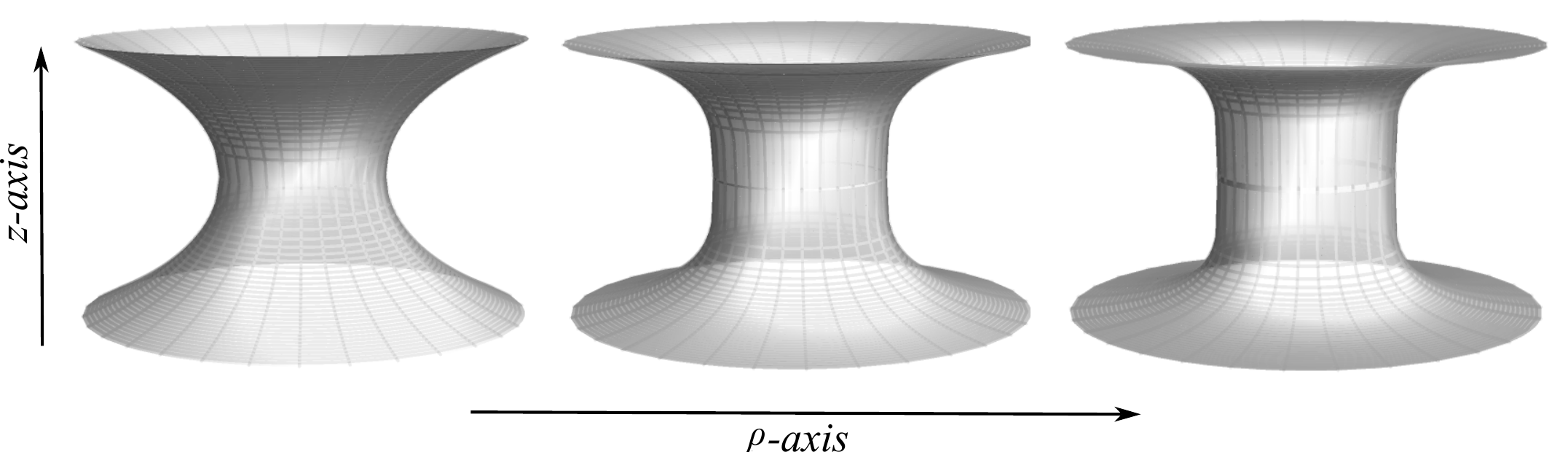} 
\caption{Embedding diagram for the GEB wormhole with 
		$a=3$ and $n=2$ (left), $n=4$ (center), and $n=6$ (right).}
    \label{fig:imersion}
\end{figure*}

   For this space-time, the function $z(x)$ defined by \eqref{z_x} is found from 
\beq    
    \frac{dz}{dx} = \pm \sqrt{1 - x^{2n-2}\bigl(x^n + a^n\bigr)^{{2}/{n}-2}}.
\eeq    
  By numerically integrating this expression, we can plot the embedding diagram for 
  this geometry, obtaining the 3D plot as described in the previous section. In 
  Fig.\,\ref{fig:imersion}, we present embedding diagrams for GEB wormholes, fixing 
  $a=3$, for different values of $n$. As stated in \cite{Crispim:2024dgd}, with increasing 
  parameter $n$, the shape of the embedding diagram becomes increasingly cylindrical, 
  significantly differing from the characteristic catenary shape of the original EB wormhole.

  For the metric \eqref{metric_EBGx}, the HMS quasilocal mass is
\bearr    \label{mass_x}
	M_{\rm HMS}(x) 
	=\frac{1}{2} \left(a^n+x^n\right)^{1/n} 
\nnn \qquad \times	
		\left(1-x^{2 n-2} \left(a^n+x^n\right)^{2/n-2}\right).
\ear  
  At the throat $x = 0$, the quasilocal HMS mass takes the value $a/2$, in accord with 
  the previous general description. It is also evident that $M(x) \equiv 0$ 
  and the ADM mass is $m =0$. 

  This modification of EB space-time has recently attracted considerable interest 
  \cite{Sharma:2021kqb,Sharma:2022dbx,Sharma:2022tiv,Godani:2023jhq,Muniz:2022aal,Nilton:2022hrp,deSouza:2022ioq,Crispim:2024dgd,Bhattacharya:2025ljc,Jana:2024pdh,Mitra:2023yjf},    
  as it represents a simple modification that offers certain advantages and freedom
  as compared to the conventional EB wormhole. In their original work, the authors in 
  \cite{Kar:1995jz} studied propagation of scalar waves in this geometry, showing that 
  at sufficiently large values of $n$, a resonance can occur at particle transmission. 
  This is mainly due to the fact that as $n$ increases, the wormhole acquires an intriguing
  almost cylindrical shape, with $dz/dx$ changing very rapidly near the throat. As stated 
  by the authors, this resonance combined with other effects related to the geometry, 
  such as gravitational lensing and motion of photons and neutrinos, may help to obtain
  information on the size of the wormhole throat, should such an object exist at 
  astrophysical scales.

  More recently, in Ref.~\cite{Crispim:2024dgd}, the authors investigated dust accretion in this space-time and also demonstrated that this geometry can be obtained as an exact solution of GR by considering a phantom scalar field coupled to nonlinear electrodynamics (NED), with either a magnetic or an electric source, in line with recent works showing that regular geometries with a rich causal structure can be supported by such fields 
\cite{Crispim:2024lzf,Rodrigues:2025plw,deSSilva:2025ncg,Bronnikov:2016xvj,
  Bronnikov:2000vy,Bronnikov:2022bud,Alencar:2024nxi,Pereira:2024rtv,Javed:2022dfn,
  Canate:2022dzb,Kamal:2018oxw,Alencar:2024yvh,Alencar:2025jvl}. 
  It is interesting to note that, unlike the usual EB wormhole, for which a free phantom scalar field is sufficient to support the geometry, the generalized case with $n>2$, within this scalar-field plus NED realization, requires both a nontrivial scalar potential and an electromagnetic contribution. This is related to the fact that the symmetry of the Einstein tensor $G^t_t = G^\theta_\theta$, present for $n=2$, is broken for $n>2$.
Moreover, as shown in Ref. \cite{Bronnikov:2022bud}, any static and spherically symmetric metric can be represented as an exact solution of the Einstein equations sourced by NED with a radial magnetic field together with a scalar field possessing, in general, a nonzero self-interaction potential. In this representation, the scalar field may exhibit the so-called ``trapped ghost'' behavior \cite{Bronnikov:2010hu}, being phantom in the strong-field region and becoming canonical outside it, with a smooth transition between the two regimes.
        
\subsection{Generalized Bardeen-like black bounce}

  To illustrate the usefulness of our approach, we now propose a specific form for the mass function \(M(x)\) introduced in Eq.\,\eqref{eq:f_def_intro}. Our construction is motivated 
  by the analysis presented in Ref.\,\cite{Lobo:2020ffi}, where the authors introduced a
  ``Bardeen-like black bounce'' geometry obtained by deforming the standard Bardeen area
  function while preserving the mass function structure. Their model combines features of
  regular black holes with those of black bounce geometries, providing a smooth transition
  between a horizonless wormhole and a regular black hole, depending on the values of 
  the model parameters. 

  Inspired by this idea, we propose here a GBL black bounce. In our case, we retain the same area function as in the GEB wormhole geometry, ensuring that the bounce structure and the minimal area remain unchanged. Thus 
  we propose the following generalized metric functions:
\beq    
	    r(x) = (x^n + a^n)^{1/n}, \quad 
	    M(x) =m\left(\frac{x}{r(x)}\right)^n,
\eeq    
  so that the metric \eqref{bbgeneral} becomes
\beq    \label{BD_GEN}
\begin{split}
     ds^2
    &= -\left(1 - \frac{2mx^n}{(x^n + a^n)^{1 + 1/n}}\right)\,dt^2
      \\
      &+\left(1 - \frac{2mx^n}{(x^n + a^n)^{1 + 1/n}}\right)^{-1}dx^2
      +  (x^n + a^n)^{2/n}\, d\Omega_2^2.
\end{split}
\eeq    
  The constant $m$ is, as before, the \Scw (ADM) mass characterizing the \grav\ field
  in the weak field regions $x \to \pm\infty$.   
\begin{figure*}[!htb]
    \centering
    \includegraphics[width=0.5\linewidth]{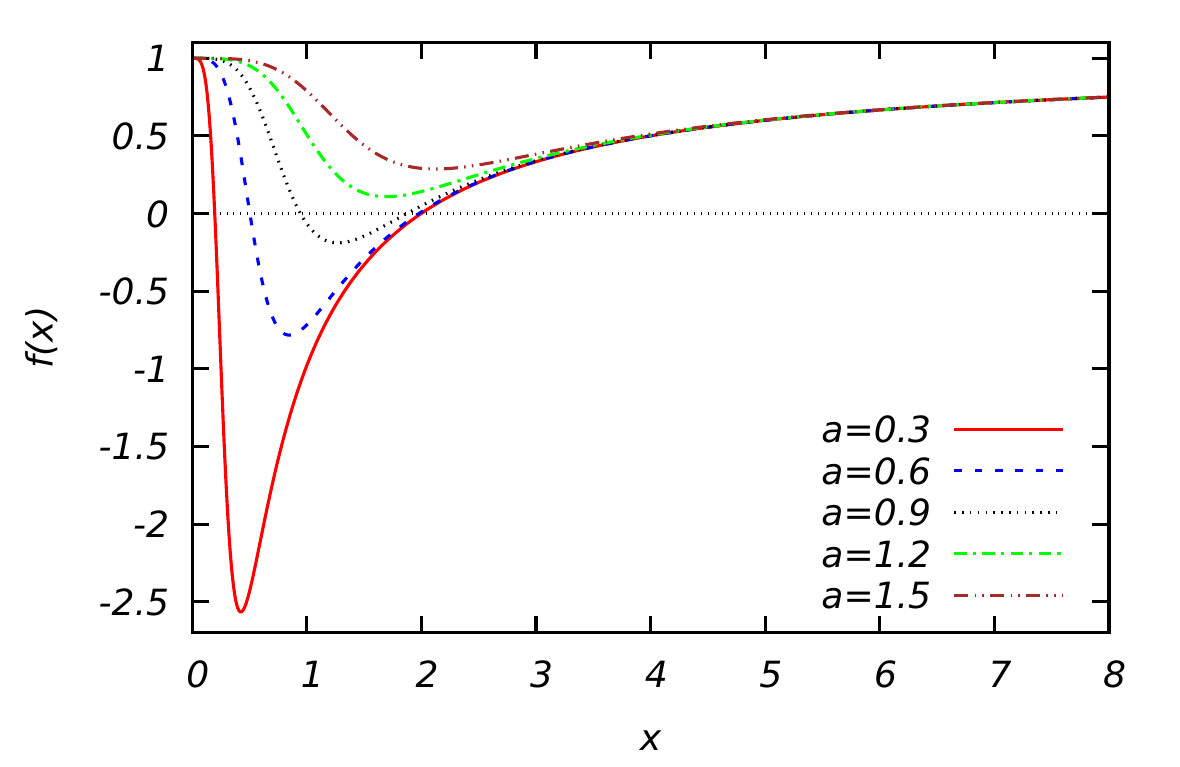}\hspace{-0.1cm}
    \includegraphics[width=0.5\linewidth]{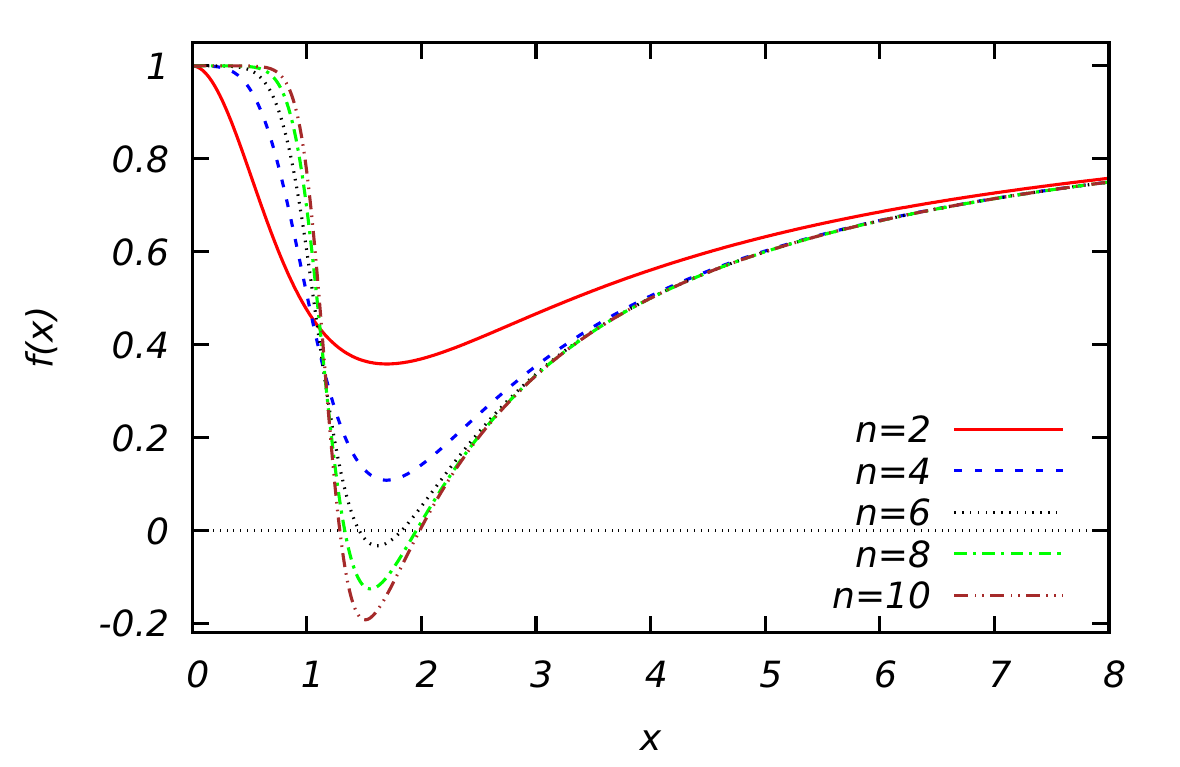}
\caption{Behavior of the function $f(x)$ with $m=1$ as a function of the radial 
    coordinate $x$ for $n=4$ and different values of $a$ (left panel), and for $a=1.2$ 
    and different values of $n$ (right panel).}
    \label{fig:f(r)}
\end{figure*}
\begin{figure}[!htb]
    \centering
    \includegraphics[width=1\linewidth]{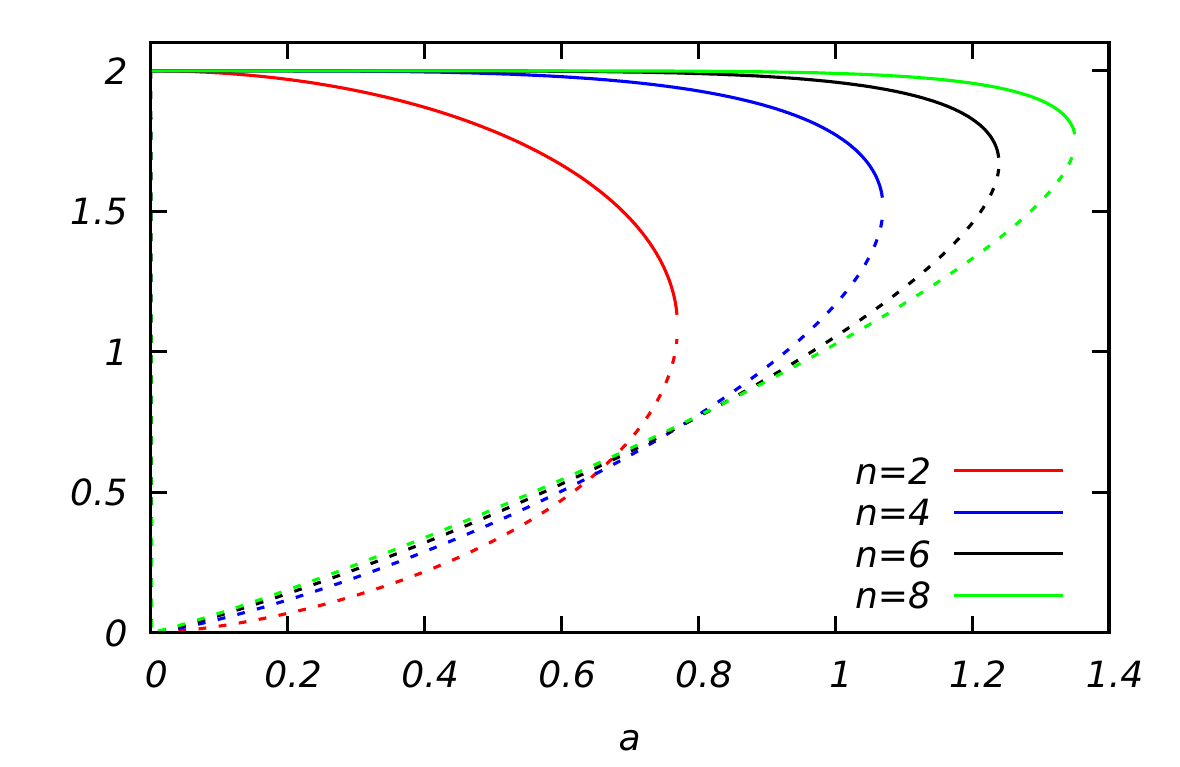}
\caption{Event horizon radius (dashed line) and Cauchy horizon radius (solid line) 
    as functions of $a$ for different values of $n$.}
    \label{fig:horizontes}
\end{figure}

  At $n=2$ this metric reduces to the original ``Bardeen-like black bounce'' of 
  \cite{Lobo:2020ffi}. At $a = 0$, the metric reduces to Schwarzschild's, while $m = 0$ 
  recovers the massless GEB wormhole. At large $|x|$, the geometry is approximately \Scw.

  To study the causal structure of this space-time, we assess the possibility of horizons via
  the condition $f(x)=0$. The behavior of $f(x)$ is depicted in Fig.\,\ref{fig:f(r)}, where we 
  notice two possible zeros of $f(x)$ at $x >0$, corresponding to an event horizon and 
  a Cauchy horizon. As $a$ increases, the points where $f(x)=0$ move closer together
  and merge at some $a = a_{\rm ext}$ at which the two horizons merge, making an 
  extremal horizon. At $a > a_{\rm ext}$, no horizons are observed, and a \wh\
  geometry is obtained. We also observe that, at a fixed value of $a$, the presence of 
  horizons depends on the chosen value of $n$. This is demonstrated in 
  Fig.\,\ref{fig:horizontes}, showing the horizon radii. The value of $a_{\rm ext}$ increases 
  with increasing $n$.
  
As is clear from \eq \rf{BD_GEN} and Fig.\,\ref{fig:f(r)}, the nonstatic regions where 
  $f(x) < 0$ do not contain the sphere $x=0$ at which $r(x)$ reaches its minimum value 
  $r=a$. Actually, $f(0) =1$ in the whole family of models under consideration, 
  inheriting the property of Bardeen's original regular \bh\ metric, \cite{Bardeen:1968pbd}
  in which $r(x) \equiv x$, and the equality $f(0) =1$ provided regularity of the geometry 
  at the center $r=0$. In the presently considered \bhs, the minimum of $r(x)$ is not 
  a black bounce in the precise meaning of the term but actually the throat of a \wh\
  region inside a \bh. So we use the term ``black bounce space-times'' in an extended 
  sense, as explained in the previous section.

\begin{figure*}[!htb]
    \centering
    \includegraphics[width=0.26\linewidth]{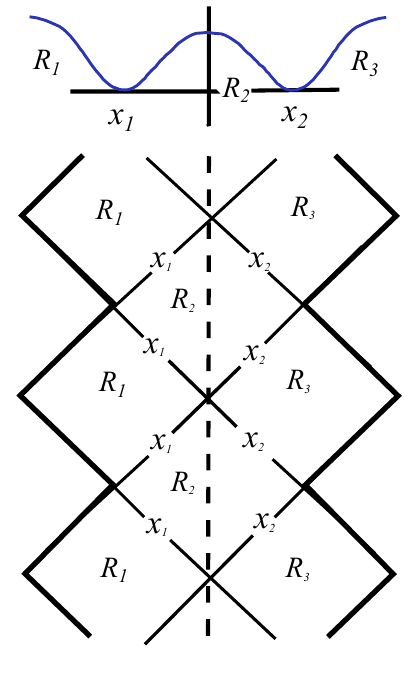}\hspace{1cm}
    \includegraphics[width=0.45\linewidth]{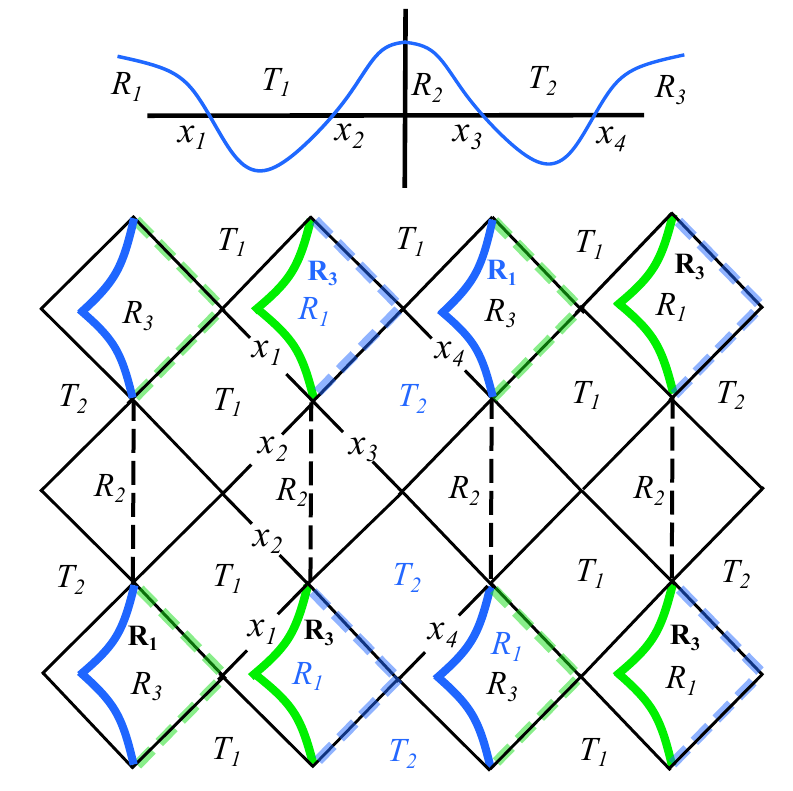}
\caption{Carter-Penrose diagrams for the GBL space-time with two extremal (left) and 
	four simple (right) horizons, the qualitative behavior of $f(x)$ being shown
	at the top with numbered horizons (where $f(x)=0$). All thick lines in the diagrams 
	correspond to $r \to \infty$, thinner tilted inner lines denote horizons. Dashed 
	thick lines mean that the corresponding spatial infinity is placed on a lower, 
	``directly invisible'' layer of the diagram, hidden under an overlapping region. 
	The letters R and T mark static (R) and nonstatic (T) regions with appropriate 
	numbers. Dashed vertical lines depict the throats at $x=0$.
	}
    \label{Penrose}
\end{figure*}

Figures \ref{fig:f(r)} and \ref{fig:horizontes} have been drawn only for $x\geq 0$ 
   since for our symmetric configurations this picture is simply repeated at negative $x$. 
   One should bear in mind, however, that the whole geometry contains two extreme 
   horizons in the case $a = a_{\rm ext}$ and four simple horizons at $a< a_{\rm ext}$. 
   This makes the global causal structure of such space-times more complex than in the
   familiar cases of one extremal or two simple horizons, known for the \RN\ space-time
   and \bhs\ with a regular center. While the Carter-Penrose diagram in the \wh\ case
   is quite simple and reduces to a single causal diamond, the one with two extreme
   horizons for $a = a_{\rm ext}$ occupies an infinitely extended strip, whereas the one 
   with four simple ones at $a< a_{\rm ext}$ covers the whole plane with a countable
   set of overlappings (\cite{Bronnikov:2021uta}, see also \cite{Lobo:2020ffi}), as shown in 
   Fig.\,\ref{Penrose}. 

   As already noted, the global regularity of the geometry follows from smoothness of 
   the metric functions $r(x)>0$ and $f(x)$ (or the mass profile $M(x)$).
   The HMS mass function is easily calculated and is also regular:
\bear
	M_{\rm HMS}(x) &=& \frac{(x^{n}+a^{n})^{1/n}}{2}-\frac{1}{2}\,x^{2n-2}\,
		(x^{n}+a^{n})^{3/n-2}
\nnn \qquad
		+m\,x^{3n-2}\,(x^{n}+a^{n})^{{2}/{n}-3}.
\ear    
  At the throat \(x=0\), one has  $M_{\rm HMS}(0)={a}/{2}$.

\section{Neutral massive particle motion}\label{sec:massive}
A useful way to probe space-time geometry is by analyzing its geodesics. In this section, we perform a detailed analysis of the geodesic motion of massive particles (the massless case is left for the next section due to its importance for shadow simulations), for the GEB space-time as well as for the GBL black bounce space-time.
\subsection{Neutral massive particle motion for GEB wormhole}
 We now study the trajectories of particles in the space-time described by the metric \eqref{metric_EBGx}. 
The Lagrangian for a test particle moving in this space-time is given by

\begin{equation}
\begin{split}
\mathcal{L} &= \tfrac{1}{2}\, g_{\mu\nu}(x)\,\dot{x}^\mu\dot{x}^\nu \\
&= \frac{1}{2}\left(-\dot{t}^2 + \dot{x}^2 + r^2(x)\dot{\theta}^2+ r^2(x)\sin^2\theta\dot{\varphi}^2\right),
\end{split}
\end{equation}
where the dot denotes a derivative with respect to the proper time $\tau$ for massive particles, and a derivative with respect to an affine parameter $\lambda$ for photons.
Because the Lagrangian does not contain the coordinates $t$ and $\varphi$ 
explicitly, the canonical momenta associated with these variables remain 
constant along the motion. Therefore, we can introduce the conserved quantities
\begin{equation}\label{energiamomento}
E \equiv -p_{t} = -\frac{\partial \mathcal{L}}{\partial \dot{t}} = \dot{t},
\quad
\ell \equiv p_{\varphi} 
= \frac{\partial \mathcal{L}}{\partial \dot{\varphi}}
= r^{2}(x)\sin^{2}\theta\,\dot{\varphi}.
\end{equation}

Moreover, the normalization condition for geodesic motion,
\begin{equation}
g_{\mu\nu}\dot{x}^{\mu}\dot{x}^{\nu} = -\delta,
\end{equation}
with $\delta=1$ for timelike particles and $\delta=0$ for null geodesics,
leads to the relation
\begin{equation}
-\dot{t}^{2} + \dot{x}^{2}
+  r^2(x)\dot{\theta}^2+ r^{2}(x)\sin^{2}\theta\,\dot{\varphi}^{2}
= -\delta.
\end{equation}

Using Eq.\eqref{energiamomento}, the above equation can be rewritten for the coordinate $x$ in equatorial plane ($\theta=\pi/2$) as follows:
\begin{equation}\label{radialx}
    E^2 = \dot{x}^2 + V_{eff}(x),
\end{equation}
where
\begin{equation}
    V_{eff}(x) = \frac{\ell^2}{r(x)^2}+ \delta.
\end{equation}

Given the potential for massive particles ($\delta=1$), we can study the motion of these particles in this space-time as a one-dimensional problem, since the potential depends only on the coordinate $x$. The potential is given by
\begin{equation}
    V_{part}(x) =\frac{\ell^2}{r(x)^2} + 1 =\frac{\ell^2}{(x^n + a^n)^{2/n}} + 1.
\end{equation}

\begin{figure}[!htb]
    \centering
    \includegraphics[width=1.\linewidth]{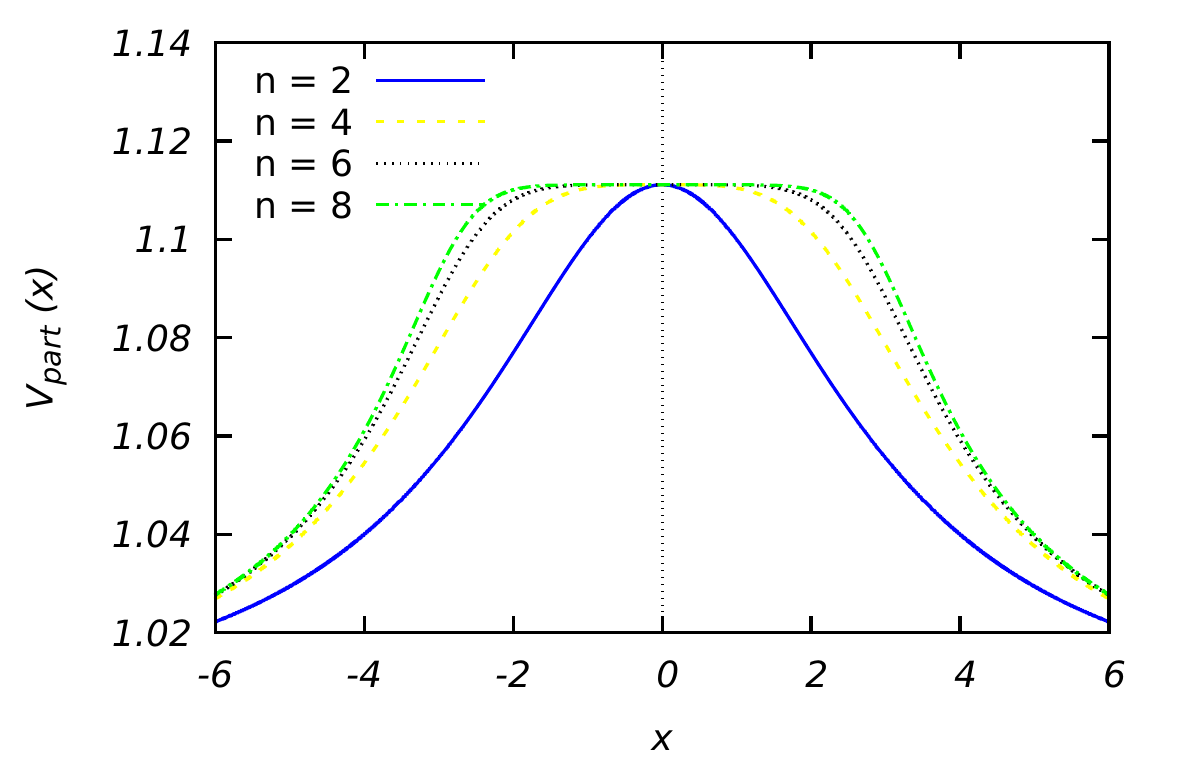}
    \caption{Effective potential for different values of $n$, fixing $\ell=1$ and $a=3$.}
    \label{fig:Vpartx}
\end{figure}

We will therefore consider the motion of neutral massive particles coming from infinity, with a given angular momentum and energy, towards the wormhole. In Fig. \ref{fig:Vpartx} we present the plot of the effective potential for massive particles, fixing $\ell=1$ and $a=3$ for different values of $n$. It is interesting to note that, as the parameter 
$n$ increases, the potential increasingly exhibits a shape similar to that of a square potential barrier.

Given the shape of the potential, it is straightforward to see that it has a maximum located at $x = 0$ ($r=a$), that is, at the throat of the wormhole. This point corresponds to an unstable circular orbit for these particles, so in this case there is a critical value of the angular momentum $\ell$ at which this occurs, given by $E^2 = V_{part}(x=0)$. For our case, this value is given by $\ell_c = \sqrt{E^2 -1}a$.

We can then classify the particle orbits into three types:
\begin{itemize}
    \item $E^2 < V_{part}(x=0)$, or $\ell > \ell_c$, the particle is unable to cross the throat and remains in the same universe;

    \item $E^2 = V_{part}(x=0)$, or $\ell = \ell_c$, the particle becomes trapped at the throat, executing an unstable circular orbit;

    \item $E^2 > V_{part}(x=0)$, $\ell < \ell_c $, the particle is able to cross the throat, proceeding towards infinity in the other universe.
\end{itemize}

To understand what actually happens, we need the equation that describes the trajectory of the particles in this geometry. Thus, by combining Eqs. \eqref{energiamomento} and \eqref{radialx}, we obtain a relation for $d\varphi/dx$
\begin{equation}
    \frac{d\varphi}{dx} = \mp \frac{\ell}{r(x)^2\sqrt{\left(E^2 - 1 - \frac{\ell^2}{r(x)^2}\right)}},
\end{equation}
where $\mp$ stand for ingoing/outgoing geodesics,
respectively. We will assume that the particle starts from position $x_i$ with an initial azimuthal angle $\varphi_i$, in such a way that the trajectory of the ingoing particles is given by the integral
\begin{equation}
    \varphi_{in}(x) = \varphi_i - \int_{x_i}^{x}\frac{\ell d\tilde{x}}{r(\tilde{x})^2\sqrt{\left(E^2 - 1 - \frac{\ell^2}{r(\tilde{x})^2}\right)}},  
\end{equation}
or
\begin{equation}
    \varphi_{in}(r) = \varphi_i - \int_{r_i}^{r}\frac{\ell d\tilde{r}\mathcal{J}(\tilde{r})}{\tilde{r}^2\sqrt{\left(E^2 - 1 - \frac{\ell^2}{\tilde{r}^2}\right)}},  
\end{equation}
where it is evaluated between the points $(r_i,\varphi_i)$ and $(r_{min},\varphi_{in}(r_{min}))$ and $\mathcal{J}$ is the Jacobian of the coordinate transformation. For the outgoing geodesics, we consider the initial condition given by $\varphi'_i=\varphi_{in}(r_{min})$, which leads to
\begin{equation}
       \varphi_{out}(r) = \varphi'_i + \int_{r_{min}}^{r}\frac{\ell d\tilde{r}\mathcal{J}(\tilde{r})}{\tilde{r}^2\sqrt{\left(E^2 - 1 - \frac{\ell^2}{\tilde{r}^2}\right)}} .
\end{equation}

In the critical case $\ell=\ell_c$, it is necessary to consider only one integral \cite{Rueda:2023val}
\begin{equation}
     \varphi(r) = \varphi_i - \int_{r_i}^{r}\frac{\ell d\tilde{r}\mathcal{J}(\tilde{r})}{\tilde{r}^2\sqrt{\left(E^2 - 1 - \frac{\ell^2}{\tilde{r}^2}\right)}}  .
\end{equation}

\begin{figure*}[!htb]
    \centering
    \includegraphics[width=0.99\linewidth]{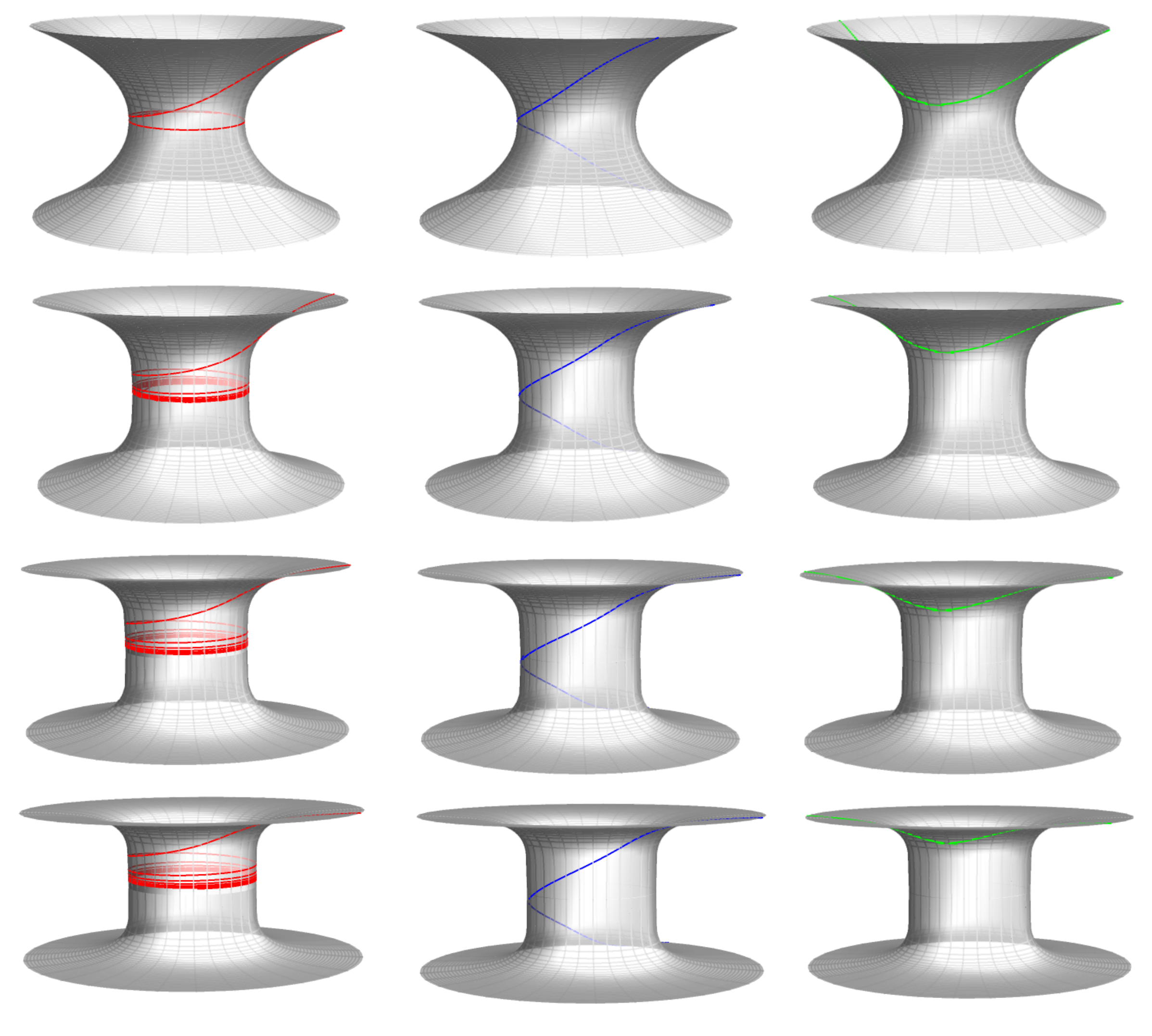}
    \caption{Timelike geodesics for $\ell = \ell_c$
  (red), $\ell = \ell_c - \epsilon$ (blue), and $\ell=\ell_c + \epsilon$ (green), with fixed parameters $a=3$, $E=1.2$, and $\epsilon=0.2$, for values of $n$ given by $n=2$ (first row), $n=4$ (second row), $n=6$(third row), and $n=8$ (fourth row).}
    \label{fig:timelike geodesics}
\end{figure*}

By numerically integrating the above equations and plotting the resulting geodesic on the corresponding embedding diagram, we obtain the result shown in Fig. \ref{fig:timelike geodesics} for different parameter values. In the image, the red lines correspond to timelike geodesics of massive particles that approach the wormhole throat and become trapped, executing an unstable circular orbital motion. The blue lines, in turn, represent timelike geodesics of massive particles that succeed in crossing the throat, thereby reaching the other universe. Finally, the green lines correspond to timelike geodesics of massive particles that approach the throat but are deflected, remaining within the same universe. The fact that for this geometry, the only unstable circular orbit is located exclusively at the throat and there is no stable orbit is of paramount importance for our purposes, and its implications for the modeling of accretion disks will be discussed in more detail in the following sections.

\subsection{Neutral massive particle motion for GBL black bounce}
Here we proceed analogously to the previous case. The Lagrangian for a particle moving in the space-time described by the line element 
\eqref{bbgeneral} is written as
\begin{equation}
    \mathcal{L}= \frac{1}{2}\left(-f\dot{t}^2 + \frac{\dot{x}^2}{f} + r(x)^2\dot{\theta}^2 + r(x)^2\sin^2\theta\dot{\varphi}^2\right),
\end{equation}
where the dot denotes a derivative with respect to the proper time $\tau$ for massive particles, and a derivative with respect to an affine parameter $\lambda$ for photons.

Since this Lagrangian does not depend explicitly on the coordinates 
$\varphi$ and $t$, the conjugate momenta associated with these 
coordinates are conserved. This allows us to define the following 
conserved quantities
\begin{equation}\label{quant_cons}
E \equiv - p_{t} = -\frac{\partial \mathcal{L}}{\partial \dot{t}} = f\dot{t},\;\;\;\ell \equiv p_{\varphi} = \frac{\partial \mathcal{L}}{\partial \dot{\varphi}} = r(x)^2\sin^2\theta\dot{\varphi}.
\end{equation}

In addition, the geodesic motion implies the condition
\begin{equation}
-f\dot{t}^2 + \frac{\dot{x}^2}{f} + r(x)^2\dot{\theta}^2 + r(x)^2\sin^2\theta\dot{\varphi}^2 = -\delta,
\end{equation}
where $\delta = 0$ for photons and $\delta = 1$ for massive particles.
We can again assume, without loss of generality, that the motion takes
place in the equatorial plane ($\theta = \pi/2$), so that the equation above can be rewritten as
\begin{equation}\label{cons_BDG}
E^2 = \dot{x}^2 + V_{eff}(x),
\end{equation}
where
\begin{equation}
    V_{eff}(x) = f(x)\left(\frac{\ell^2}{r(x)^2} + \delta\right).
\end{equation}

We can compute the radial acceleration felt by these particles through the relation
\begin{equation}
    \Ddot{x}=-\frac{1}{2}\frac{dV_{eff}(x)}{dx}.
\end{equation}

\begin{figure*}
    \centering
    \subfigure[]{\includegraphics[width=.5\linewidth]{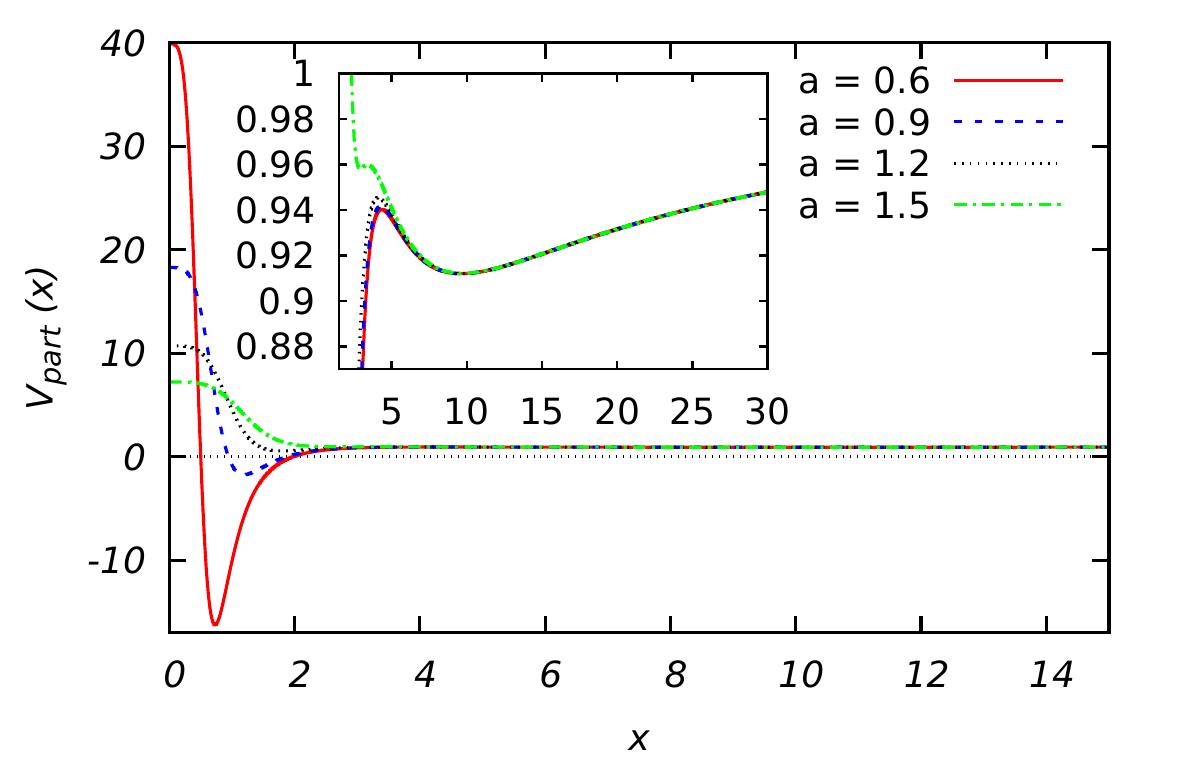}}\hspace{-0.1cm}
    \subfigure[]{\includegraphics[width=.5\linewidth]{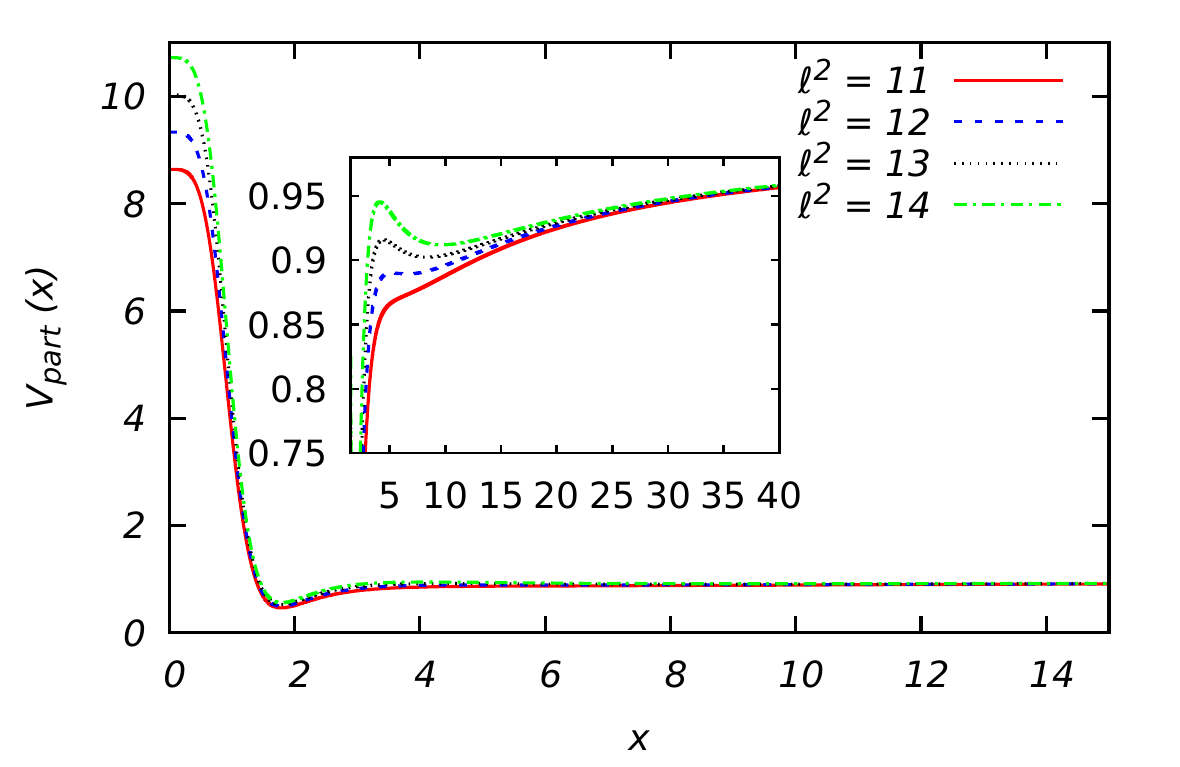}}
    \subfigure[]{\includegraphics[width=.5\linewidth]{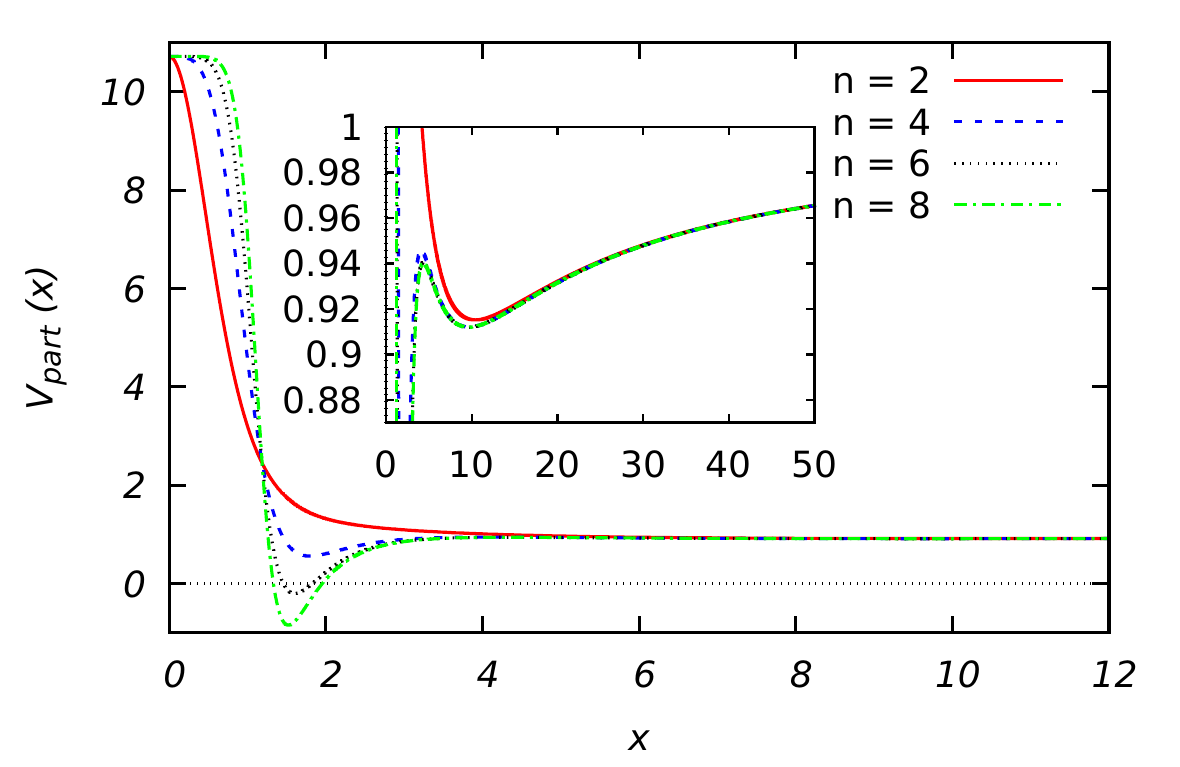}}
    \caption{Behavior of the effective potential for massive particles in the space-time \eqref{BD_GEN}, with $m=1$. In panel (a), we fix $\ell^{2}=14$ and $n=4$ for different values of $a$. In panel (b), we fix $n=4$ and $a=1.2$ for different values of $\ell^{2}$. In panel (c), we fix $a=1.2$ and $\ell^{2}=14$ for different values of $n$. Although the plots are restricted to $x>0$, the effective potential is symmetric under $x \to -x$.}
    \label{fig:V_m_BD}
\end{figure*}

For massive particles ($\delta=1$), the effective potential is given by
\begin{equation}
\begin{split}
    V_{part}=&f(x)\left(\frac{\ell^2}{r(x)^2}+1\right)\\=&\left(1-\frac{2mx^n}{\left(x^n+a^n\right)^{1+1/n}}\right)\left(\frac{\ell^2}{\left(x^n+a^n\right)^{2/n}}+1\right).
\end{split}
\end{equation}
When $a < a_{ext}$, the potential vanishes at the horizons ($f(x)=0$). We also have the following limits:
\begin{equation}
    \lim_{x \to \infty} V(x) = 1, \quad \mbox{and} \quad \lim_{x \to 0} V(x) =1+\frac{\ell^2}{a^2}.
\end{equation}
In Fig.~\ref{fig:V_m_BD}, we show the behavior of the effective potential for different parameter values. Depending on the choice of parameters, there can be up to four critical points in the effective potential. We observe that $x=0$ is always a maximum. When horizons are present, the effective potential is negative between the horizons; the maximum at $x=0$ and a minimum of the potential are hidden behind the horizon. When no horizons exist, these points may still be present. We also see that at $x=0$ the potential takes different values for different $\ell$ and $a$, and it does not depend on $n$, as verified by the limiting behavior at that point. Recall that maxima of the effective potential correspond to unstable circular orbits for massive particles, whereas minima correspond to stable circular orbits.

\begin{figure*}[htb]
    \centering
    \includegraphics[width=0.36\linewidth]{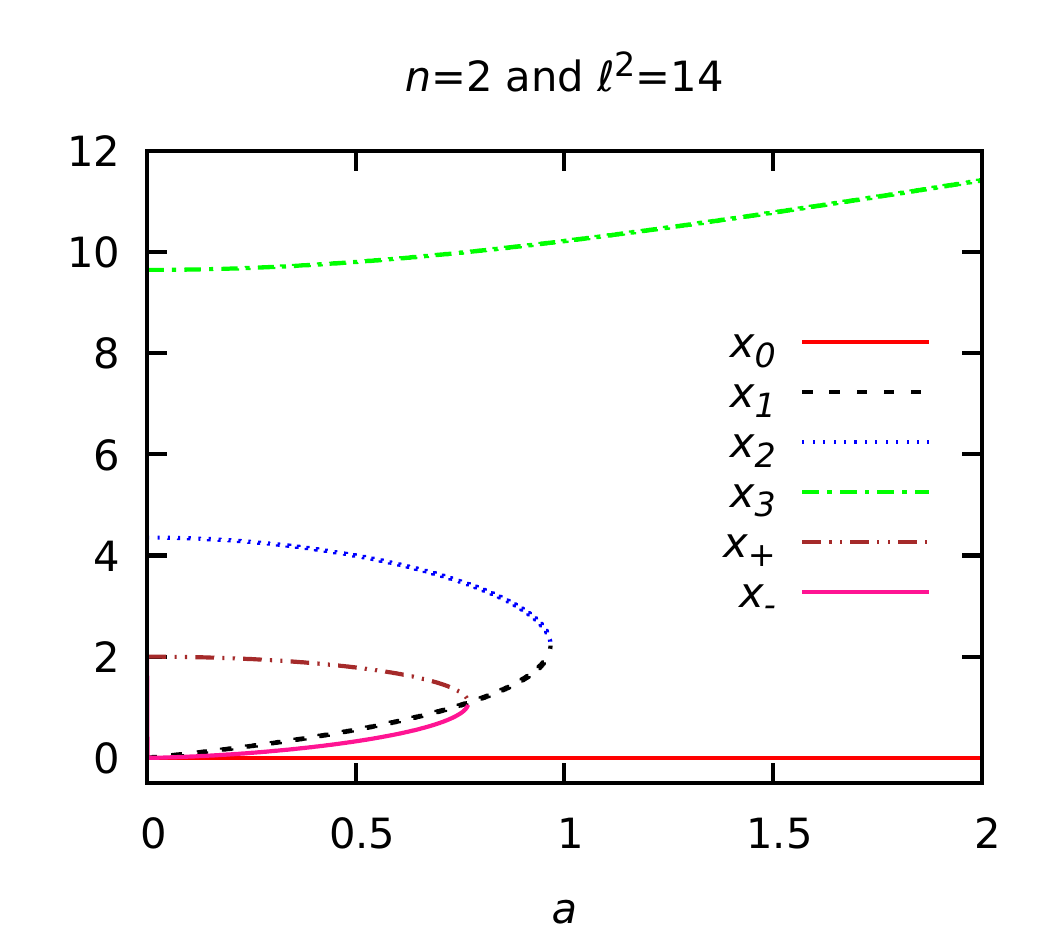}\hspace{-0.9cm}
    \includegraphics[width=0.36\linewidth]{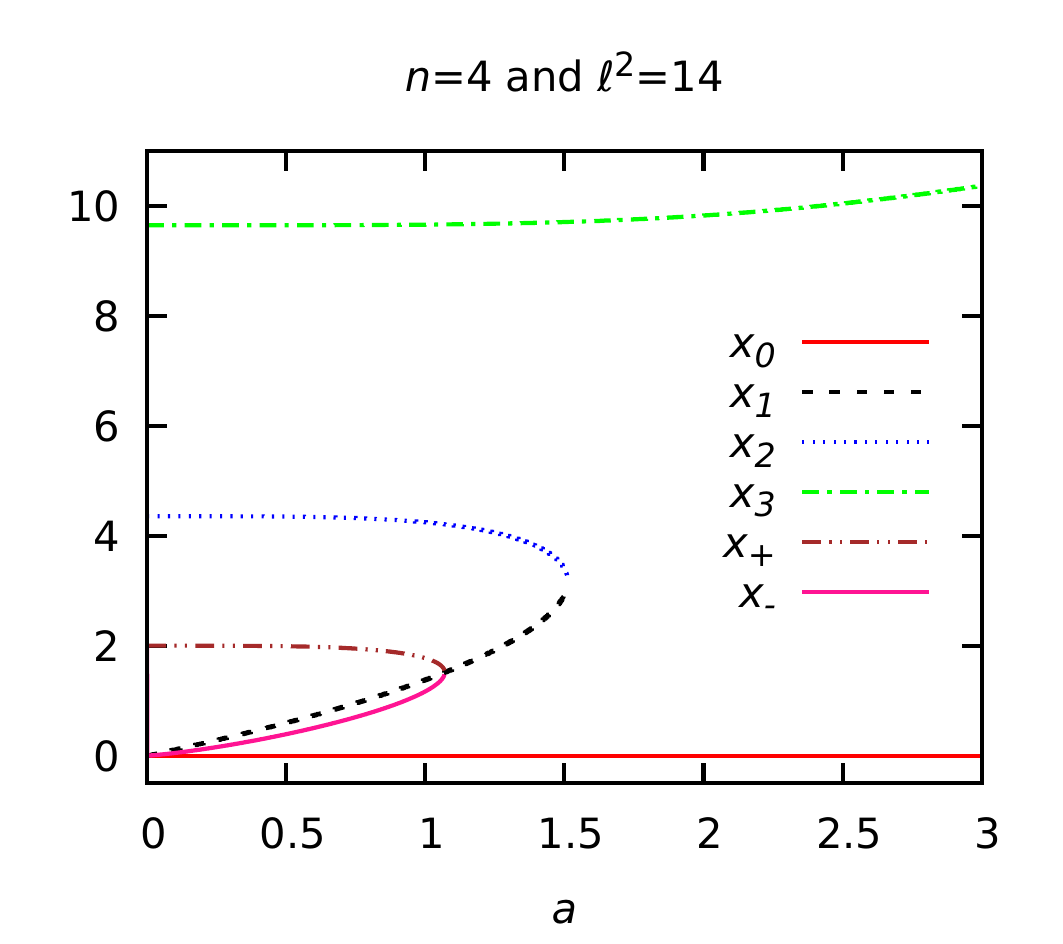}\hspace{-0.9cm}
    \includegraphics[width=0.36\linewidth]{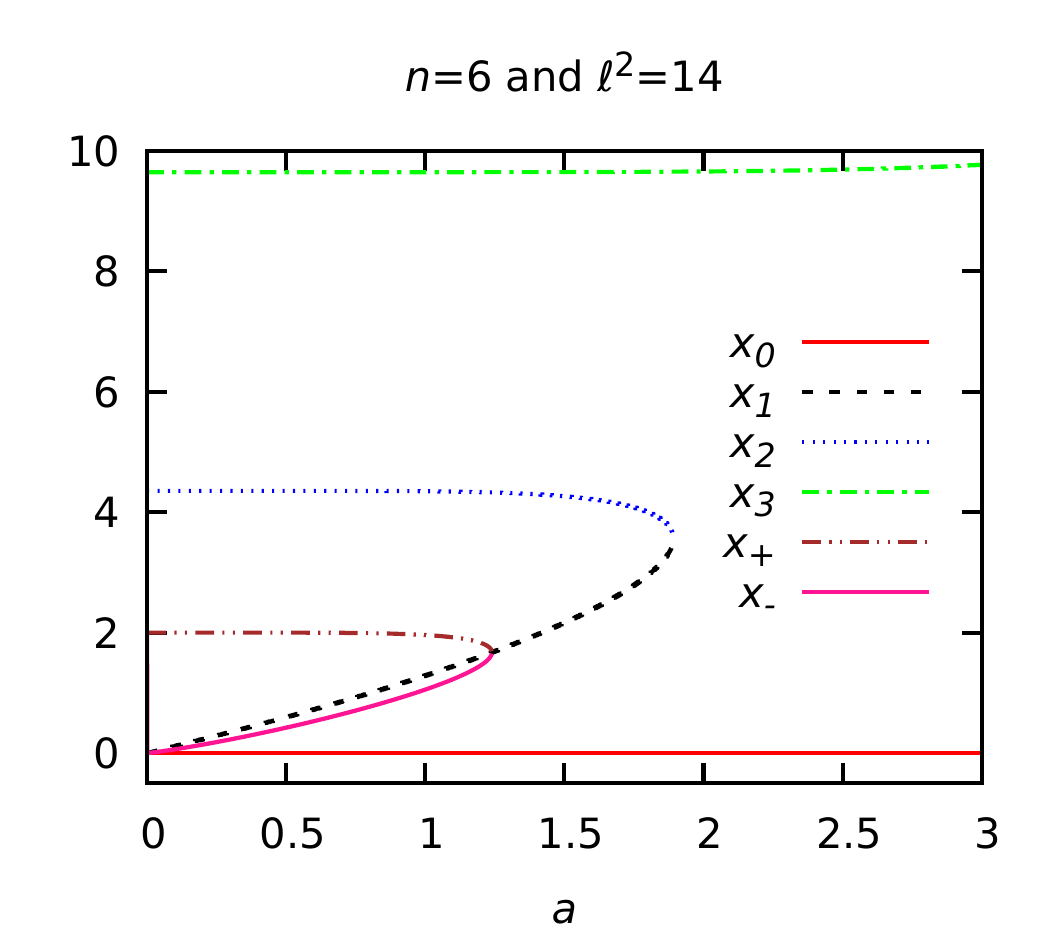}
    \includegraphics[width=0.36\linewidth]{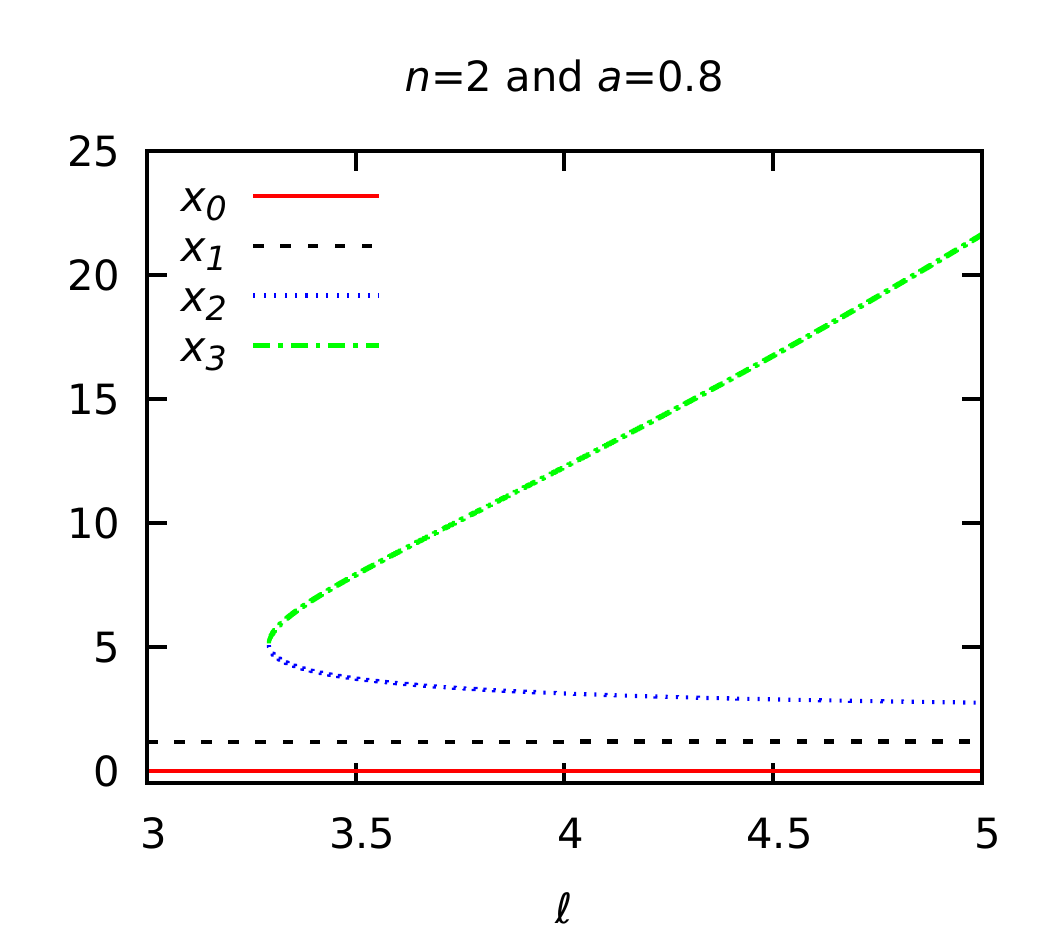}\hspace{-0.9cm}
    \includegraphics[width=0.36\linewidth]{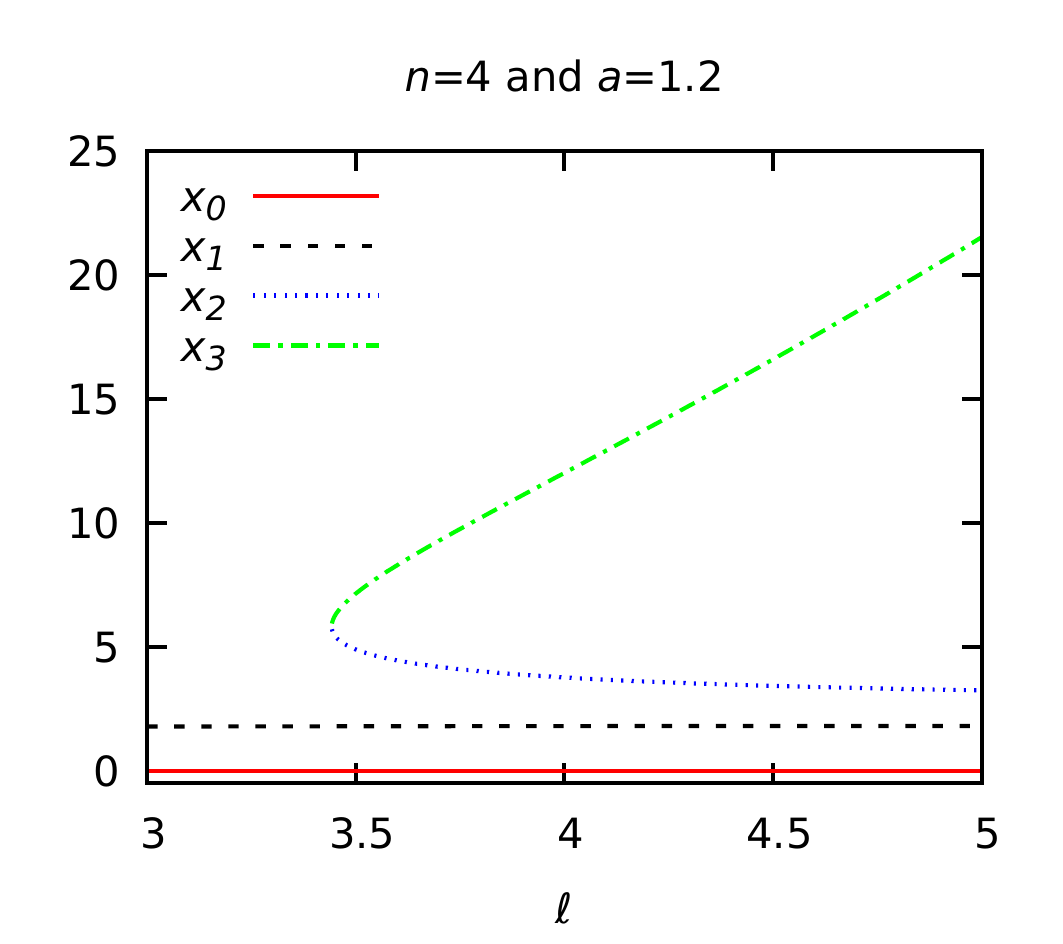}\hspace{-0.9cm}
    \includegraphics[width=0.36\linewidth]{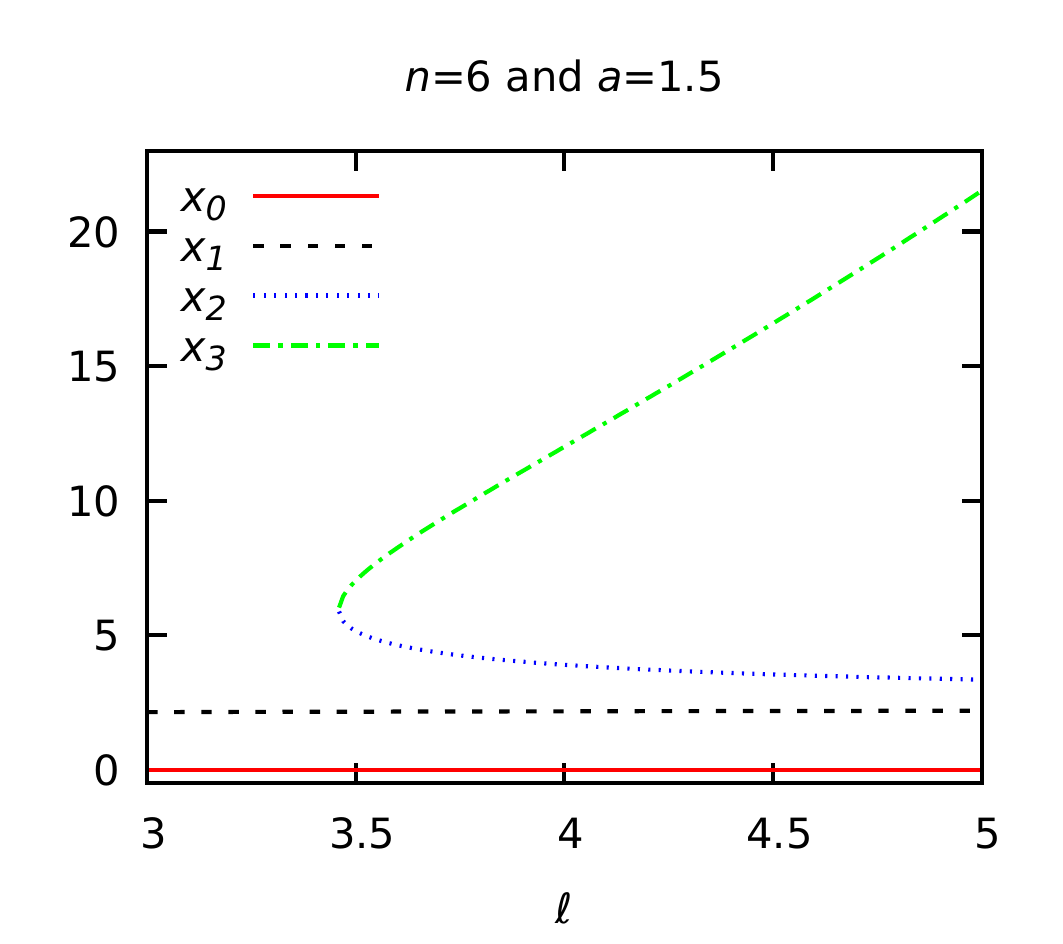}
    \caption{Behavior of the circular orbit radius for massive particles in the space-time \eqref{BD_GEN} with $m=1$. We denote $x_{+}$ as the event horizon radius, $x_{-}$ as the Cauchy horizon radius, $x_{0}$ as the radius of the inner unstable circular orbit, $x_{1}$ as that of the inner stable circular orbit, $x_{2}$ as that of the outer unstable circular orbit, and $x_{3}$ as that of the outer stable circular orbit. In the top set of panels, we fix $\ell^{2}=14$ and vary the parameter $a$ for different values of $n$. In the bottom set, we vary $\ell$ for different values of $n$ and $a$. The value of $a$ in the bottom set is chosen to ensure the existence of circular orbits (at least over a certain range of $\ell$) and the absence of horizons.}
    \label{fig:radius_massive_BD}
\end{figure*}

For circular orbits, we have $\dot{x}=\ddot{x}=0$. With these conditions, the angular momentum and energy corresponding to each circular orbit are given by
\begin{equation}
\begin{split}
    \ell^2_c=\frac{r (x_c)^3 f'(x_c)}{2 f(x_c) r '(x_c)-r (x_c) f'(x_c)}, \\ E_c^2=\frac{2 f(x_c)^2 r '(x_c)}{2 f(x_c) r '(x_c)-r (x_c) f'(x_c)},
\end{split}
\end{equation}
where $x_c$ is the radius of the circular orbits, whether stable or unstable. Due to the complexity of our model, it is not feasible to analyze the behavior of the circular orbit radius analytically. Thus, using Fig.~\ref{fig:radius_massive_BD}, we can verify graphically how these radii behave as we vary the model parameters. In general, circular orbits exist for $x \ge 0$. The point $x=0$ is always an unstable circular orbit, as it corresponds to a maximum of the potential, and the radius of this orbit does not change. For values of $a$ for which horizons exist, the orbits at $x=0$ and at $x_1$ are hidden inside the horizons. Even in the horizonless case, $x_1$ may still exist over a certain range of $a$. As $a$ increases, $x_1$ and $x_2$ approach the same value and, beyond that, only $x_0$ and $x_3$ remain. We also note that the larger the value of $n$, the more slowly these radii vary as $a$ increases. Thus, we note that both the horizons and $x_1$ and $x_2$ persist over a larger range of $a$ as the value of $n$ increases. For the case where we vary $\ell$, it is clear that $x_0=0$ will not change, while $x_1$ is nearly constant. The orbits most affected by changes in $\ell$ are $x_2$ and $x_3$, for which there exists a minimum value of $\ell$ such that $x_2=x_3=x_{\text{ISCO}}$ and $\ell=\ell_{\text{ISCO}}$. This value of $x$ is commonly known as the radius of the innermost stable circular orbit (ISCO). For $\ell<\ell_{\text{ISCO}}$, the orbits $x_2$ and $x_3$ cease to exist, leaving only $x_0$ and $x_1$.

\begin{figure}
    \centering
    \includegraphics[width=1\linewidth]{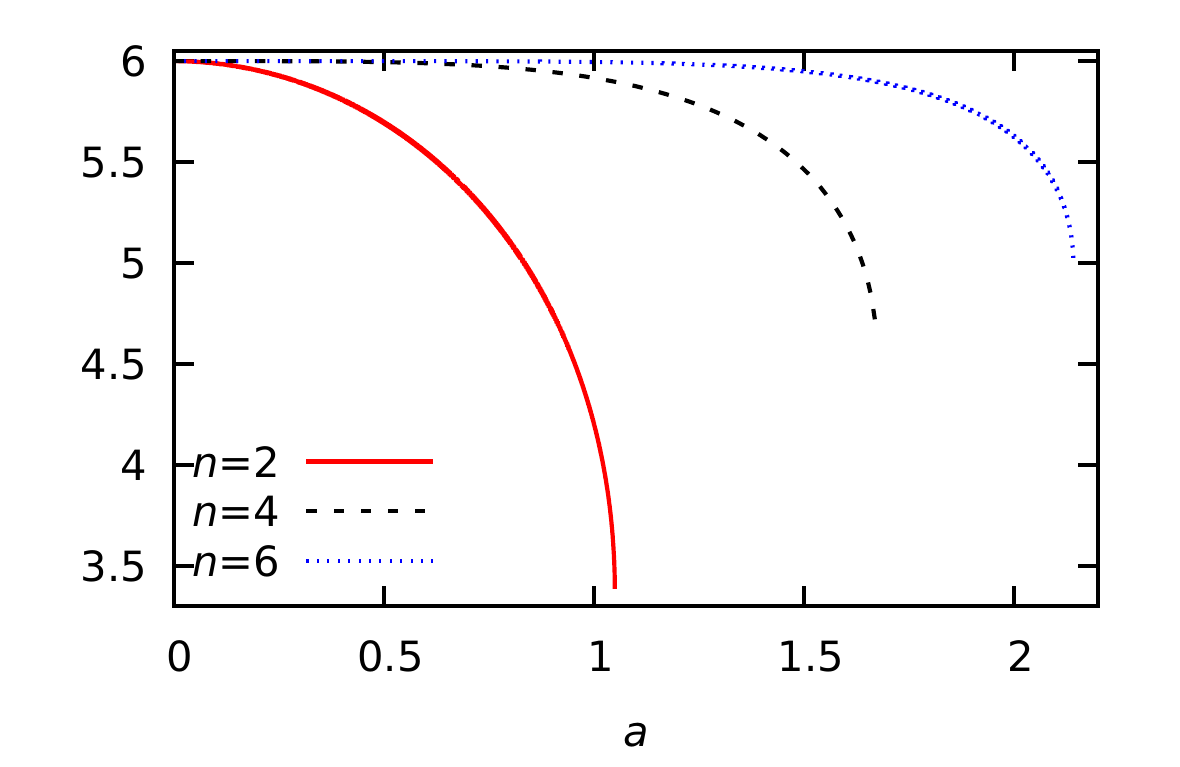}
    \caption{Behavior of the ISCO radius as $a$ is varied for different values of $n$ and $m=1$.}
    \label{fig:x_isco}
\end{figure}

The location of the ISCO is crucial in the study of astrophysical black holes, especially for problems involving the accretion disk that may exist around these space-times. As $\ell \to \ell_{\text{ISCO}}$, the radii $x_2$ and $x_3$ approach the same point, and the minimum and maximum of the effective potential coalesce into an inflection point. In this situation, the ISCO radius can be obtained from the condition $V''(x_{ISCO},\ell_c,E_c)=0$ or
\begin{equation}\label{isco_condition}
   -2 f r  f'' r '+2 f r f' r ''+4 r  f'^2 r '-6 f f' r '^2=0.
\end{equation}
In Fig.~\ref{fig:x_isco}, we display the behavior of the ISCO radius as the parameter $a$ varies for different values of $n$. As $a \to 0$, we recover $x_{\text{ISCO}}=6$, which is the Schwarzschild result. The larger $a$ becomes, the smaller $x_{\text{ISCO}}$ is, and there is a maximum value of $a$ for which $x_{\text{ISCO}}$ still exists. For larger $n$, one can sustain larger $a$ while keeping $x_{\text{ISCO}}$; conversely, the smaller $n$ is, the smaller $x_{\text{ISCO}}$ tends to be as $a$ increases.

In contrast to the simple wormhole case, the effective potential of massive particles in the GBL black bounce may exhibit multiple critical points. Besides the extremum located at the throat $x=0$ ($r=a$), additional extrema may appear at finite radii, corresponding to outer stable circular orbits and
an intermediate unstable one.
As a consequence, the particle motion cannot be characterized by a single
critical angular momentum or energy. Depending on the values of $E$ and $\ell$, particles may become trapped either
near the throat or around the outer stable/unstable orbit, or cross the throat and reach the other asymptotic region in the wormhole case.

In the previous case, the only existing orbit was located at the wormhole throat, which hosted an unstable circular orbit for massive particles. In that scenario, the geodesic trajectories were distinguished according to the value of the critical angular momentum.

In the present case, however, the effective potential for massive particles exhibits a much richer structure, displaying multiple local maxima and minima depending on the values of the model parameters. In order to properly understand the nature of geodesic motion in this space-time, we focus on the wormhole case ($a > a_{ext}$) and the situation in which the effective potential possesses four extrema: two corresponding to unstable circular orbits (one of which remains located at the throat) and two associated with stable circular orbits.

Accordingly, we fix the value of the angular momentum of the particles and adjust their energy such that the condition for a circular orbit is satisfied, namely $E_c^2 = V_{\text{part}}(x_c)$, where $x_c$ denotes the position of the circular orbit and the condition $\ddot{x}=0$ is satisfied. Since there also exists an unstable circular orbit at $x=0$, this location satisfies ${E_c^{(0)}}^2 = V_{\text{part}}(x=0)$.

For particles launched from an initial position $x_i$, the stable circular orbits correspond to confined configurations and therefore do not arise dynamically from generic initial conditions. As a consequence, the physically relevant trajectories are governed by the unstable extrema of the effective potential. In this case, the geodesic motion may exhibit four qualitatively distinct behaviors:

Depending on the relation between the particle energy and the local extrema of the effective potential, four distinct dynamical regimes arise. Here, $x_c$ denotes the position of the unstable circular orbit located away from the throat. Specifically,
\begin{itemize}
    \item $E_c < E < E_c^{(0)}$: the particle is reflected by the potential barrier associated with the throat and is scattered back to the same asymptotically flat region from which it was launched.
    
    \item $E = E_c< E_c^{(0)}$: the particle asymptotically approaches the unstable circular orbit located at $x = x_c$, corresponding to a critical trajectory separating scattering and transmission regimes.
    
    \item $E = E_c^{(0)}>E_c$: the particle asymptotically approaches the unstable circular orbit located at the wormhole throat, $x = 0$, which acts as a separatrix between reflection and throat-crossing trajectories.
    
    \item $E > E_c^{(0)}>E_c$: the particle overcomes both potential barriers and crosses the wormhole throat, reaching the opposite asymptotically flat region of the space-time.
\end{itemize}

As in the GEB case, we can combine Eqs. \eqref{quant_cons} and \eqref{cons_BDG}, to obtain the relation $d\varphi/dx$, which is given by
\begin{equation}
    \frac{d\varphi}{dx} = \mp \frac{\ell}{r(x)^2\sqrt{E^2 - f(x)\left(1 + \frac{\ell^2}{r(x)^2}\right)}}.
\end{equation}

In analogy with the GEB case, we assume that the particle is launched from the position $x_i$ with initial azimuthal angle $\varphi_i$, such that the ingoing particle trajectories are described by the integral
\begin{equation}
    \varphi_{in}(x) = \varphi_i - \int_{x_i}^x \frac{\ell d\tilde{x}}{r(\tilde{x})^2\sqrt{E^2 - f(\tilde{x})\left(1 + \frac{\ell^2}{r(\tilde{x})^2}\right)}},
\end{equation}
or
\begin{equation}
    \varphi_{in}(r) = \varphi_i - \int_{r_i}^r \frac{\ell d\tilde{r}\mathcal{J}(\tilde{r})}{\tilde{r}^2\sqrt{E^2 - f(\tilde{r})\left(1 + \frac{\ell^2}{\tilde{r}^2}\right)}},
\end{equation}
where it is again evaluated between the points $(r_i,\varphi_i)$ and $(r_{\min},\varphi_{\min})$. For the outgoing geodesics, we consider again the initial condition $\varphi_i'=\varphi_{\min}$, which leads to
\begin{equation}
    \varphi_{out}(r) = \varphi_i' + \int_{r_{min}}^r \frac{\ell d\tilde{r}\mathcal{J}(\tilde{r})}{\tilde{r}^2\sqrt{E^2 - f(\tilde{r})\left(1 + \frac{\ell^2}{\tilde{r}^2}\right)}}.
\end{equation}

In both critical cases, when $E = E_c$ or $E = E_c^{(0)}$, only a single integration is performed
\begin{equation}
     \varphi(r) = \varphi_i - \int_{r_i}^r \frac{\ell d\tilde{r}\mathcal{J}(\tilde{r})}{\tilde{r}^2\sqrt{E^2 - f(\tilde{r})\left(1 + \frac{\ell^2}{\tilde{r}^2}\right)}}.
\end{equation}

\begin{figure*}[!htb]
    \centering
    \includegraphics[width=.373\linewidth]{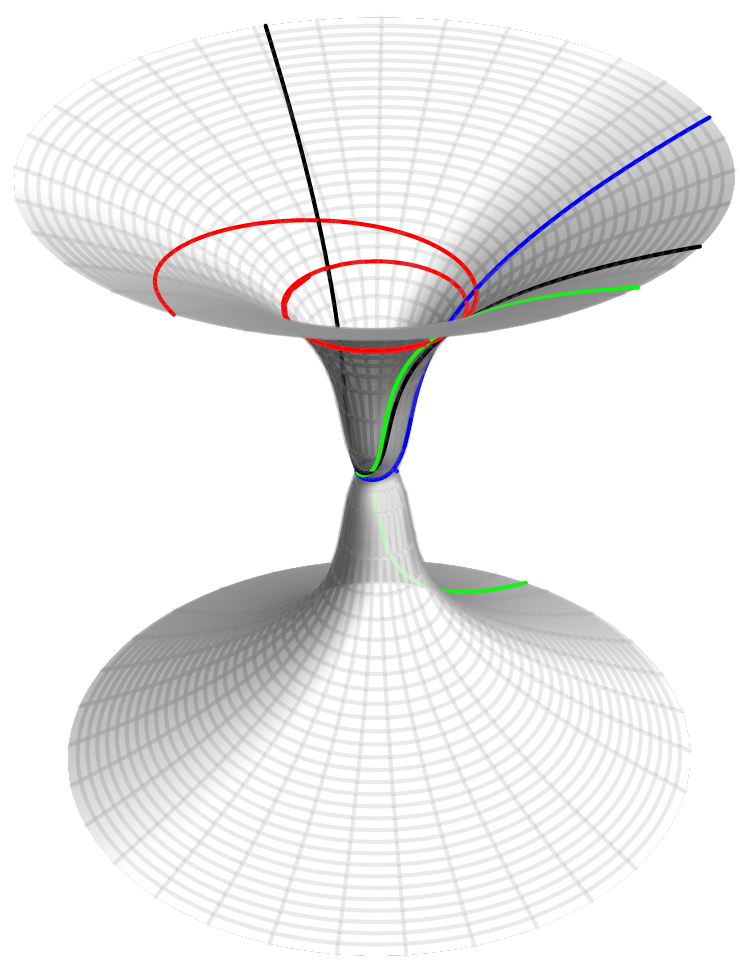}\includegraphics[width=.5\linewidth]{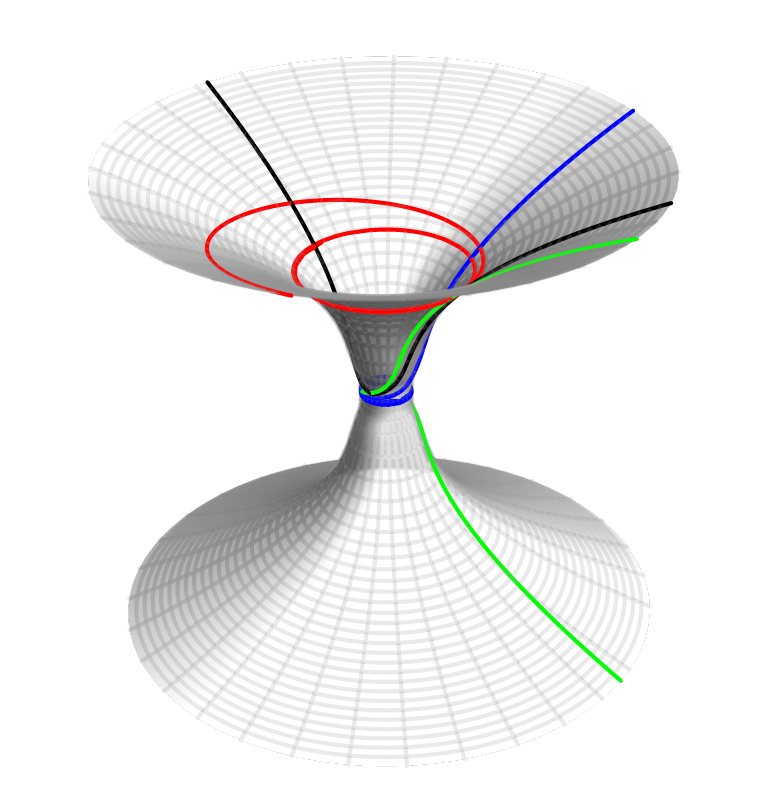}

    \includegraphics[width=.5\linewidth]{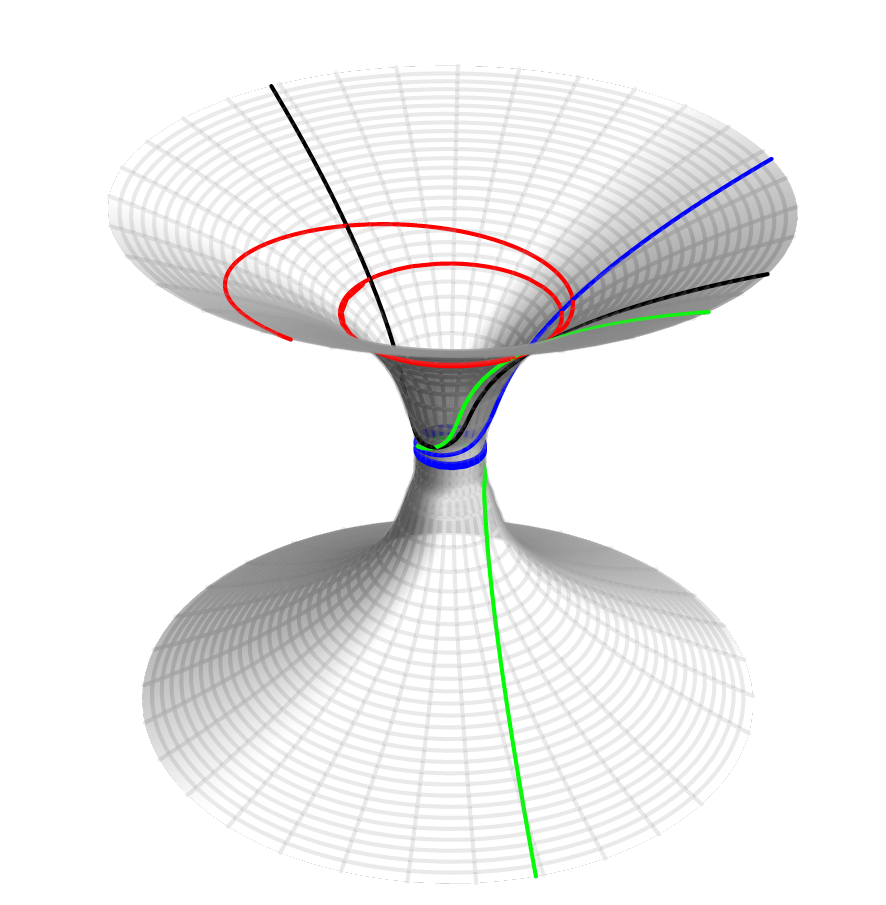}\includegraphics[width=.5\linewidth]{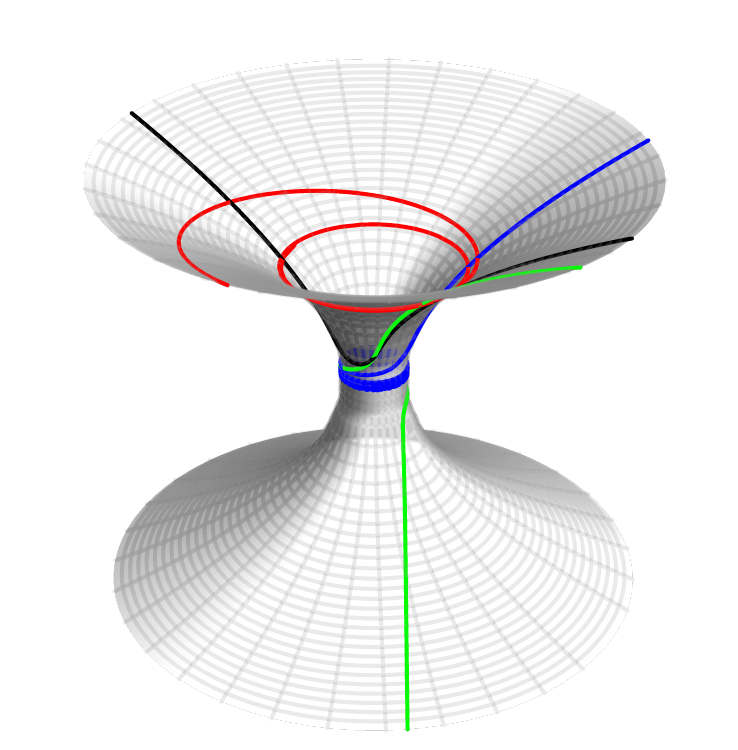}
    \caption{Timelike geodesics projected onto the embedding diagram for fixed angular momentum $\ell^2=14$ and $m=1$. From left to right and top to bottom, the panels correspond to $n=2$ ($a=0.8$), $n=4$ ($a=1.2$), $n=6$ ($a=1.4$), and $n=8$ ($a=1.6$), respectively.
}
    \label{fig:massivebb}
\end{figure*}

By numerically integrating the above equations and projecting the resulting geodesics onto the corresponding embedding diagram, we obtain the configurations displayed in Fig.~\ref{fig:massivebb} for different parameter values. The red curves represent unstable circular orbits located away from the throat, while the blue curve corresponds to the unstable circular orbit at the throat. The green trajectories describe particles that cross the wormhole throat and reach the opposite asymptotically flat region, whereas the black trajectories correspond to scattering orbits that are deflected and return to the same asymptotic region.

The main difference between this case and the GEB scenario lies in the existence of circular orbits located away from the throat. These additional orbits substantially enrich the structure of the effective potential and introduce new critical energies and separatrices that govern the timelike geodesic motion, allowing for scattering, asymptotic trapping, and throat-crossing trajectories. This richer dynamical structure has direct consequences for the optical appearance of the space-time, since the presence of unstable circular orbits outside the throat modifies the lensing properties and the propagation of matter and radiation emitted by accretion disks. In particular, these external orbits can enhance or suppress specific photon and particle trajectories, altering the brightness distribution, the shape of the observed images, and the formation of characteristic features such as rings or sharp intensity transitions when compared to the GEB case.

\section{Theoretical setup for shadow simulation}\label{massless}
We shall now analyze the photon trajectories and their implications for the shadow simulations of the compact object.
\subsection{Photon motion and ray-tracing procedure for GEB}
For photons ($\delta=0)$, it is clear that the effective potential for the massless case closely resembles that of massive particles. The potential is given by
\begin{equation}
    V_{ph}(x) =\frac{\ell^2}{r(x)^2}= \frac{\ell^2}{(x^n + a^n)^{2/n}}.
\end{equation}

\begin{figure}[!htb]
    \centering
    \includegraphics[width=1\linewidth]{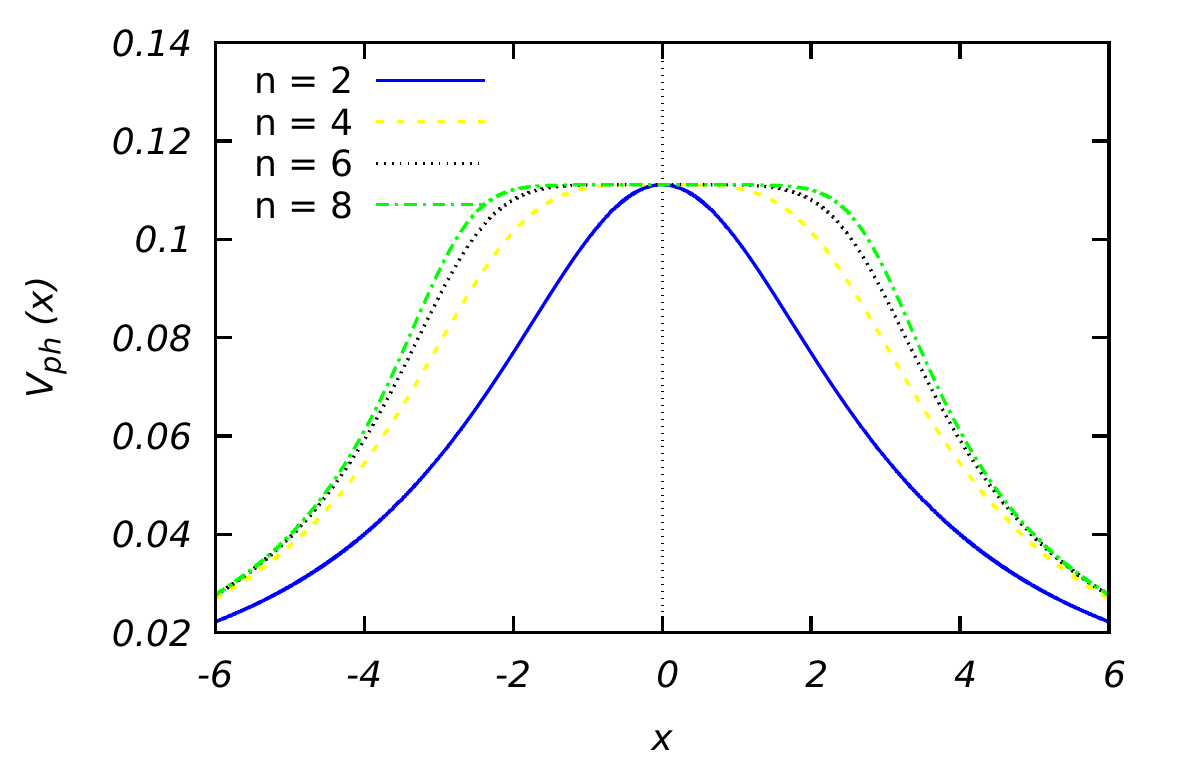}
    \caption{Effective potential for photons for different values of $n$, fixing $\ell=1$ and $a=3$.}
    \label{fig:vpho}
\end{figure}

In Fig. \ref{fig:vpho}, we present the plot of the effective potential for photons, considering different values of the parameter $n$. From the graph, it is evident that the potentials exhibit qualitatively very similar behaviors, despite the differences in the nature of the two cases. 

Similarly, here photons also originate from a position $x_i$ with a given impact parameter $b$, approaching the wormhole until they are deflected and return to infinity. However, there exists a critical value of the impact parameter for which photons enter an unstable circular orbit. This point is characterized by: $1/b_c^2 = V_{ph}(x_c), \, V_{ph}'(x_c)=0$. Given the shape of the potential, we conclude that this point is characterized by the critical impact parameter $b_c=a$, for which photons enter an unstable orbit located at $x=0$

We can then classify the photon orbits into three types:
\begin{itemize}
    \item $b > b_c$, the photon is unable to cross the throat. It reaches a turning point at some $r_{\min}>a$ and is deflected back to infinity in the same universe;

    \item $b = b_c$, the photon becomes trapped at the throat, executing an unstable circular orbit located at $r=a$;

    \item $b < b_c$, the photon crosses the throat and proceeds towards infinity in the other universe connected by the wormhole.
\end{itemize}
For the photon trajectories, we can proceed in a similar way. Thus,  we can write the equation for the azimuthal angle as a function of the impact parameter $b$, defined as $b \equiv \ell/E$. This leads to
\begin{eqnarray}\label{raytracing}
    \frac{d\varphi}{dx} = \mp \frac{b}{ r(x)^2\sqrt{\left(1 - \frac{b^2}{r(x)^2}\right)}}.
\end{eqnarray}
We assume that the photon starts at a position $x_i$ with initial azimuthal angle $\varphi_i$, so that the trajectory of the ingoing branch is described by
\begin{equation}
    \varphi_{in}(x) 
    = \varphi_i 
    - \int_{x_i}^{x}
    \frac{b}{r(\tilde{x})^{2}\sqrt{
    \left(1 - \frac{b^{2}}{r(\tilde{x})^{2}}\right)}}\, d\tilde{x},
\end{equation}
or
\begin{equation}
    \varphi_{in}(r) 
    = \varphi_i 
    - \int_{r_i}^{r}
    \frac{b\mathcal{J}(\tilde{r})}{\tilde{r}^{2}\sqrt{
    \left(1 - \frac{b^{2}}{\tilde{r}^{2}}\right)}}\, d\tilde{r},
\end{equation}
where the integral is evaluated between the points  
\((r_i,\varphi_i)\) and \((r_{min},\varphi_{in}(r_{min}))\).  

For the outgoing photons, we choose the initial condition  
\(
\varphi'_i = \varphi_{in}(r_{min}),
\)
which yields
\begin{equation}
    \varphi_{out}(r) 
    = \varphi'_i 
    + \int_{r_{min}}^{r}
    \frac{b\mathcal{J}(\tilde{r})}{\tilde{r}^{2}\sqrt{
    \left(1 - \frac{b^{2}}{\tilde{r}^{2}}\right)}}\, d\tilde{r}.
\end{equation}

\begin{figure*}[htb]
    \centering
    \includegraphics[width=0.33\linewidth]{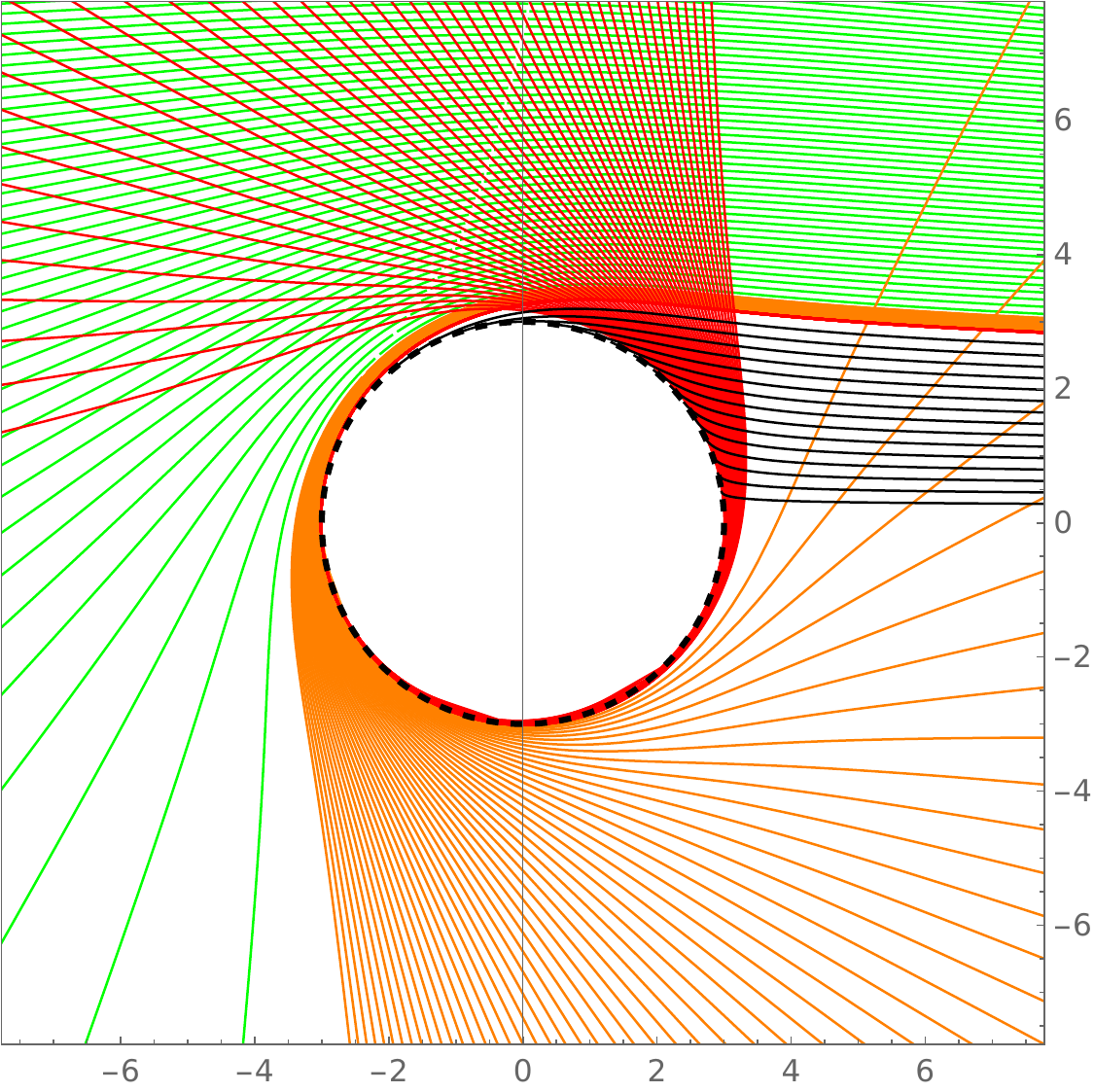}\includegraphics[width=0.33\linewidth]{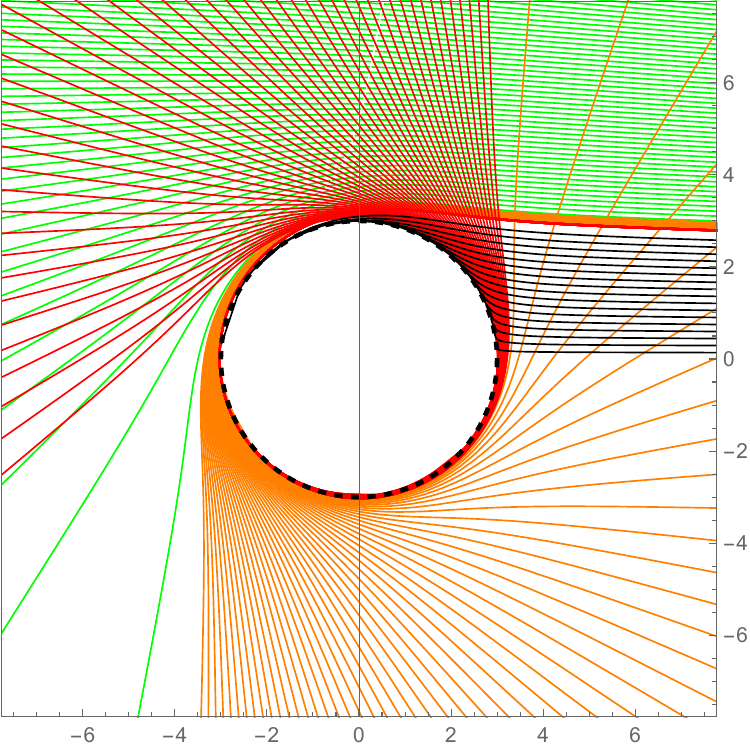}\hspace{0.2cm}\includegraphics[width=0.33\linewidth]{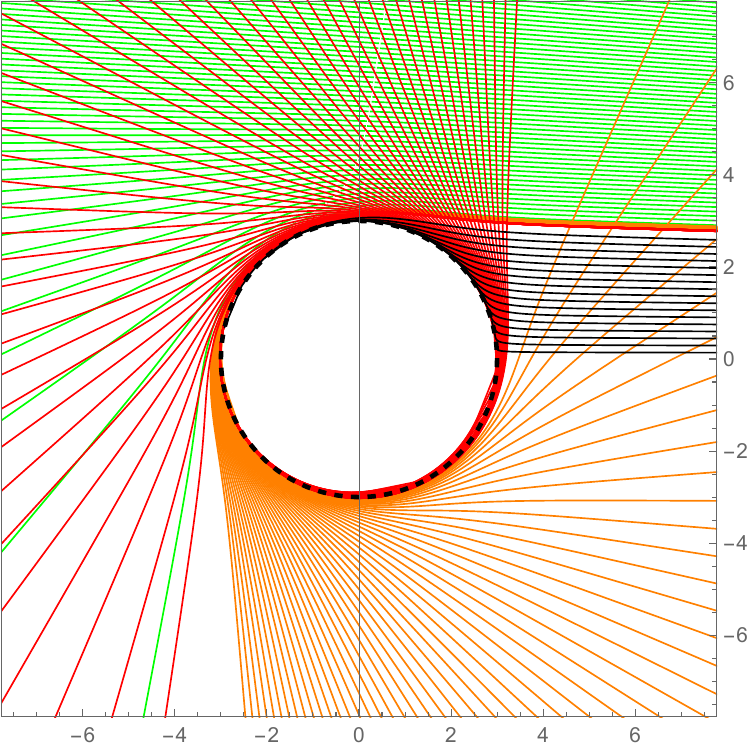}
   \includegraphics[width=0.33\linewidth]{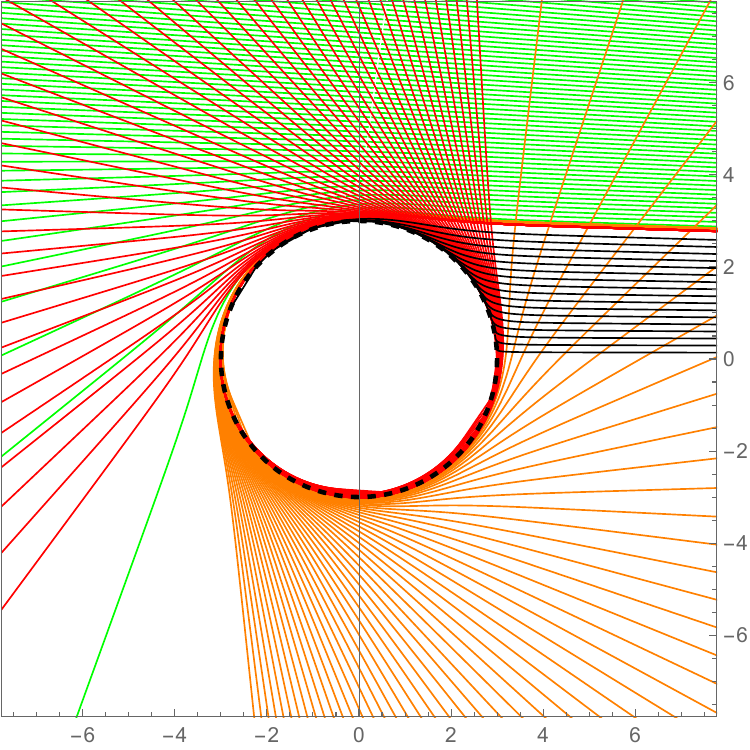}
    \caption{Equatorial light-ray trajectories in GEB space-time for $b \in (0, 8)$ fixing $a=3$. Panels (left to right and top to bottom) show $n=2,4,6$, and $8$ . Green, orange, and red curves indicate emissions with $n_{\varphi} = 0, 1, 2$, respectively. The black circle indicates the throat at $r=a$.}
    \label{fig:null_geodesics}
\end{figure*}

In the critical case $b = b_c$, only a single integral is needed:
\begin{equation}
    \varphi(r)
    = \varphi_i
    - \int_{r_i}^{r}
    \frac{b_c\mathcal{J}(\tilde{r})}{\tilde{r}^{2}\sqrt{
    \left(1 - \frac{b_c^{2}}{\tilde{r}^{2}}\right)}}\, d\tilde{r}.
\end{equation}
We can now employ the ray-tracing method, which consists of backtracking the light rays seen by the observer using the equation for the azimuthal angle presented above. This allows us to determine the emission point of the photons. Thus, we consider a scenario analogous to the massive particle case in which we integrate the in-going and out-going geodesics, Fig \ref{fig:null_geodesics}. However, in this case, we plot the trajectory only within our universe, that is, a 2D plot showing how the light ray is deflected when passing near the wormhole.

The photon trajectories are categorized by the number of half-orbits executed around the compact object, defined by the index $n_\varphi = \lfloor \varphi/\pi \rfloor$. In the resulting Fig \ref{fig:null_geodesics}, we employ a chromatic scale to distinguish the image morphology: $n_\varphi=0$ (green) represents direct light, $n_\varphi=1$ (orange) identifies the primary lensed ring, and $n_\varphi=2$ (red) corresponds to the secondary photon ring.
\subsection{Photon motion and ray-tracing procedure for GBL}
For the massless case, the effective potential to the GBL black bounce is given by
\begin{equation}
V_{ph}=\frac{f(x)\ell^2}{r(x)^2}=\left(1-\frac{2mx^n}{\left(x^n+a^n\right)^{1+1/n}}\right)\frac{\ell^2}{\left(x^n+a^n\right)^{2/n}}.
\end{equation}
Analytically, we can state a few things about the shape of the effective potential. Unlike the massive case, the shape of the potential does not depend on $\ell^{2}$; the fundamental difference is only an increase or decrease in the potential’s magnitude. As in the massive case, at the points where horizons exist, i.e., when $f(x)=0$, the effective potential vanishes. We also have the following limits
\begin{equation}
    \lim_{x \to \infty} V(x) = 0, \quad \mbox{and} \quad \lim_{x \to 0} V(x) =\frac{\ell^2}{a^2}.
\end{equation}
It is interesting to note that, for a given type of geodesic, the GEB wormhole and the GBL geometry have identical limiting values both at spatial infinity and at the throat. For timelike geodesics, the effective potential approaches $1$ at infinity and $1+\ell^2/a^2$ at the throat in both geometries. Likewise, for null geodesics, it approaches $0$ at infinity and $\ell^2/a^2$ at the throat. This correspondence follows from the fact that the mass deformation adopted here satisfies $f(x)\to1$ in both limits.

\begin{figure*}
    \includegraphics[width=.5\linewidth]{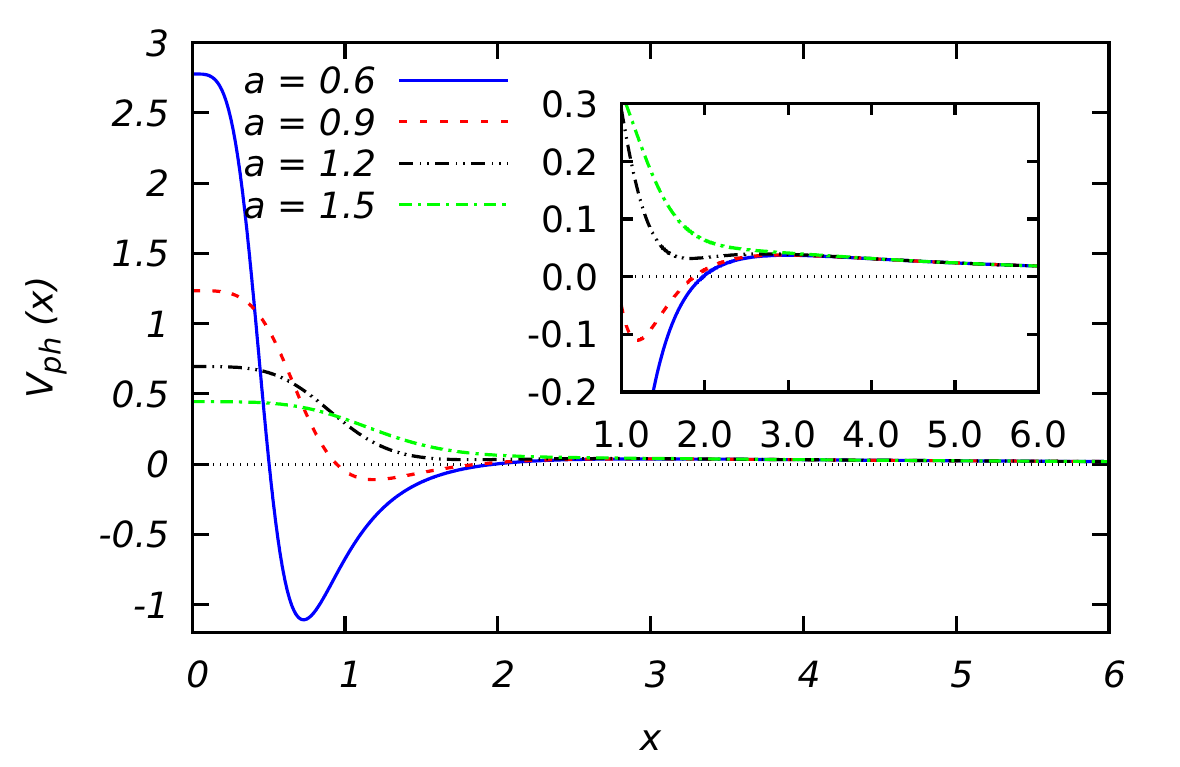}\hspace{-0.5cm}
    \includegraphics[width=.5\linewidth]{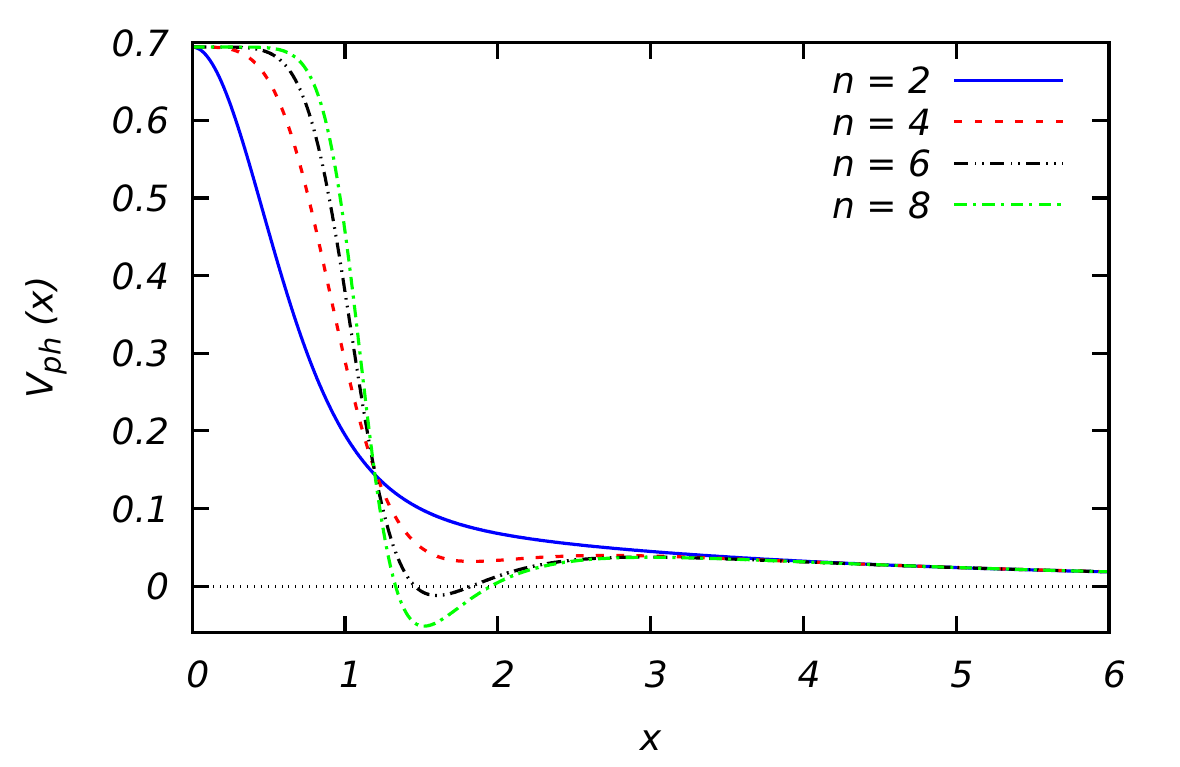}
    \caption{Behavior of the effective potential in terms of the radial coordinate for $n=4$ and different values of $a$ (left panel), and for $a=1.2$ and different values of $n$ (right panel). We fixed $\ell=m=1$. The potential’s behavior is symmetric under $x \to -x$; therefore, we focus only on the region $x>0$.}
    \label{fig:pot_massless_BD}
\end{figure*}

In Fig. \ref{fig:pot_massless_BD}, we show how the effective potential for photons behaves as the model parameters are varied. Unlike the GEB wormhole, which had only one maximum at the throat, here there is also a minimum and another maximum. When horizons are present, the minimum and one of the maxima are hidden behind the horizons, so that only a single maximum is observable. We also see that the potential at $x \to 0$ does not depend on the value of $n$, but it does depend on $a$, as we had previously shown analytically.

\begin{figure*}
    \centering
    \includegraphics[width=0.36\linewidth]{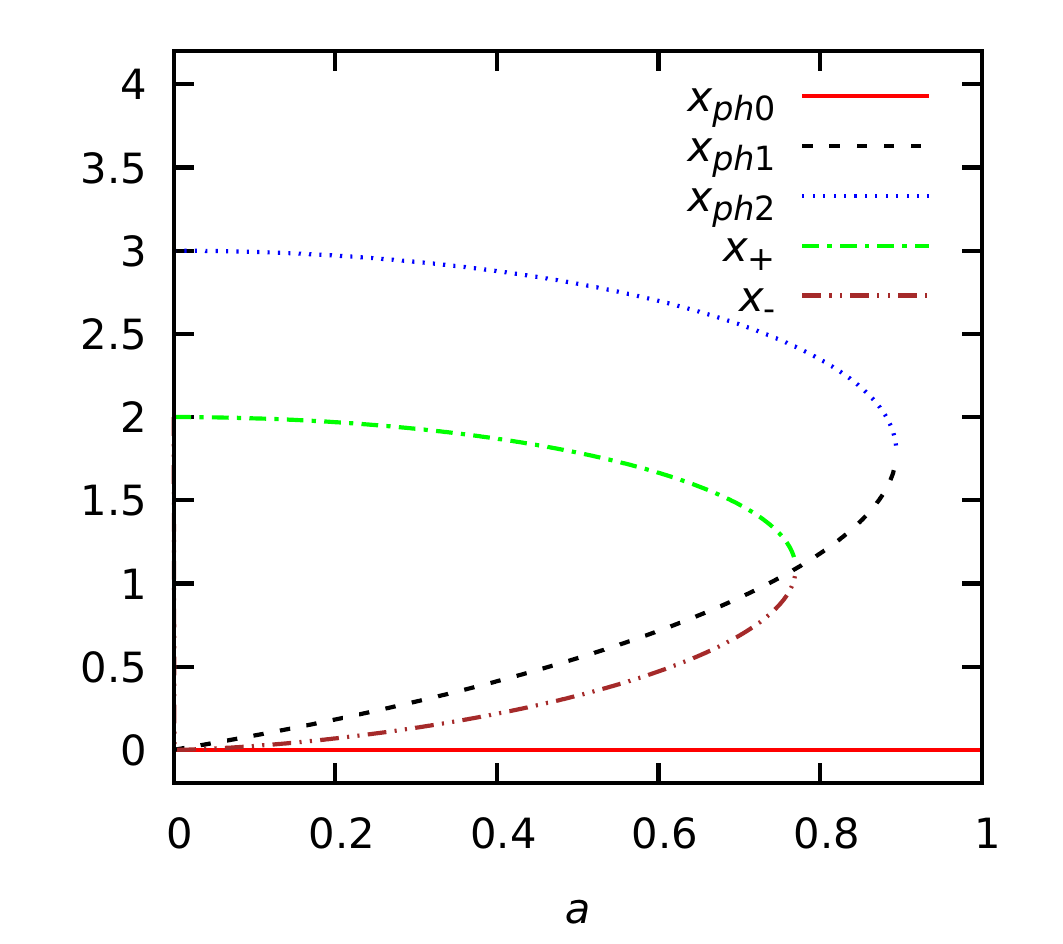}\hspace{-0.9cm} \includegraphics[width=0.36\linewidth]{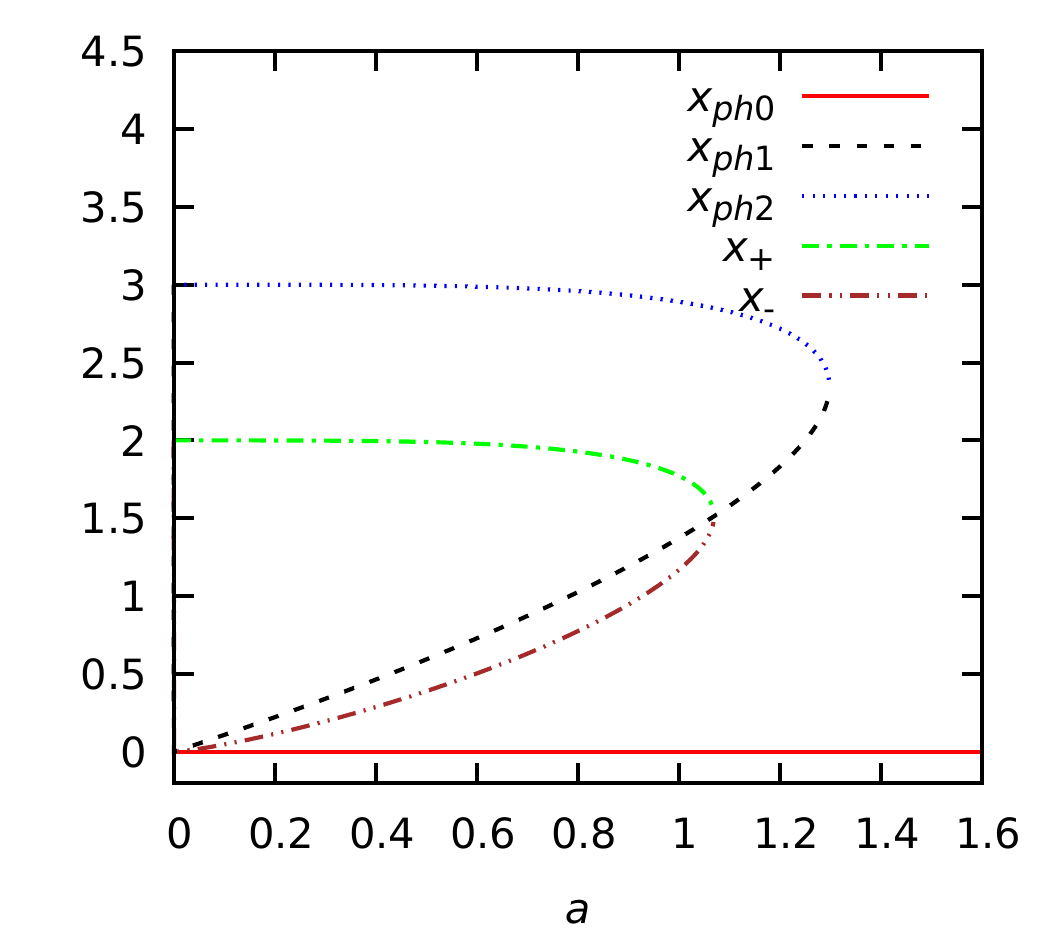}\hspace{-0.9cm}    \includegraphics[width=0.36\linewidth]{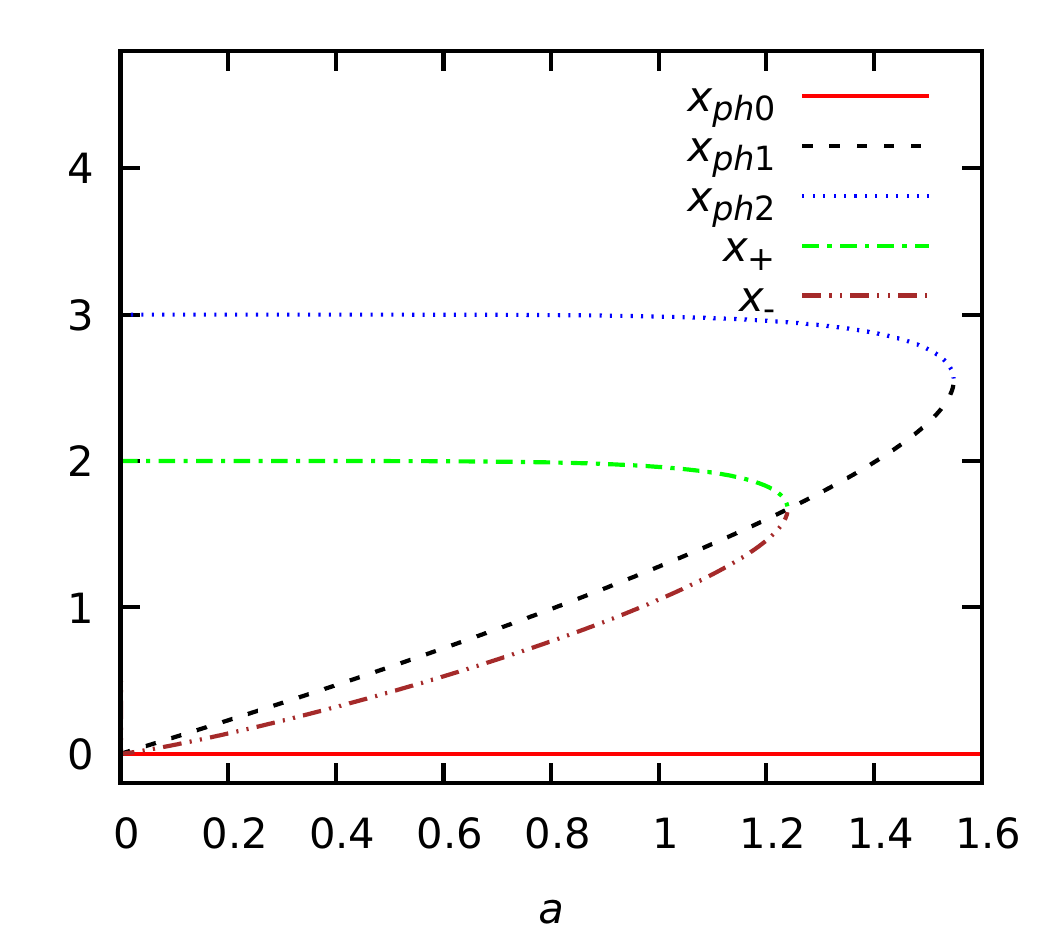}
    \caption{Behavior of the light-ring radius with $m=1$ as $a$ varies. The values of $n$ are: $n=2$ (left panel), $n=4$ (middle panel), and $n=6$ (right panel). The nomenclature is: $x_{-}$, Cauchy horizon radius; $x_{+}$, event horizon radius; $x_{ph0}$, radius of the unstable orbit at the throat; $x_{ph1}$, radius of the stable orbit; and $x_{ph2}$, radius of the unstable orbit.}
    \label{fig:xph_BD}
\end{figure*}

From the extrema of the effective potential, we can analyze the behavior of circular photon orbits (photon spheres). At the extrema, $V'_{ph}=0$, while the sign of $V_{ph}''$ determines stability: stable orbits have $V_{ph}''>0$ and unstable orbits have $V_{ph}''<0$. In both cases, we have $\dot{x}=\ddot{x}=0$. The critical impact parameter to this case is given by
\begin{equation}
    b_c^2=\frac{\ell^2_c}{E_c^2}=\frac{r^2(x_{ph})}{f(x_{ph})},
\end{equation}
where $x_{ph}$ is the radius of the photon sphere, which is obtained from the condition $V'_{ph}(x_{ph})=0$, which results in
\begin{equation}
    2f(x_{ph})r'(x_{ph})-f'(x_{ph})r(x_{ph})=0.
\end{equation}
In Fig. \ref{fig:xph_BD}, we show how the photon orbit radius behaves in different cases. The orbit located at the throat, at $x=0$ ($r=a$), always exists, as expected for symmetric wormholes \cite{Xavier:2024iwr}. As $a \to 0$, we recover the Schwarzschild value for the unstable orbit, $x=3m$. There is always a maximum value of $a$ for which three possible values of $x_{ph}$ exist; above that value, only the orbits at $x=0$ remain. It can also be seen that, while horizons are present, the orbits $x_{ph1}$ and $x_{ph0}$ are hidden, but there is a certain range of $a$ for which all three orbits are observable. The larger $n$ is, the larger the value of $a$ that allows for the existence of $x_{ph1}$ and $x_{ph2}$.

\begin{figure*}
    \centering
    \includegraphics[width=0.33\linewidth]{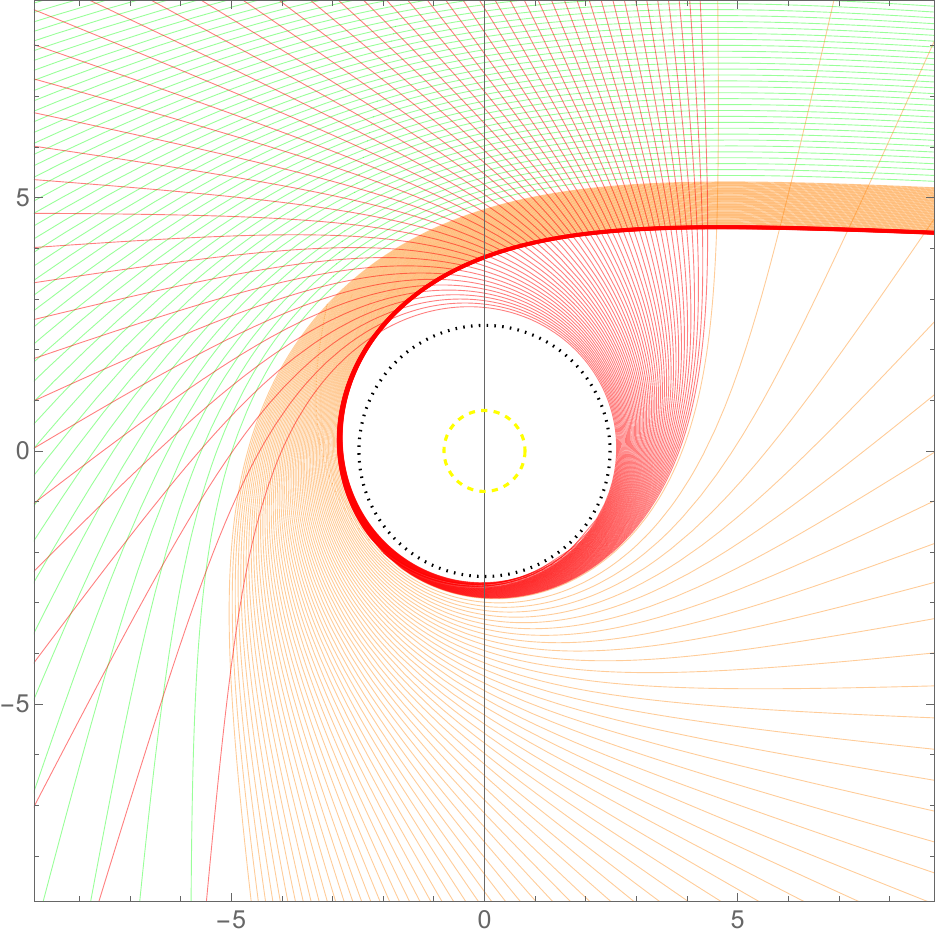}\includegraphics[width=0.33\linewidth]{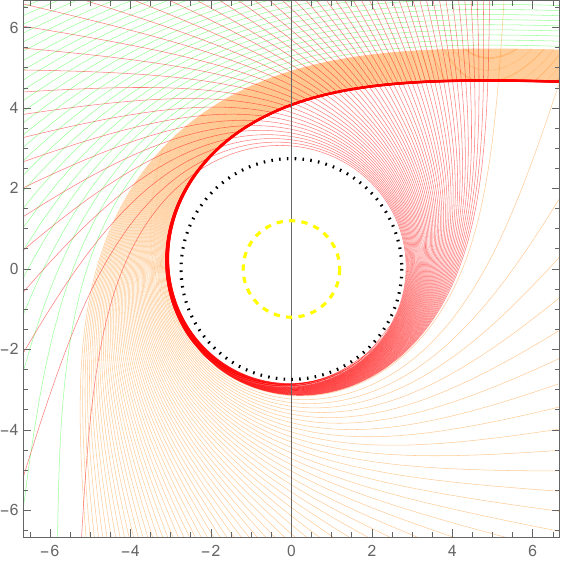}\includegraphics[width=0.33\linewidth]{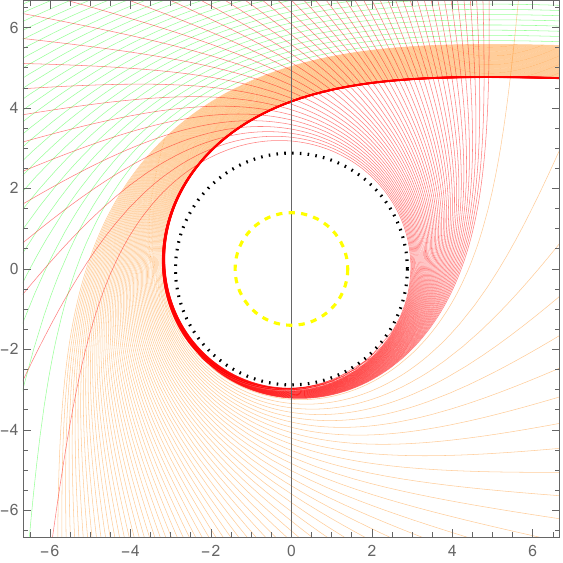}

    \includegraphics[width=0.33\linewidth]{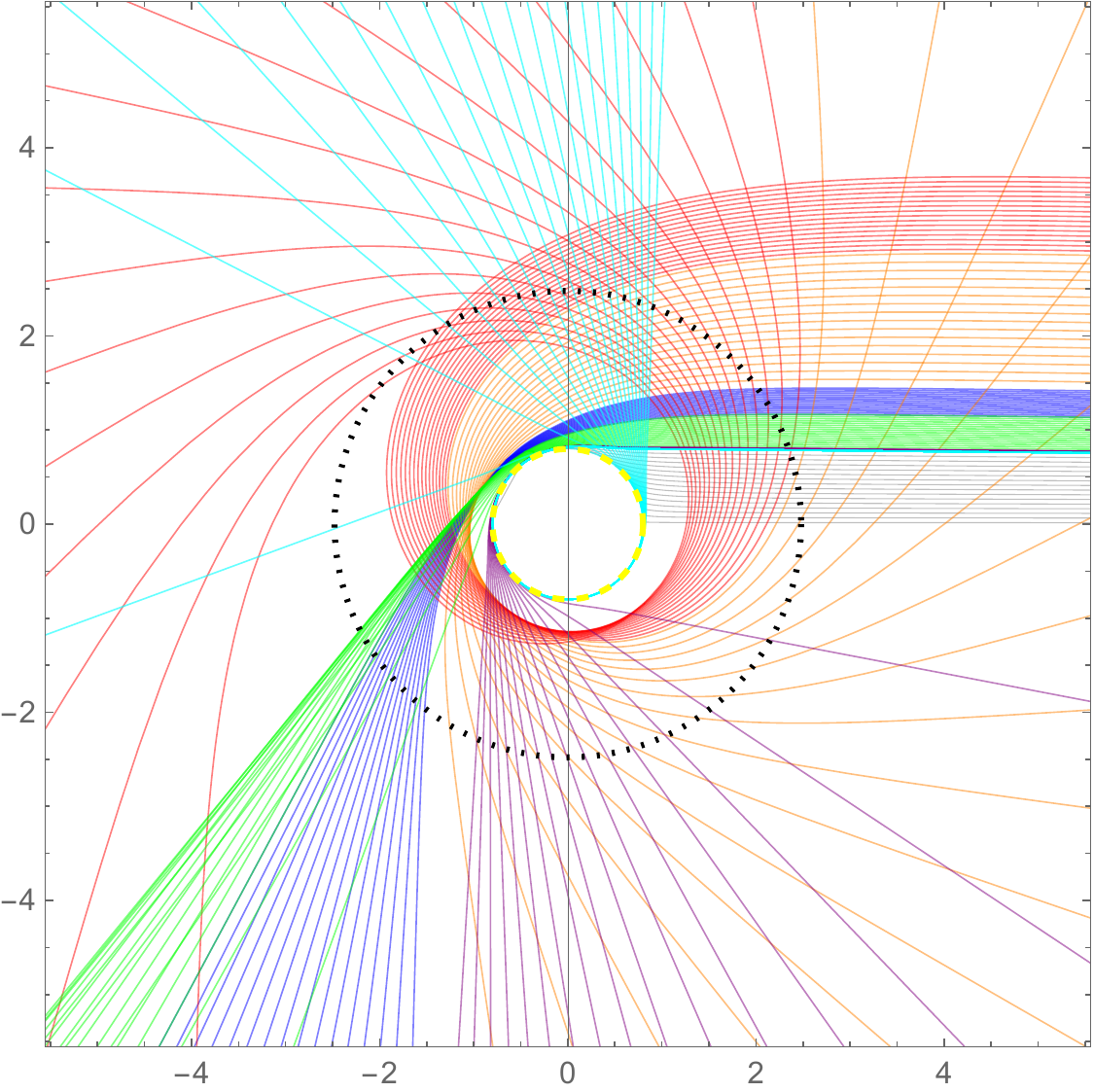}\includegraphics[width=0.33\linewidth]{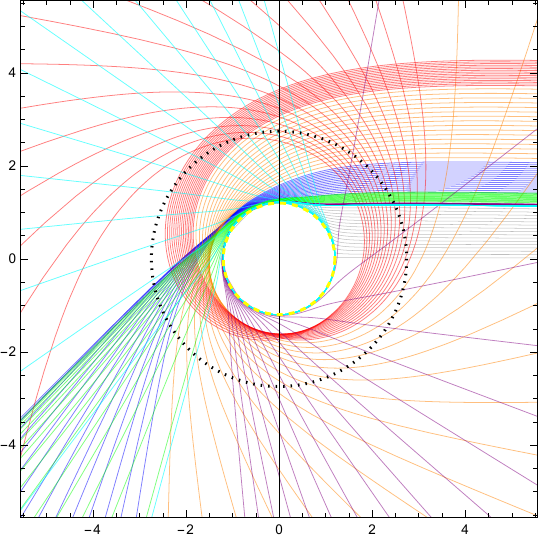}\includegraphics[width=0.33\linewidth]{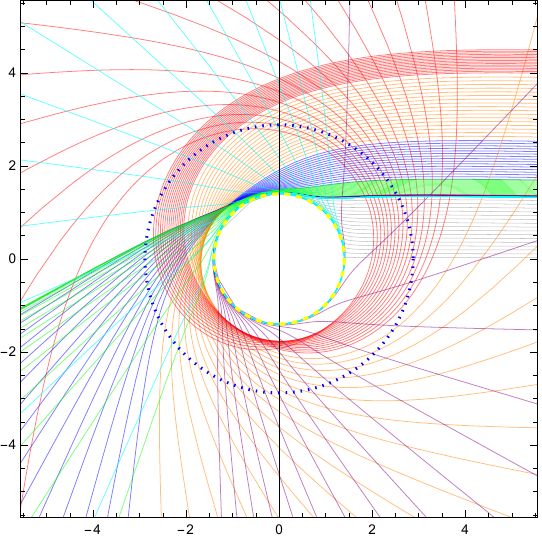}
    \caption{
Equatorial light-ray trajectories in the GBL BB space-time for traversable wormhole configurations with two photon spheres, fixing $m = 1$. The top row displays trajectories with impact parameters above the outer critical value ($b > b_c^{(2)}$), whereas the bottom row corresponds to trajectories with impact parameters below the outer critical value ($b < b_c^{(2)}$), covering both the intermediate region ($b_c^{(1)} < b < b_c^{(2)}$) and the inner shadow region ($b < b_c^{(1)}$). Panels from right to left show parameters $(n, a) = (2, 0.8m)$, $(4, 1.2m)$, and $(6, 1.4m)$. The yellow circle indicates the wormhole throat at $r = a$, while the black circle denotes the outer photon sphere. For $b > b_c^2$ (top row), green, orange, and red curves represent direct ($n_\varphi = 0$), lensed ($n_\varphi = 1$), and photon ring ($n_\varphi \geq 2$) trajectories, respectively. In the bottom row ($b < b_c^{(2)}$), for $b_c^{(1)} < b < b_c^{(2)}$ we find the outer photon ring (red), lensed (orange), and direct (blue) trajectories, as well as the inner direct (green), lensed (purple), and photon ring (cyan) trajectories. Below $b < b_c^{(1)}$, we observe the inner (blank) shadow, which is populated by light rays passing through the throat from the other side of the wormhole.
}

    \label{raytracingWH}
\end{figure*}

\begin{figure*}
    \centering
    \includegraphics[width=0.33\linewidth]{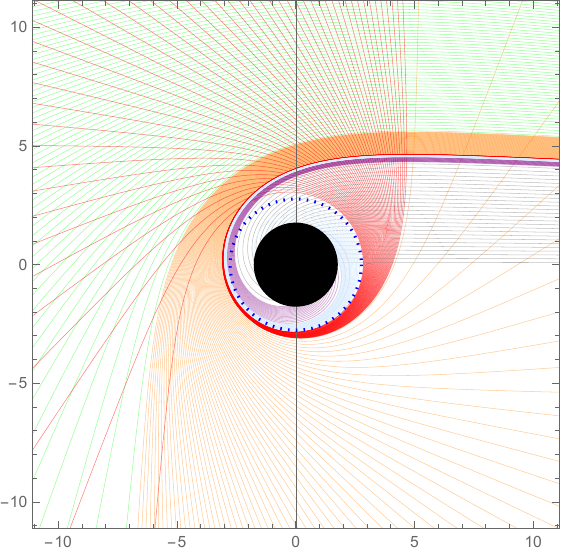}\includegraphics[width=0.33\linewidth]{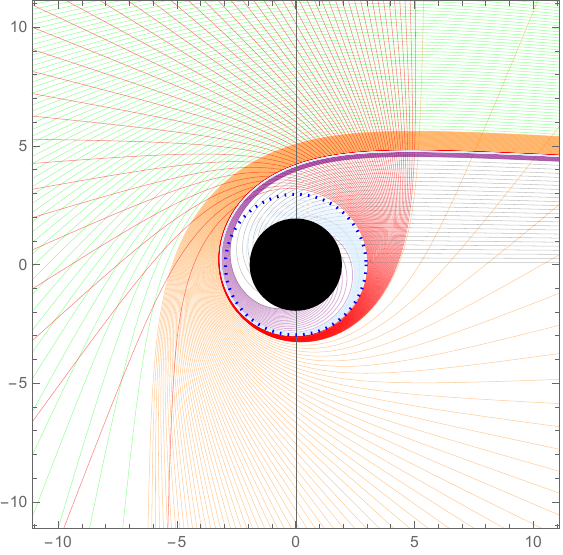}\includegraphics[width=0.33\linewidth]{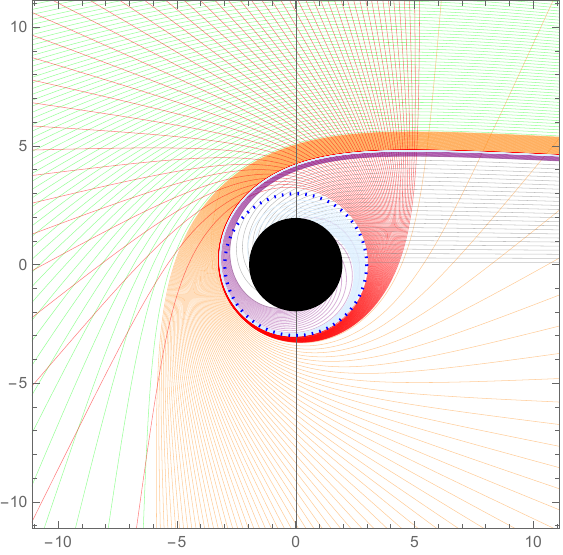}

\caption{
Equatorial light-ray trajectories in the GBL BB space-time for $b \in (0, 10)$. Panels (left to right) correspond to $(n, a) = (2, 0.6m)$, $(4, 0.8m)$, and $(6, 1m)$ (regular black hole). Green, orange, and red curves represent emissions with $n_{\varphi} = 0, 1, 2$, respectively. Gray, purple, and light-blue curves represent retro-trajectories (trajectories emerging out of the photon sphere). The central black disk represents the event horizon, while the blue dashed circle indicates the unstable photon orbit.}

    \label{raytracingBH}
\end{figure*}

Figs. \ref{raytracingWH} and \ref{raytracingBH} illustrate the ray-tracing results for the GBL space-time in its wormhole and black hole regimes, respectively. By mapping the orbit index $n_\varphi$, we observe that the shadow edges are consistently traced by the accumulation of higher-order rings ($n_\varphi \geq 2$), which converge toward the critical impact parameters. Particularly for the traversable wormhole regime shown in Fig. \ref{raytracingWH}, the geometry admits two photon spheres, leading to a much richer structure of null geodesics characterized by two critical impact parameters, $b_c^{(1)}$ and $b_c^{(2)}$. In this plot, the yellow curve represents the wormhole throat, while the black curve denotes the outer photon sphere. Trajectories hovering in the region between these two critical values ($b_c^{(1)} < b < b_c^{(2)}$) exhibit a complex, frenzied behavior driven by the combined influence of the inner and outer critical curves. Furthermore, the absence of an event horizon in this regime allows light rays coming from the other side to flow through the throat, forming an inner shadow. Conversely, Fig. \ref{raytracingBH} displays the characteristic shadow of a regular black hole, where the $n_\varphi=0$ (green) region is sharply truncated by the capture of photons.

\subsection{A brief comment about photon rings}

The optical appearance of a compact object illuminated by an accretion disk is dominated primarily by the brightest ring of light \cite{Olmo:2025ctf,Gralla:2019yty,Guerrero:2021ues}, which corresponds to the direct emission ($n_\varphi=0$) and its higher-order images ($n_\varphi=1, 2, \dots$). This feature corresponds to the \textit{photon ring}.

In this context, photons with an impact parameter slightly deviating from the critical value ($b \gtrsim b_c$) are particularly significant, as they correspond to trajectories that execute multiple orbits around the object before reaching the observer. These paths are therefore termed \textit{quasi-bound orbits} (Observed in orange and red in Figs.\ref{fig:null_geodesics} and \ref{raytracingWH}-\ref{raytracingBH}), that is, trajectories slightly displaced from the photon sphere. Thus, if we consider that the trajectory followed by the photon deviates marginally from the photon sphere radius, i.e., $x(\lambda) = x_{ph} + \delta x(\lambda)$,
the trajectory is determined perturbatively by Eq~\eqref{cons_BDG}, such that, in terms of the affine parameter, we obtain:
\begin{equation}
  \delta \dot{x} = \sqrt{-\frac{1}{2}V''(x_{ph})}\;\delta x
    \label{eq:affine_solution}
\end{equation}
Using relation~\eqref{quant_cons}, we can finally express the photon trajectory as:
\begin{equation}\label{photontrajectory}
    \delta x(n_\varphi) = \delta x_0\, e^{\gamma_{ph}n_\varphi}, 
\end{equation}
where
\begin{equation}
    \gamma_{ph} = \pi\sqrt{f_{ph}r_{ph}r_{ph}'' + f_{ph}r_{ph}'^2 - \frac{1}{2}f_{ph}''r_{ph}^2}
\end{equation}
is the \textit{lensing Lyapunov exponent}, a quantity that expresses the degree of instability of the system. This parameter governs the rate of exponential divergence of the photon trajectories from the critical orbit \cite{Olmo:2025ctf,Cardoso:2008bp}, determining how rapidly a photon spirals away from the photon sphere to either escape to infinity or be captured by the object.

The geometric structure of the photon ring is governed by a strict self-similarity, where the width and separation of consecutive sub-rings are demagnified by a factor determined by the Lyapunov exponent, $\gamma_{ph}$ \cite{Johnson:2019ljv}. Theoretically, this implies that the ratio of impact parameters and flux densities between successive images ($n_\varphi$ and $n_\varphi+1$) should follow a canonical exponential decay proportional to $e^{-\gamma_{ph}}$. 

From an observational standpoint, this scaling behavior provides a critical test for the background space-time geometry. Future extensions of Very Long Baseline Interferometry (VLBI) aimed at resolving the $n_\varphi=1$ sub-ring could, in principle, utilize these geometric correlations to extract $\gamma_{ph}$ and constrain deviations from Kerr paradigm \cite{Gralla:2020srx}, potentially serving, in principle, as a test for more exotic geometries. 

However, translating this geometric prediction into radiometric observables presents significant challenges. The idealized relationship assumes that photons from higher-order images traverse identical emission regions. In realistic accretion scenarios, the turbulent and inhomogeneous nature of the plasma implies that photons executing additional orbits sample distinct azimuthal sectors of the disk \cite{EventHorizonTelescope:2019pgp}. Consequently, local variations in emissivity introduce stochastic deviations in the observed flux ratios, breaking the perfect exponential hierarchy predicted by the vacuum geodesic equations.

\subsection{Shadow radius}
Once we have the radius of the photon sphere, we may investigate the shadow size of our black bounce model with the EHT observations of Sgr~A*. 
We constrain the parameter $a$ of our black bounce model by comparing its predicted shadow size with the EHT results for Sgr~A*. The EHT images the bright emission ring rather than the shadow edge itself, so the shadow angular scale is inferred by combining the ring measurement with independent determinations of the mass $m$ and distance $d$ \cite{EventHorizonTelescope:2022xqj}. These quantities are obtained from near-infrared monitoring of S-stars orbiting Sgr~A*, with the star S0-2 playing a central role due to its short period and strong astrometric and spectroscopic variations \cite{Do:2019txf,GRAVITY:2020gka}. 

Besides the ratio $d/m$, one must account for a calibration factor that relates the observed ring size to the true shadow radius \cite{Vagnozzi:2022moj}. We parameterize this offset as \cite{Vagnozzi:2022moj,EventHorizonTelescope:2022xqj}
\begin{equation}\label{eq:delta_new}
\delta \equiv \frac{r_s}{r_s^{\rm schw}}-1,
\qquad
r_s^{\rm schw}=3\sqrt{3}\,m .
\end{equation}
Using the Keck- and VLTI/GRAVITY-based estimates, $\delta_{\rm Keck}=-0.04^{+0.09}_{-0.10}$ and $\delta_{\rm VLTI}=-0.08^{+0.09}_{-0.09}$ \cite{Vagnozzi:2022moj,EventHorizonTelescope:2022xqj}, and combining them as in \cite{Vagnozzi:2022moj}, we adopt $\delta\simeq -0.060\pm 0.065$. This corresponds to
\begin{equation}
-0.125 \lesssim \delta \lesssim 0.005 \ (1\sigma),
\qquad
-0.19 \lesssim \delta \lesssim 0.07 \ (2\sigma),
\end{equation}
and therefore implies, from $r_s/m=3\sqrt{3}\,(1+\delta)$, the observational ranges \cite{Vagnozzi:2022moj}
\begin{equation}\label{eq:rs_over_M_ranges_new}
4.55 \lesssim \frac{r_s}{m} \lesssim 5.22 \ (1\sigma),
\qquad
4.21 \lesssim \frac{r_s}{m} \lesssim 5.56 \ (2\sigma).
\end{equation}

\begin{figure}
    \centering
    \includegraphics[width=1\linewidth]{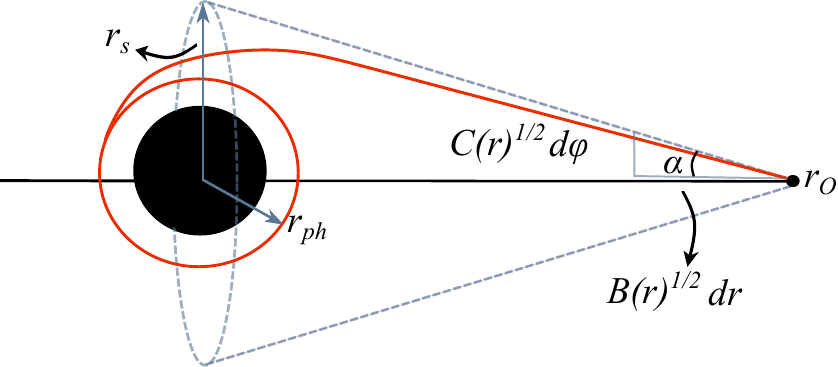}
    \caption{Representation of a light ray emitted from the point $r_O$ and entering a circular orbit around the black bounce.}
    \label{fig:shapeshadow}
\end{figure}

Following \cite{Perlick:2021aok}, we compute the apparent shadow radius for a static and spherically symmetric line element,
\begin{equation}
ds^2=-A(r)dt^2+B(r)dr^2+C(r)d\Omega_2^2.
\end{equation}
 The geometrical meaning of the relevant angles is illustrated in Fig.~\ref{fig:shapeshadow}: a photon emitted (or received) at $r=r_O$ is sent along the critical direction such that it asymptotically approaches the unstable circular orbit at $r=r_{ph}$. The emission angle $\alpha$ can be related to the metric functions as
\begin{equation}
\label{eq:cotalpha_new}
\cot \alpha
=\frac{\sqrt{g_{rr}}}{\sqrt{g_{\varphi\varphi}}}
\left.\frac{dr}{d\varphi}\right|_{r=r_O}
=\frac{\sqrt{B(r)}}{r}
\left.\frac{dr}{d\varphi}\right|_{r=r_O}.
\end{equation}
After rewriting $dr/d\varphi$ in terms of the impact parameter $b$ (and performing the corresponding coordinate transformation), one obtains
\begin{equation}
\sin^2\alpha=\frac{b^2A(r_O)}{r_O^2}.
\end{equation}

\begin{figure*}
    \centering
    \includegraphics[width=0.52\linewidth]{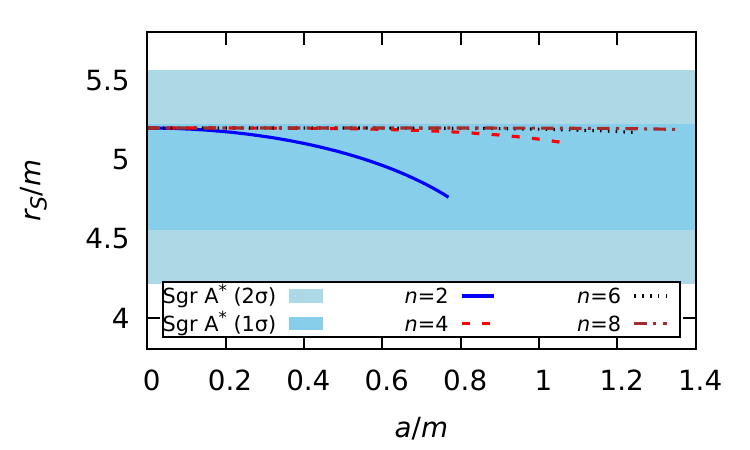}\hspace{-1.cm}
    \includegraphics[width=0.52\linewidth]{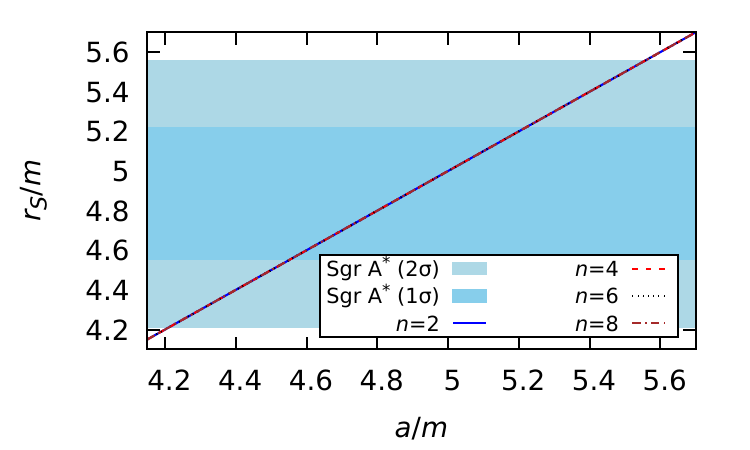}
    \caption{Behavior of the apparent shadow radius of the compact object as $a$ varies for different values of $n$. We consider two cases: (i) when the solution still has horizons (left panel) and (ii) when we have the wormhole case (right panel). For the wormhole, we consider the photon sphere located at the throat, $x_{ph}=0$ (or $r_{ph}=a$).}
    \label{fig:shadow_radius}
\end{figure*}

In the small angle regime, $\sin\alpha\simeq \alpha$ and $\alpha\simeq r_s/r_O$, so that for the critical photon orbit ($b=b_c$) the apparent shadow radius reads
\begin{equation}\label{eq:rS_new}
r_s=r_{ph}\sqrt{\frac{A(r_O)}{A(r_{ph})}}.
\end{equation}
where $r_{ph}$ is the radius of the unstable circular photon orbit. In the next step, we compute $r_s/M$ from Eq.~\eqref{eq:rS_new} as a function of the model parameter $a$ and impose consistency with the EHT-inferred intervals in Eq.~\eqref{eq:rs_over_M_ranges_new}, thereby obtaining observational bounds on $a$.
In Fig.~\ref{fig:shadow_radius}, we show how the shadow radius of this space-time behaves as the throat radius varies. It is noteworthy that, for the black bounce case, the radius remains within the $1\sigma$ band, and the larger $n$ is, the smaller the variation in the radius when $a$ is changed. For the wormhole case, the situation changes because the photon sphere is located at the throat, so $r_{ph}=a$, independently of the value of $n$. This helps explain why the results in the wormhole case are so similar even for different values of $n$. Within $1\sigma$, we can constrain the throat radius, to two decimal places, to $4.55 \lesssim \frac{a}{m} \lesssim 5.22$; within $2\sigma$, we obtain $4.21 \lesssim \frac{a}{m} \lesssim 5.56 $.

\section{Optical appearance from geometrically thin accretion disks}
\label{sec:OA}

We now examine the optical appearance of the geometries under consideration. Although the critical curves and photon trajectories determine the main geometrical scales of the image, its complete intensity distribution also depends on the emission profile of the surrounding matter. Consequently, geometries with similar critical curves may exhibit distinct brightness profiles and higher-order images. This distinction is particularly relevant when comparing black holes and wormholes. In a black hole space-time, the event horizon provides a capture channel for photons and may therefore produce a dark central region. In a traversable wormhole, by contrast, photons crossing the throat can continue toward the opposite asymptotic region. The brightness of the central region therefore depends on the assumed illumination and emission conditions.

We model the emitting matter as a geometrically and optically thin equatorial disk viewed face-on by a distant observer. The emission is assumed to be isotropic in the local rest frame of the static emitters, while absorption and scattering are neglected to isolate the effects of the space-time geometry. The local frequency-integrated emission is described by a radial intensity profile $I^{\rm em}(x)$. For the wormhole configurations, both the disk and the observer are restricted to the asymptotic region $x>0$, and no independent emitting component is included in the opposite universe. Under this prescription, photons crossing the throat encounter no emitting matter in the other asymptotic region.

Using the ray-tracing framework established previously, we integrate the null geodesics backwards from the observer's screen, with the impact parameter $b$ serving as the radial screen coordinate. The disk intersections are described by the transfer functions $x_{n_\phi}(b)$, where $n_\phi=0$, $1$, and $2$ correspond to the direct, lensed, and first photon ring contributions, respectively. The quantity $|dx_{n_\phi}/db|$ characterizes the radial demagnification of each contribution. A steep transfer function maps an extended interval of the disk coordinate $x$ into a narrow range of impact parameters, producing strong demagnification, whereas a slowly varying branch produces a broader contribution on the observer's screen. The numerical behavior of these functions therefore enables a direct comparison of the relative widths and demagnification of the image contributions in the GEB and GBL geometries.

We describe the intrinsic disk emission using the phenomenological Standard Unbound (SU) profile,
\begin{equation}
I^{\rm em}_{S_U}(x;\mu,\sigma,\gamma)=\frac{\exp\left[-\dfrac{1}{2}\left(\gamma+\operatorname{arcsinh}\left(\frac{x-\mu}{\sigma}\right)\right)^2\right]}{\sqrt{(x-\mu)^2+\sigma^2}
}.
\label{SU_profile}
\end{equation}
Here, $\mu$ is a location parameter, $\sigma$ controls the radial width of the emitting region, and $\gamma$ determines the asymmetry of the profile. For each configuration, the intensity is normalized such that $\max I^{\rm em}(x)=1$. We consider a small set of representative profiles whose maxima are associated with the characteristic radial scales of each geometry. Their specific parameters will be presented in the corresponding subsections.

According to Liouville's theorem, $I_\nu/\nu^3$ is conserved along a null geodesic, so that $I_{\nu_{\rm O}}^{\rm obs}=g^3I_{\nu_{\rm e}}^{\rm em}$. For a static emitter located at $x=x_{n_\varphi}(b)$ and a static observer at $x=x_{\rm O}$, the frequency-shift factor is
\begin{equation}
g_{n_\varphi}(b)=\sqrt{\frac{f\left(x_{n_\phi}(b)\right)}{f(x_{\rm O})}}.
\end{equation}
After integration over frequency, each contribution is weighted by $g_{n_\phi}^4$. The total observed intensity is therefore obtained by summing over the retained image orders,
\begin{equation}
I^{\rm obs}(b)=\sum_{n_\phi=0}^{N}\left[\frac{f\left(x_{n_\phi}(b)\right)}{f(x_{\rm O})}
\right]^2I^{\rm em}\left(x_{n_\phi}(b)\right).
\label{total_obs_intensity}
\end{equation}
Unless otherwise stated, we set $N=2$, retaining the direct, lensed, and first photon ring contributions. Higher-order terms may be included to assess the convergence of the intensity distribution, particularly for geometries containing degenerate throat circular photon orbits.

Finally, owing to spherical symmetry and the face-on orientation of the disk, the observed intensity depends only on the radial screen coordinate. Denoting the Cartesian screen coordinates by $(X,Y)$, the two-dimensional image is constructed as
\begin{equation}
I_{\rm image}(X,Y)=I^{\rm obs}\left(\sqrt{X^2+Y^2}\right).
\label{two_dim_image}
\end{equation}
This framework will now be applied first to the massless GEB wormhole and subsequently to the GBL black bounce geometry.

\subsection{Optical appearance of the GEB wormhole}
\label{subsec:OA_GEB}
We begin with the massless GEB wormhole. As discussed in the previous section, the photon orbit located at the throat is nondegenerate for $n=2$ and becomes degenerate for $n>2$. This change in the near throat photon dynamics may affect the higher-order image contributions even when the corresponding critical curves have the same characteristic scale.

The GEB geometry has $f(x)=1$. Consequently, there is no gravitational frequency shift between the static emitters and the observer, and $g_{n_\phi}(b)=1$ for every disk intersection. Equation~\eqref{total_obs_intensity} therefore reduces to
\begin{equation}
I_{\rm GEB}^{\rm obs}(b)=
\sum_{n_\varphi=0}^{N}
I^{\rm em}
\left(x_{n_\phi}(b)\right),
\label{GEB_total_obs_intensity}
\end{equation}
where only intersections with the emitting disk in the region $x>0$ are retained. The optical appearance is therefore determined entirely by the transfer functions $x_{n_\phi}(b)$ and the adopted emission profile. This allows us to isolate the influence of the throat geometry and of the circular photon orbit degeneracy on the direct and higher-order image contributions.

\begin{figure}[!htb]
    \centering
    \includegraphics[width=1\linewidth]{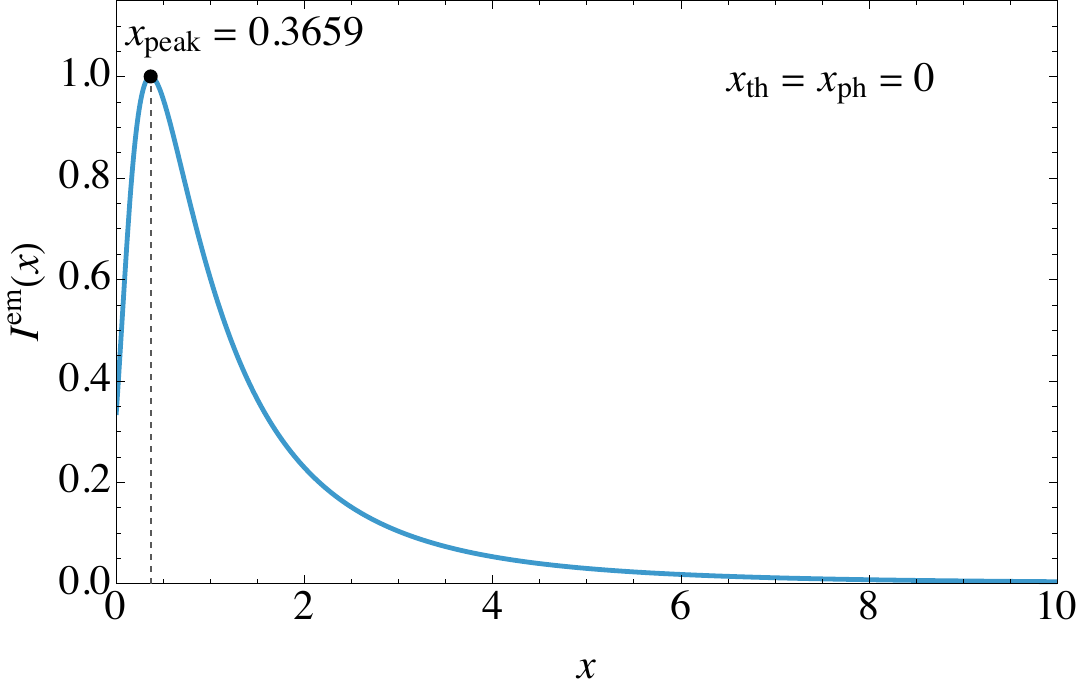}
    \caption{Normalized emitted intensity profile $I^{\rm em}(x)$ adopted for the GEB wormhole with $a=3$. The parameters are $\mu=0$, $\sigma=1/4$, and $\gamma=-2$. The black point and the vertical dashed line mark the maximum at $x_{\rm peak}\simeq0.3659$, while the throat and the throat photon sphere are both located at $x_{th}=x_{ph}=0$. The same emission profile is used for $n=2$, $4$, and $6$.}
    \label{fig:Perfil_em_GEB}
\end{figure}

Since the massless GEB wormhole possesses neither an event horizon nor an ISCO, emission profiles associated with these characteristic positions are not applicable. Its natural inner reference scale is instead the throat at $x=0$, which also coincides with the unstable circular orbits of null and massive particles. We therefore set $\mu=0$ and fix $a=3$ throughout this subsection. For $\gamma=-2$ and $\sigma=1/4$, the normalized emitted intensity reaches its maximum at $x_{\rm peak}\simeq0.3659$, describing an emission profile concentrated slightly outside the throat, in the same asymptotic region as the observer. The resulting profile is displayed in Fig.~\ref{fig:Perfil_em_GEB} and is used for all three cases, $n=2$, $4$, and $6$.

\begin{figure*}[!htb]
    \centering
    \includegraphics[width=0.45\linewidth]{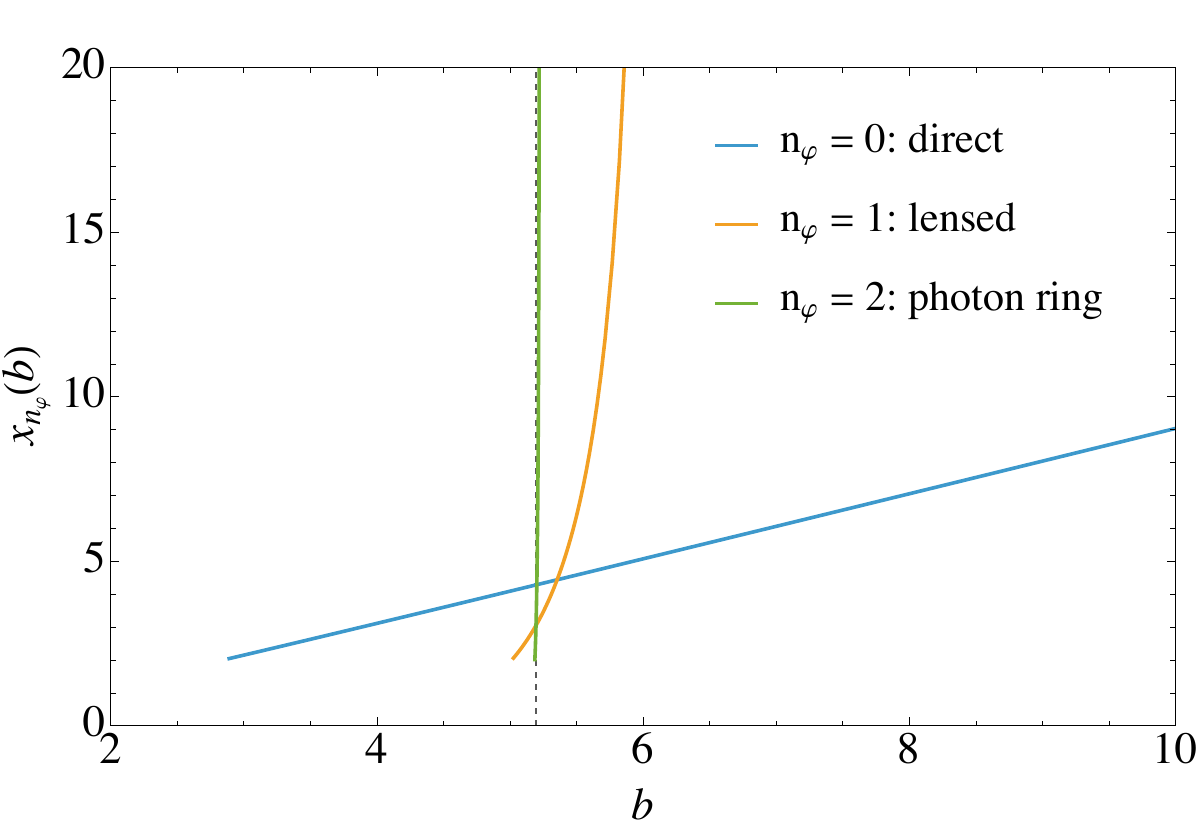}
    \includegraphics[width=0.45\linewidth]{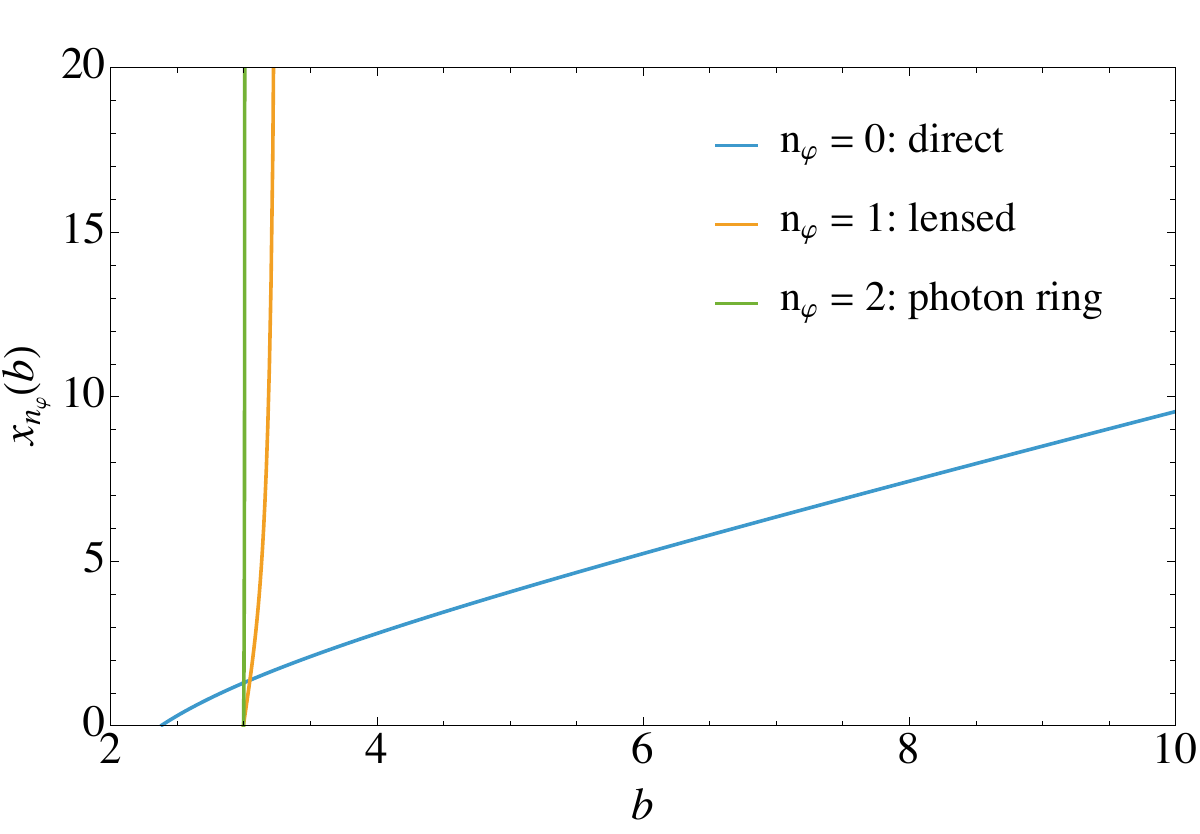}
    \includegraphics[width=0.45\linewidth]{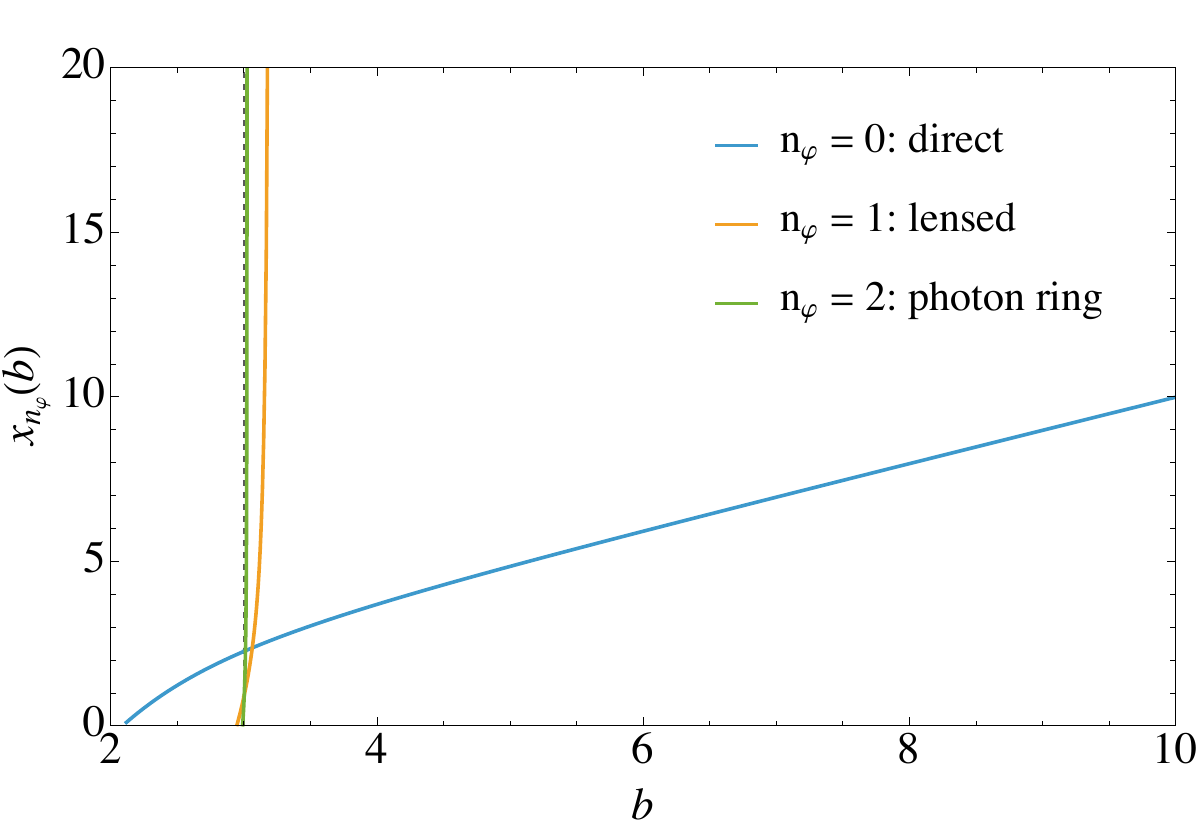}
    \includegraphics[width=0.45\linewidth]{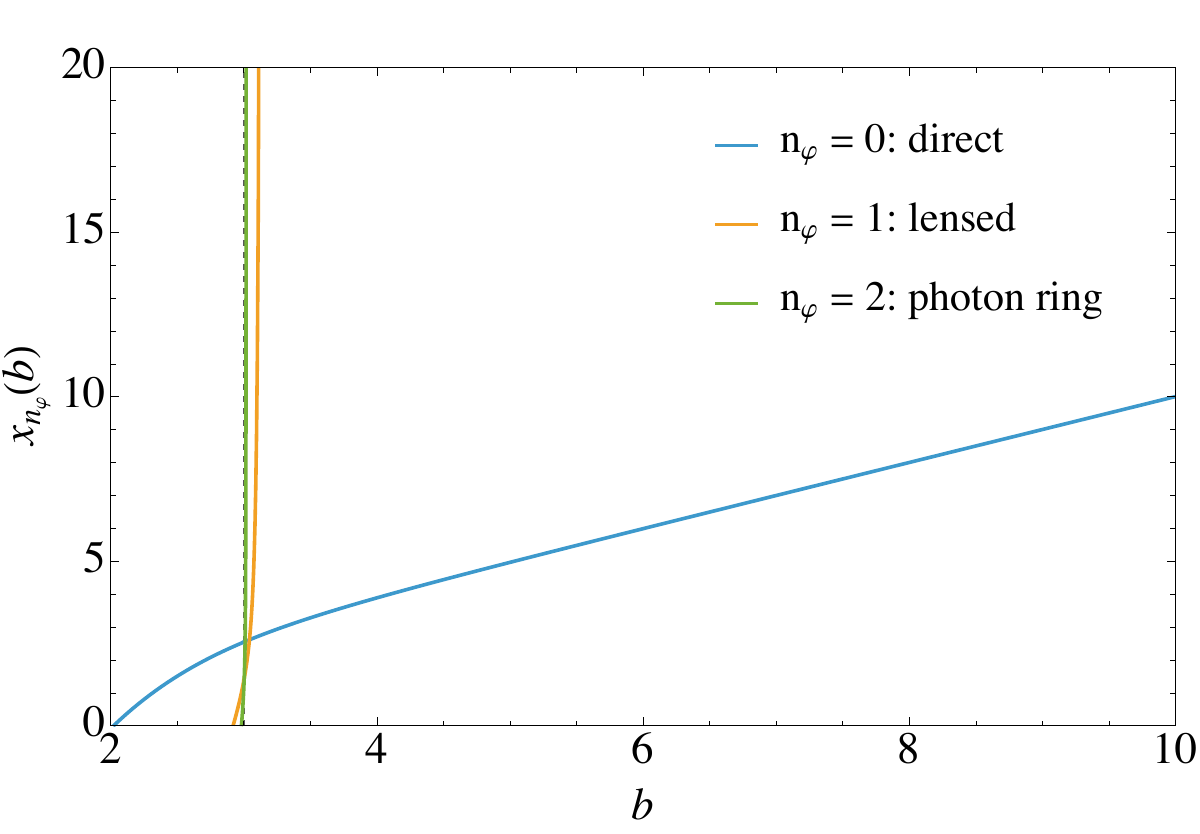}
    \caption{Transfer functions $x_{n_\phi}(b)$ for a Schwarzschild black hole with $M=1$ and for the massless GEB wormhole with $a=3$. From the upper left to the lower right, the panels correspond to Schwarzschild and to the GEB geometries with $n=2$, $4$, and $6$, respectively. The blue, orange, and green curves represent the direct emission ($n_\phi=0$), the lensed emission ($n_\phi=1$), and the first photon ring contribution ($n_\phi=2$), respectively. The vertical dashed lines indicate the corresponding critical impact parameters, $b_c=3\sqrt{3}$ for Schwarzschild and $b_c=a=3$ for the GEB wormholes.}
    \label{fig:transfer_GEB}
\end{figure*}

We now examine how the different regions of the disk are mapped onto the observer's screen. Fig.~\ref{fig:transfer_GEB} shows the transfer functions for the GEB wormhole with $n=2$, $4$, and $6$. The Schwarzschild result is also included as a reference. In all cases, the direct branch extends over a broad interval of impact parameters and varies relatively slowly, indicating that it undergoes the weakest radial demagnification. By contrast, the lensed and photon ring branches are considerably steeper and are concentrated near the corresponding critical impact parameter. These branches therefore map extended portions of the disk into narrow intervals on the observer's screen. This compression is particularly strong for the first photon ring contribution, whose transfer function lies very close to the critical curve.

Although all three GEB configurations have the same critical impact parameter, $b_c=a=3$, their transfer functions are not identical. The direct branches approach the same behavior at large $b$, as expected from their common asymptotically flat limit, while the differences become more evident close to the throat. In particular, changing $n$ modifies the curvature and radial extent of the lensed and photon ring branches near $b_c$. The comparison between $n=2$ and the degenerate cases $n=4$ and $6$ therefore shows that the position of the critical curve alone does not completely determine the mapping of the higher-order images.

\begin{figure*}[!htb]
    \centering
    \includegraphics[width=0.9\linewidth]{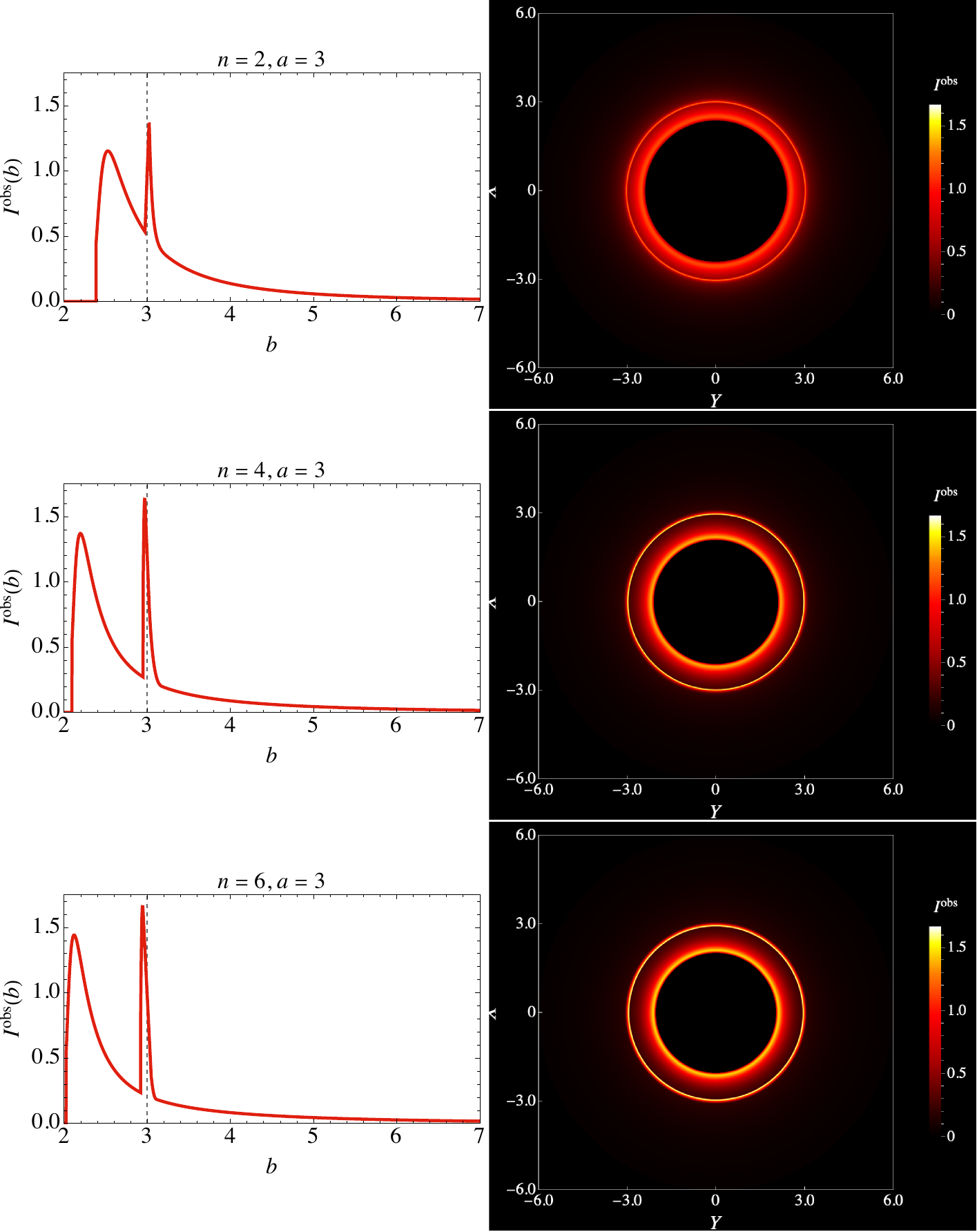}
   
    \caption{Observed intensity profiles and corresponding optical images of the massless GEB wormhole with $a=3$. From top to bottom, the rows correspond to $n=2$, $4$, and $6$, respectively. The left panels show the total observed intensity $I^{\rm obs}$ obtained by summing the direct, lensed, and first photon ring contributions, while the right panels display the corresponding face-on images. The vertical dashed lines indicate the critical impact parameter $b_c=a=3$. All configurations employ the same normalized emission profile with $\mu=0$, $\sigma=1/4$, and $\gamma=-2$, and only disk intersections in the region $x>0$ are retained. Thus, the central dark region results from the absence of emission from the opposite asymptotic region and should not be interpreted as photon capture by an event horizon.}
    \label{fig:Shadow_GEB}
\end{figure*}

The resulting observed intensity profiles and corresponding two-dimensional optical images are presented in Fig.~\ref{fig:Shadow_GEB}. Although the three GEB configurations have the same critical impact parameter, $b_c=a=3$, and employ the same emission profile, their radial intensity distributions are clearly different. This confirms that the critical curve fixes the characteristic position of the higher-order structures but does not determine their brightness or radial distribution.

All three profiles exhibit a broad maximum at $b<b_c$, followed by a narrow and more intense structure around the critical impact parameter. The broad component is primarily associated with the direct image, whereas the narrow feature results from the concentration and overlap of the lensed and first photon ring contributions near $b_c$. The observed intensity can therefore exceed unity even though the emitted profile is normalized according to $\max I^{\rm em}(x)=1$, since $I^{\rm obs}(b)$ is obtained by summing the radiation from different disk intersections. For $b>b_c$, the intensity decreases smoothly as the direct transfer function samples increasingly distant regions of the disk, where the adopted emission profile rapidly decays.

Increasing $n$ shifts the onset and the broad maximum of the observed intensity toward smaller impact parameters. Consequently, the broad emission annulus extends farther inward and the radius of the central dark region decreases from $n=2$ to $n=6$. By contrast, the narrow outer ring remains located at $b_c=3$ in all three images. Its peak intensity depends nonmonotonically on $n$, reaching its largest value for $n=4$ and decreasing again for $n=6$. Thus, the transition from the nondegenerate throat photon sphere at $n=2$ to the degenerate cases with $n>2$ modifies the relative brightness and width of the image components without changing the radius of the critical curve.

The central dark region is not produced by photon capture, since the GEB wormhole has no event horizon. Instead, it follows from restricting the emitting disk to the same asymptotic region as the observer and excluding radiation originating from the opposite side of the throat.

\subsection{Optical appearance of the GBL geometry}
\label{subsec:OA_GBL}
We now turn to the optical appearance of the GBL geometry. In contrast to the massless GEB wormhole, the GBL space-time possesses a nontrivial metric function $f(x)$ and contains the mass parameter $M$. Consequently, the gravitational frequency shift between the disk and the observer must be retained, and the observed intensity is determined by the combined effects of the transfer functions, the redshift factor, and the adopted emission profile.

Accordingly, the total observed intensity for the GBL geometry is given by
\begin{equation}
I_{\rm GBL}^{\rm obs}(b)=\sum_{n_\phi=0}^{N}\left[\frac{f\left(x_{n_\phi}(b)\right)}{f(x_{\rm O})}\right]^2I^{\rm em}\left(x_{n_\phi}(b)\right).
\label{GBL_total_obs_intensity}
\end{equation}
Unlike the GEB result, each contribution is weighted by the gravitational redshift between the corresponding disk intersection and the observer. Since the observer is located at spatial infinity, $f(x_{\rm O})\rightarrow1$.

Depending on the value of the bounce parameter $a$, the GBL geometry may describe either a black hole or a traversable wormhole. These two branches must be considered separately because they possess different causal structures and inner emitting regions. In the black hole configurations, only intersections outside the outer event horizon, satisfying $x_{n_\phi}(b)>x_h$, contribute to the observed intensity, while photons reaching the horizon are captured. In the wormhole configurations, both the disk and the observer are placed in the region $x>0$, and no radiation is assumed to originate from the opposite asymptotic region. Accordingly, only intersections satisfying $x_{n_\phi}(b)>0$ are retained.

The characteristic emission models also differ between the two branches. For the black hole configurations, we consider profiles associated with the event horizon, the unstable photon sphere, and the ISCO. Since the wormhole configurations possess no event horizon, the corresponding inner emission model is replaced by a profile concentrated near the throat, while the light ring and ISCO profiles are retained. For each causal branch, we examine $n\in{2,4,6}$ to determine how the generalization parameter modifies the direct and higher-order image contributions.

To construct the representative emission models, we set $M=1$ and use the same parameter combinations adopted in the geodesic analysis. For each configuration, the location parameter $\mu$ of the SU distribution is associated with one of the relevant geometrical scales. In the black hole branch, we set $\mu=x_h$, $x_{\rm ph}$, and $x_{\rm ISCO}$ for the event horizon, light ring, and ISCO profiles, respectively. In the wormhole branch, the event horizon profile is replaced by the choice $\mu=x_{\rm th}=0$, while the other two profiles are associated with the outer unstable circular photon orbit and the ISCO. The same values $\gamma=-2$ and $\sigma=1/4$ are used for all configurations so that the comparison is not affected by changes in the intrinsic shape of the emission model. For these parameters, the maximum does not coincide exactly with $\mu$. This accounts for the displacement between the vertical dashed lines, which indicate the characteristic geometrical positions, and the colored points marking the maxima of the profiles. Each profile is normalized independently such that $\max_x I^{\rm em}(x)=1$.

\begin{figure*}
    \centering
    \includegraphics[width=1\linewidth]{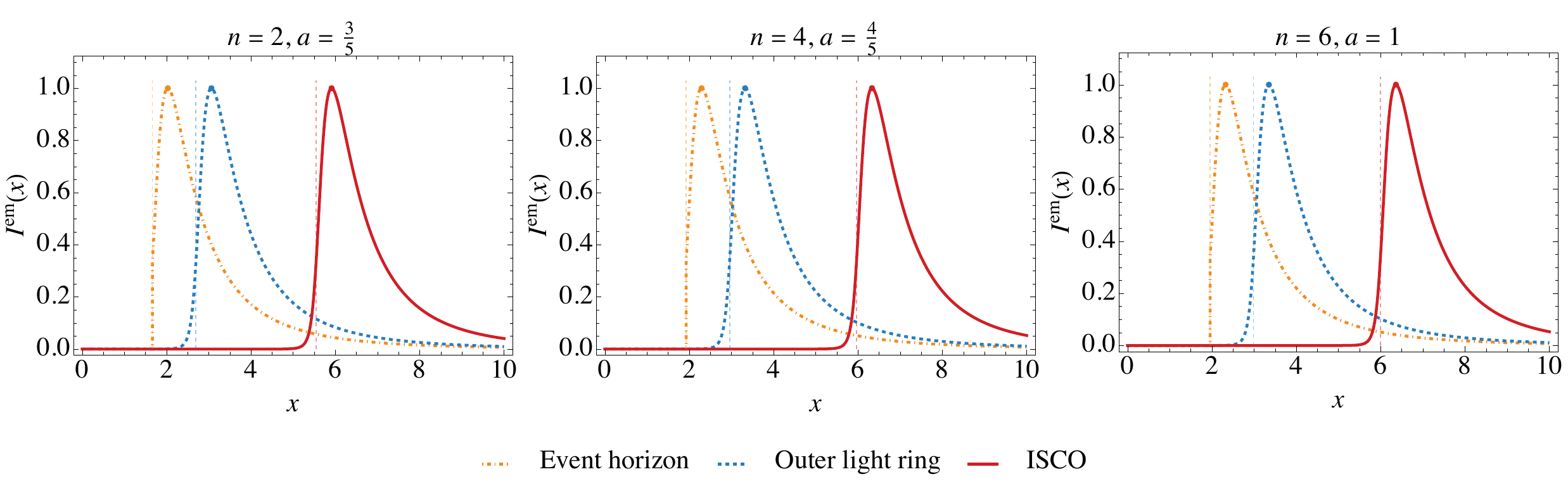}
    \caption{Normalized SU emitted intensity profiles $I^{\rm em}(x)$ adopted for the GBL black bounce configurations with $M=1$. From left to right, the panels correspond to $(n,a)=(2,3/5)$, $(4,4/5)$, and $(6,1)$. The orange dot-dashed, blue dashed, and red solid curves represent the profiles associated with the event horizon, the unstable circular photon orbit, and the ISCO, respectively. The vertical dashed lines indicate the corresponding characteristic positions, while the colored points mark the maxima of the normalized profiles. In all cases, $\gamma=-2$ and $\sigma=1/4$.
}
    \label{fig:Perfil_em_BB}
\end{figure*}

\begin{figure*}
    \centering
    \includegraphics[width=1\linewidth]{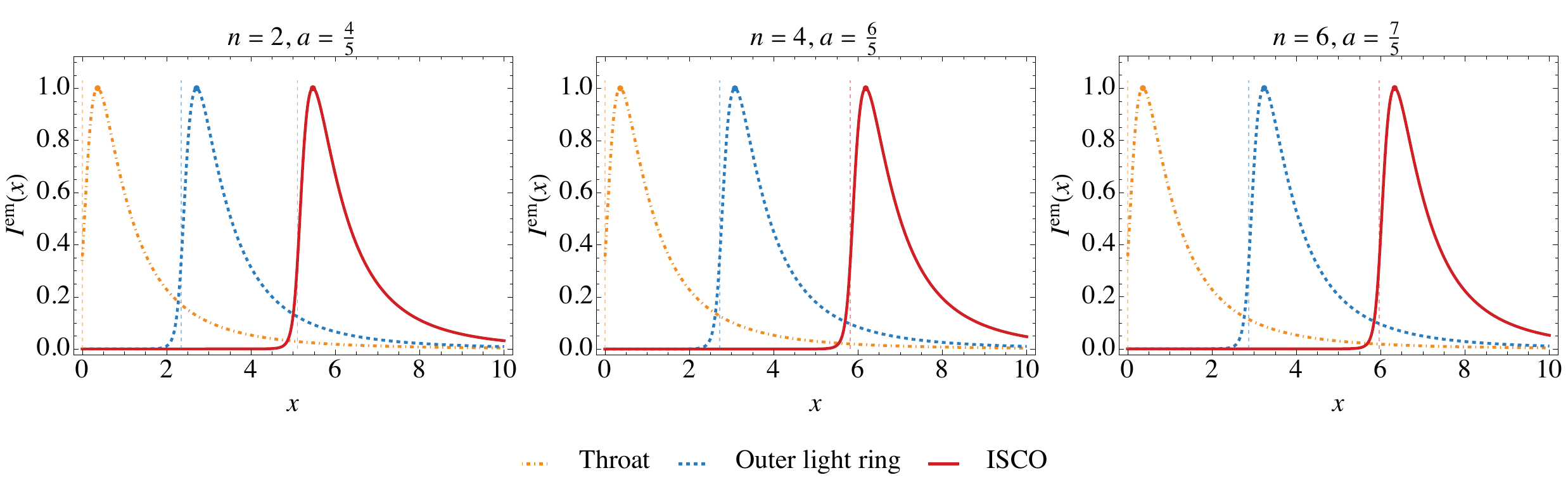}
    \caption{Normalized SU emitted intensity profiles $I^{\rm em}(x)$ adopted for the GBL wormhole configurations with $M=1$. From left to right, the panels correspond to $(n,a)=(2,4/5)$, $(4,6/5)$, and $(6,7/5)$. The orange dot-dashed, blue dashed, and red solid curves represent the profiles associated with the throat, the outer unstable circular photon orbit, and the ISCO, respectively. The vertical dashed lines indicate the corresponding characteristic positions, while the colored points mark the maxima of the normalized profiles. In all cases, $\gamma=-2$ and $\sigma=1/4$.
}
    \label{fig:Perfil_em_WH}

\end{figure*}

The resulting emitted intensity profiles for the black hole and wormhole branches are shown in Figs.~\ref{fig:Perfil_em_BB} and \ref{fig:Perfil_em_WH}, respectively. In the black hole configurations, the emission is restricted to the region outside the event horizon, and only the unstable circular photon orbit accessible to the external observer is considered. In the wormhole configurations, the disk is restricted to $x>0$, and the profile associated with the throat provides the innermost emission model. The common choice of $\gamma$ and $\sigma$ allows changes in characteristic positions to be examined in the selected configurations $(n,a)$ without simultaneously changing the shape of the local emission profile.

\begin{figure*}[!htb]
    \centering
    \includegraphics[width=1\linewidth]{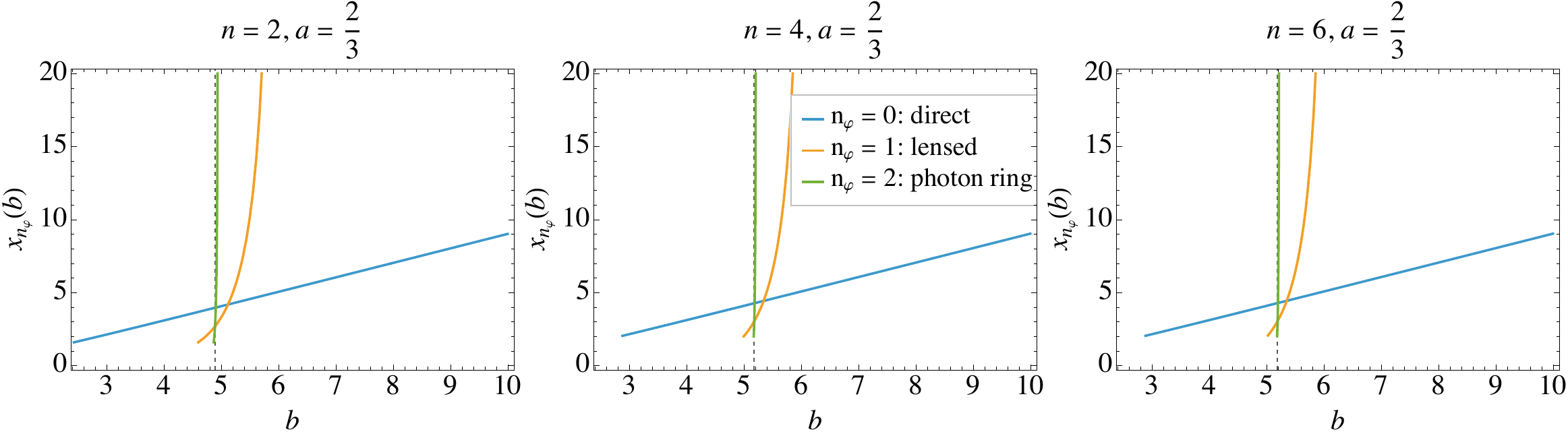}
    \includegraphics[width=1\linewidth]{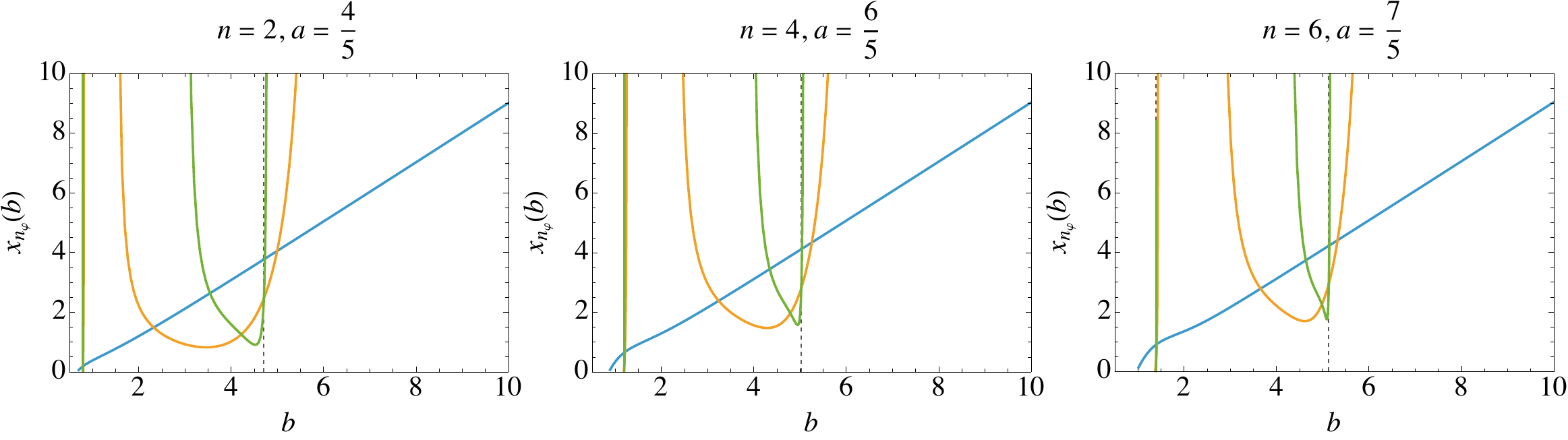}
    \includegraphics[width=1\linewidth]{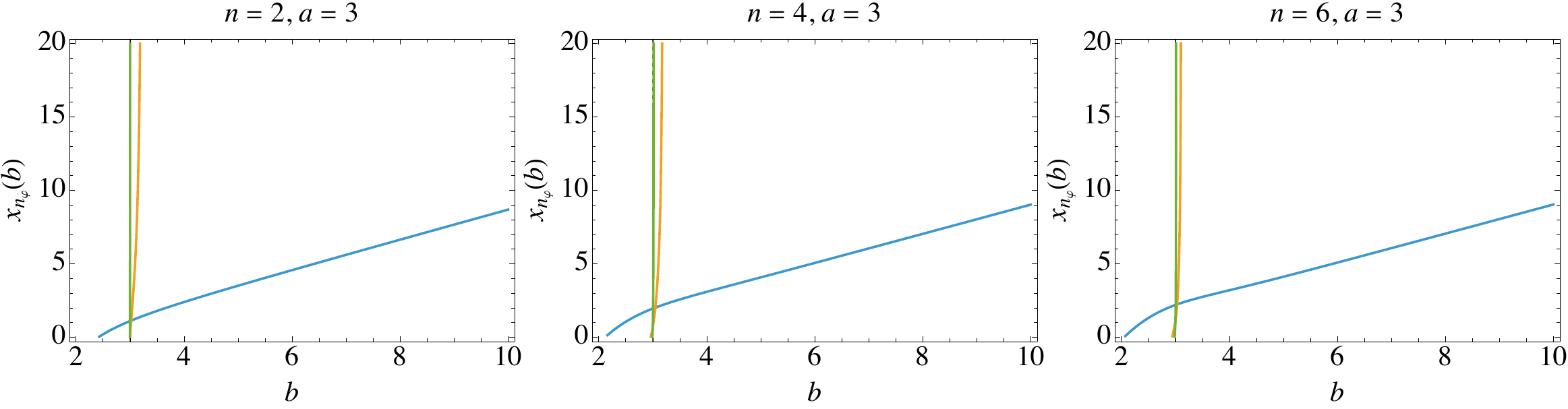}
    \caption{Transfer functions $x_{n_\varphi}(b)$ for the GBL geometry with $M=1$. From left to right, the columns correspond to $n=2$, $4$, and $6$. The upper row shows the black hole configurations with $a=2/3$. The middle row shows wormhole configurations with $a=4/5$, $6/5$, and $7/5$, which possess a throat, an intermediate stable, and an outer unstable photon sphere. The lower row corresponds to wormholes with $a=3$, for which only the throat circular photon orbit remains. The blue, orange, and green curves represent the direct, lensed, and first light ring contributions, associated with $n_\varphi=0$, $1$, and $2$, respectively. The vertical dashed lines indicate the critical impact parameters associated with the unstable circular photon orbits accessible from the observer’s asymptotic region.
}
    \label{fig:transfer_GBL}
\end{figure*}

The transfer functions for the representative GBL black hole and wormhole configurations are shown in Fig.~\ref{fig:transfer_GBL}. For the black hole configurations, the direct branch varies smoothly over a broad interval of impact parameters, whereas the lensed and photon ring branches become increasingly steep and accumulate near the critical impact parameter associated with the outer unstable circular photon orbit. These higher-order contributions are therefore mapped into narrow regions of the observer's screen and are strongly demagnified. Increasing $n$ produces only moderate shifts in the critical impact parameter and in the shapes of the transfer functions.

A qualitatively richer structure appears for the wormholes containing multiple light rings. In this case, the photon sphere in the throat and outer circular photon orbit define two critical scales, while the intermediate stable photon sphere modifies the photon trajectories between them. Consequently, the lensed and photon ring transfer functions develop nonmonotonic behavior and extend over distinct intervals of $b$. Some portions of these branches are considerably less steep than their black hole counterparts, indicating that the corresponding higher-order contributions may occupy broader regions of the observer's screen. The locations and widths of these features also vary with $n$, showing that the generalization parameter has a more pronounced optical effect when multiple light rings are present.

For $a=3$, the nonzero photon spheres disappear and only the one in the throat remains. The transfer functions then recover a considerably simpler structure: the direct contribution remains broad, while the lensed and photon ring branches are compressed around the throat critical impact parameter $b_c=a=3$. This behavior qualitatively resembles that found for the GEB wormhole, although the nontrivial metric function of the GBL geometry still modifies photon propagation. These configurations therefore provide a useful reference for assessing how the mass-dependent geometry and the gravitational redshift affect the observed intensity.

\begin{figure*}[!htb]
    \includegraphics[width=.785\linewidth]{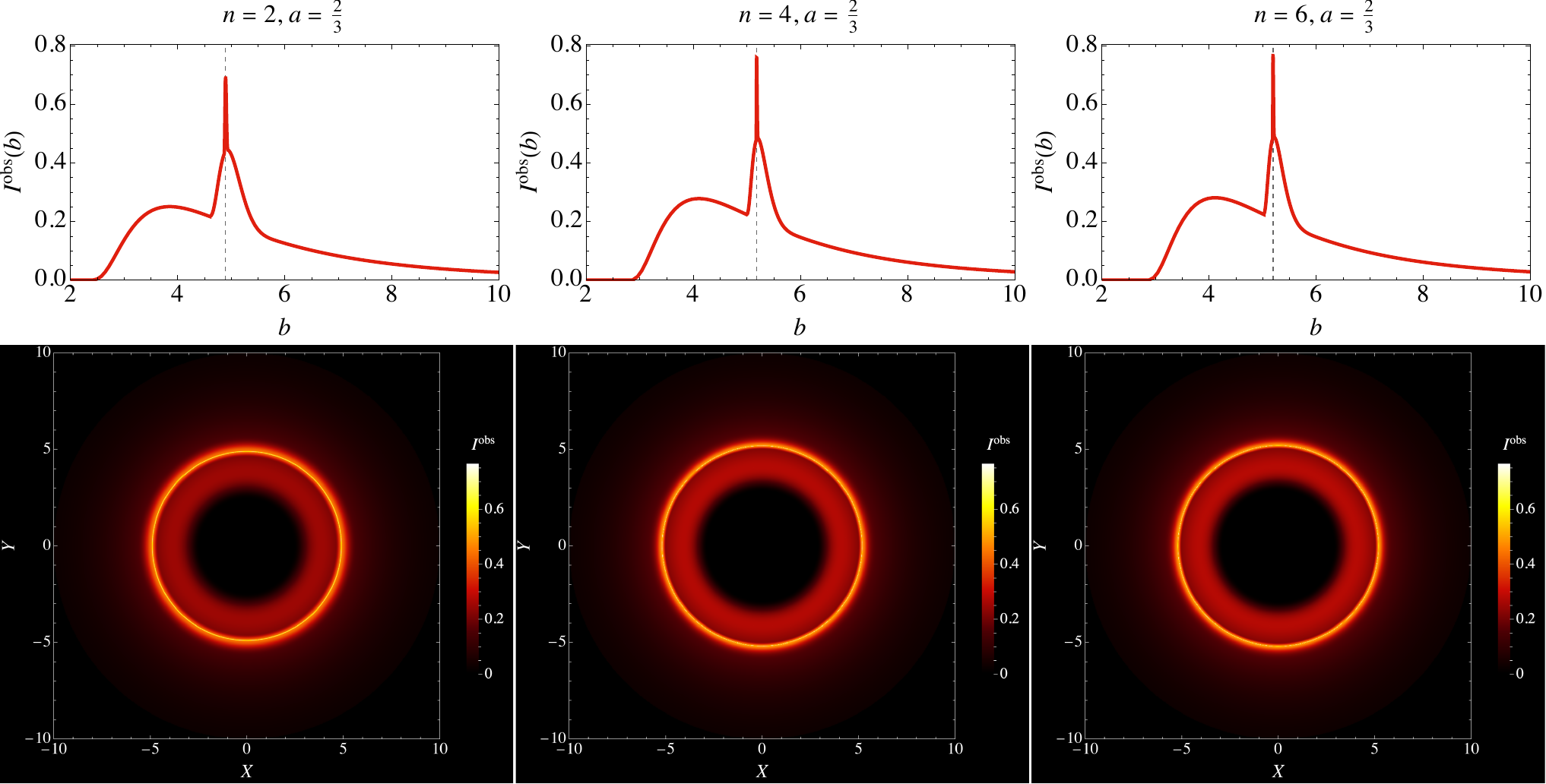}
    \includegraphics[width=.785\linewidth]{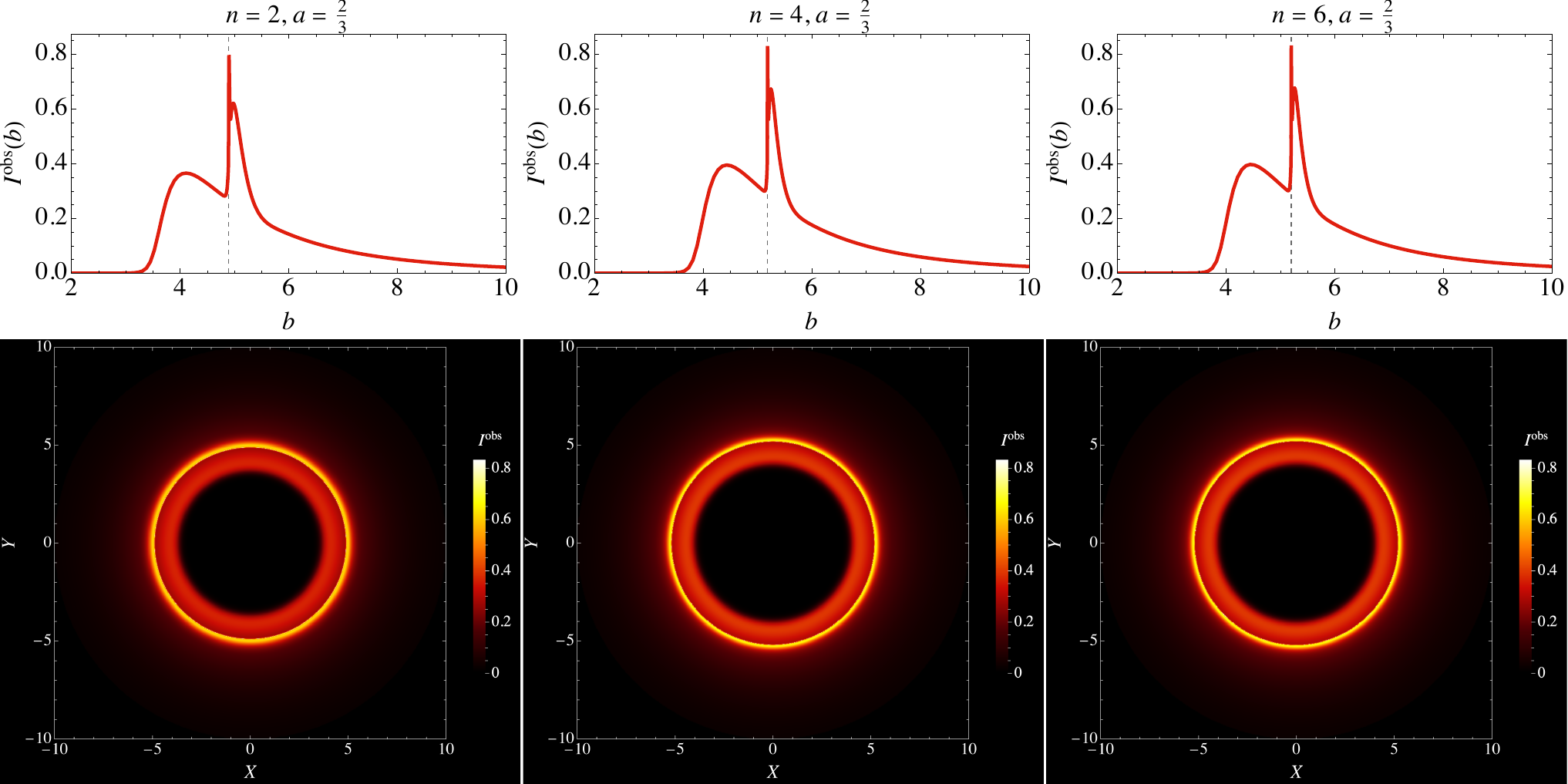}
    \includegraphics[width=.785\linewidth]{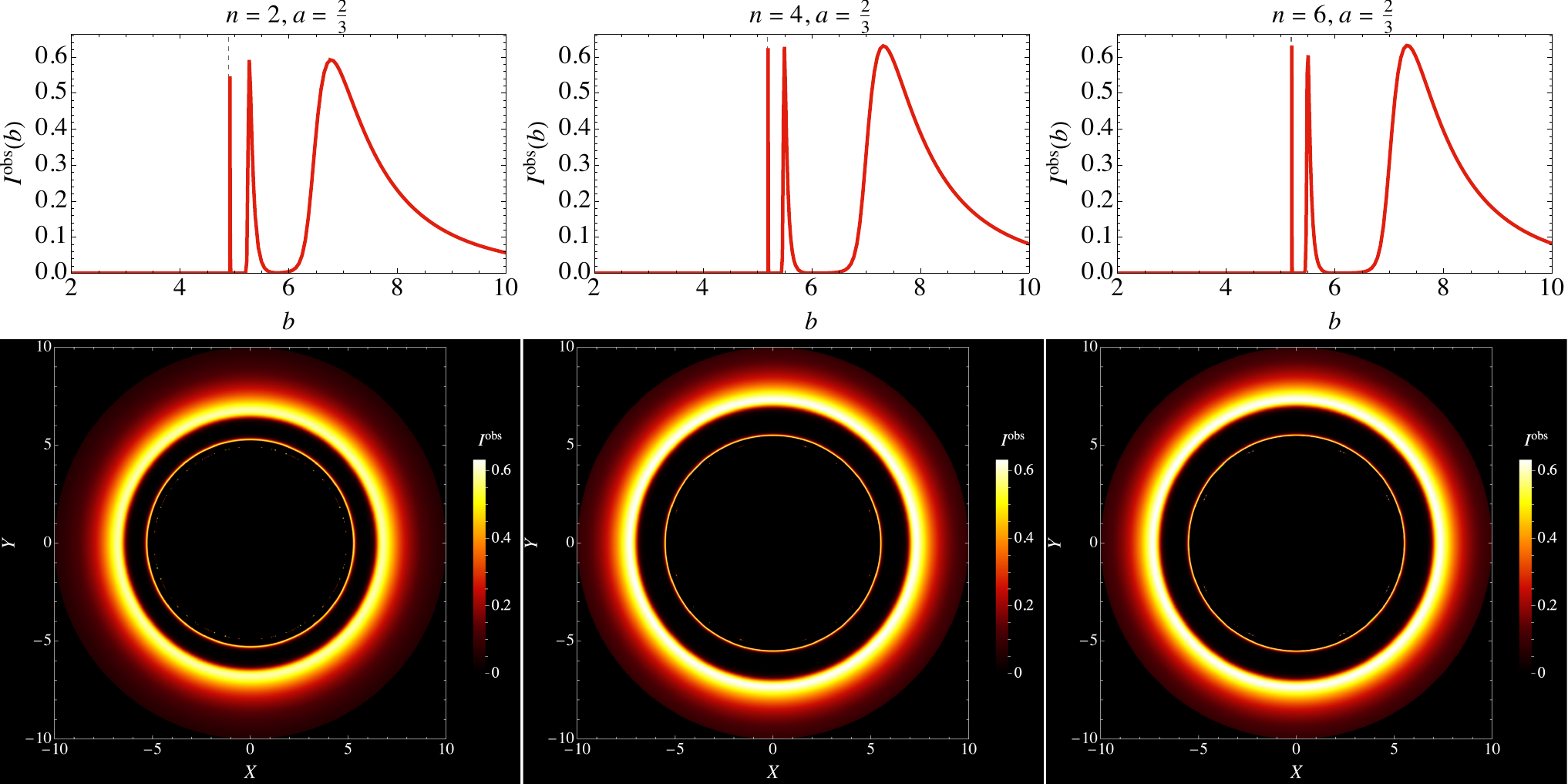}
    \caption{Observed intensity profiles and corresponding two-dimensional optical images for the GBL black hole configurations with $M=1$ and $a=2/3$. From left to right, the columns correspond to $n=2$, $4$, and $6$. The first, second, and third pairs of rows correspond, respectively, to the emission profiles associated with the event horizon, the outer unstable light ring, and the ISCO. Within each pair, the upper panels show the total observed intensity $I^{\rm obs}$, while the lower panels display the corresponding face-on images. The vertical dashed lines indicate the critical impact parameter $b_c$.
}
    \label{fig:Obs_GBL_BH}
\end{figure*}

The optical appearance of the GBL black bounce branch is comparatively simple and remains qualitatively similar for the three values of $n$ shown in Fig.~\ref{fig:Obs_GBL_BH}. For the emission profiles associated with the event horizon and the outer unstable light ring, the observed intensity contains a broad contribution produced mainly by the direct image, together with a narrow enhancement around $b_c$ generated by the accumulation of the lensed and first photon ring contributions. In the two-dimensional images, these components appear as a diffuse annulus surrounding a thin and brighter ring located near the critical curve. Although $b_c$ determines the characteristic radius of the critical curve, it does not coincide in general with the visible boundary of the central dark region. Indeed, photons with $b<b_c$ may still intersect the disk before eventually falling through the event horizon, thereby partially illuminating the interior of the critical curve. Consequently, the apparent size and brightness distribution of the central depression depend on the adopted emission profile. For the ISCO emission profile, the dominant direct contribution is displaced toward larger impact parameters, producing a broad outer annulus. At the same time, a narrow ring remains close to $b_c$, separated from the outer emission region by a comparatively dim interval. Thus, changing the characteristic emitting radius modifies the radial brightness distribution without significantly altering the critical scale. Increasing $n$ produces only moderate changes in the position and relative intensity of these structures, and the overall black hole morphology remains nearly unchanged.

\begin{figure*}[!htb]
    \includegraphics[width=.785\linewidth]{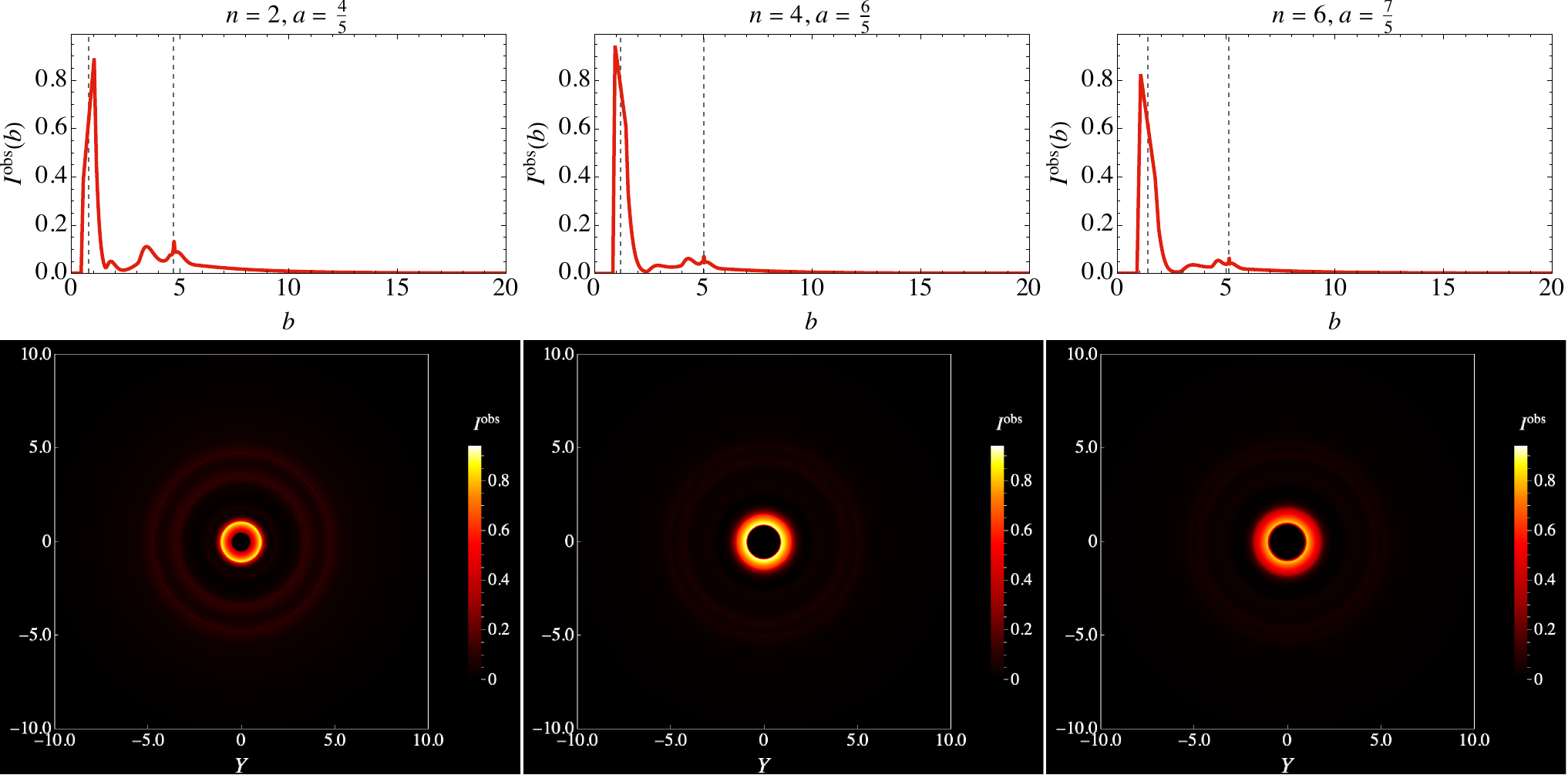}
    \includegraphics[width=.785\linewidth]{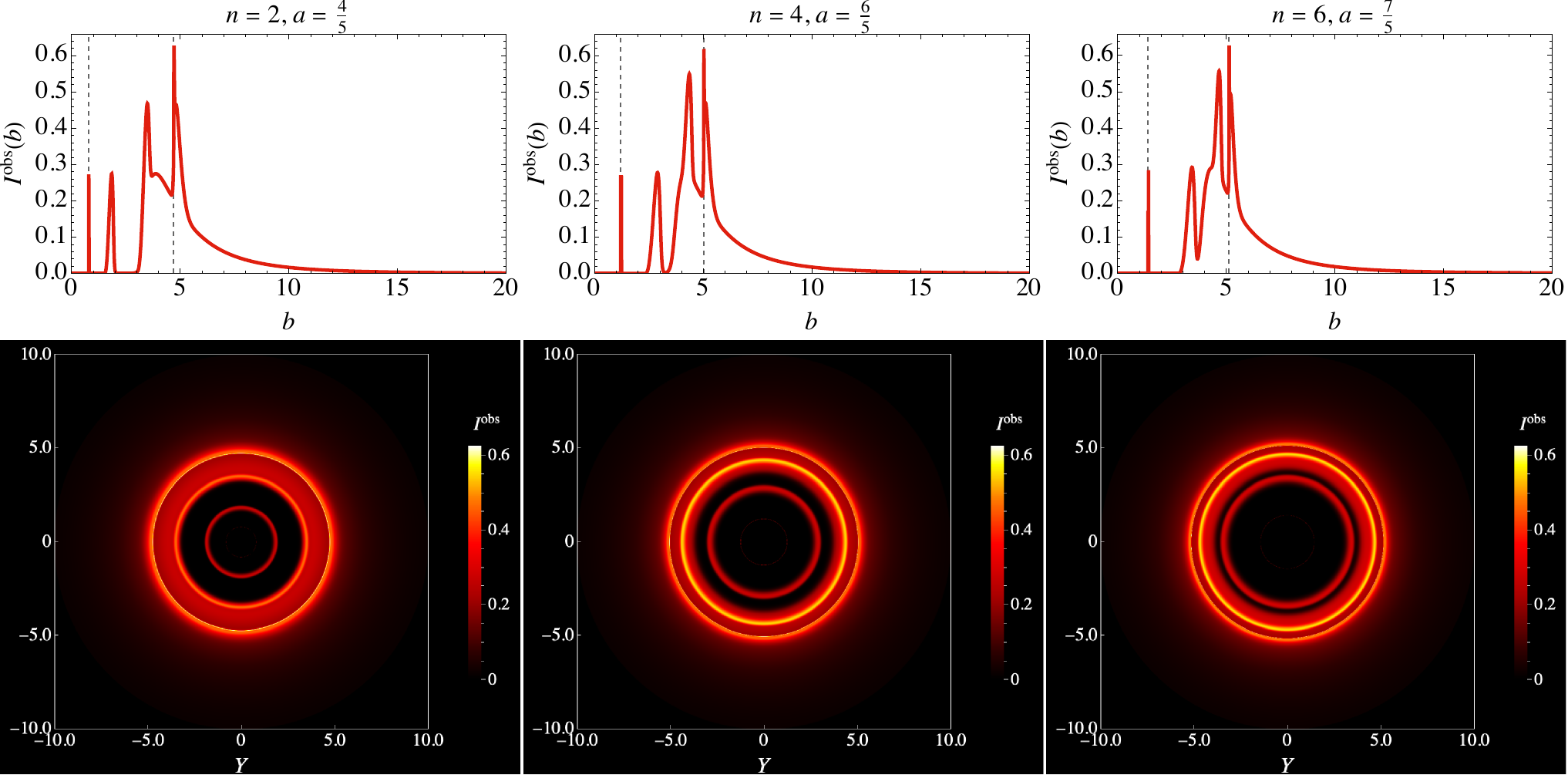}
    \includegraphics[width=.785\linewidth]{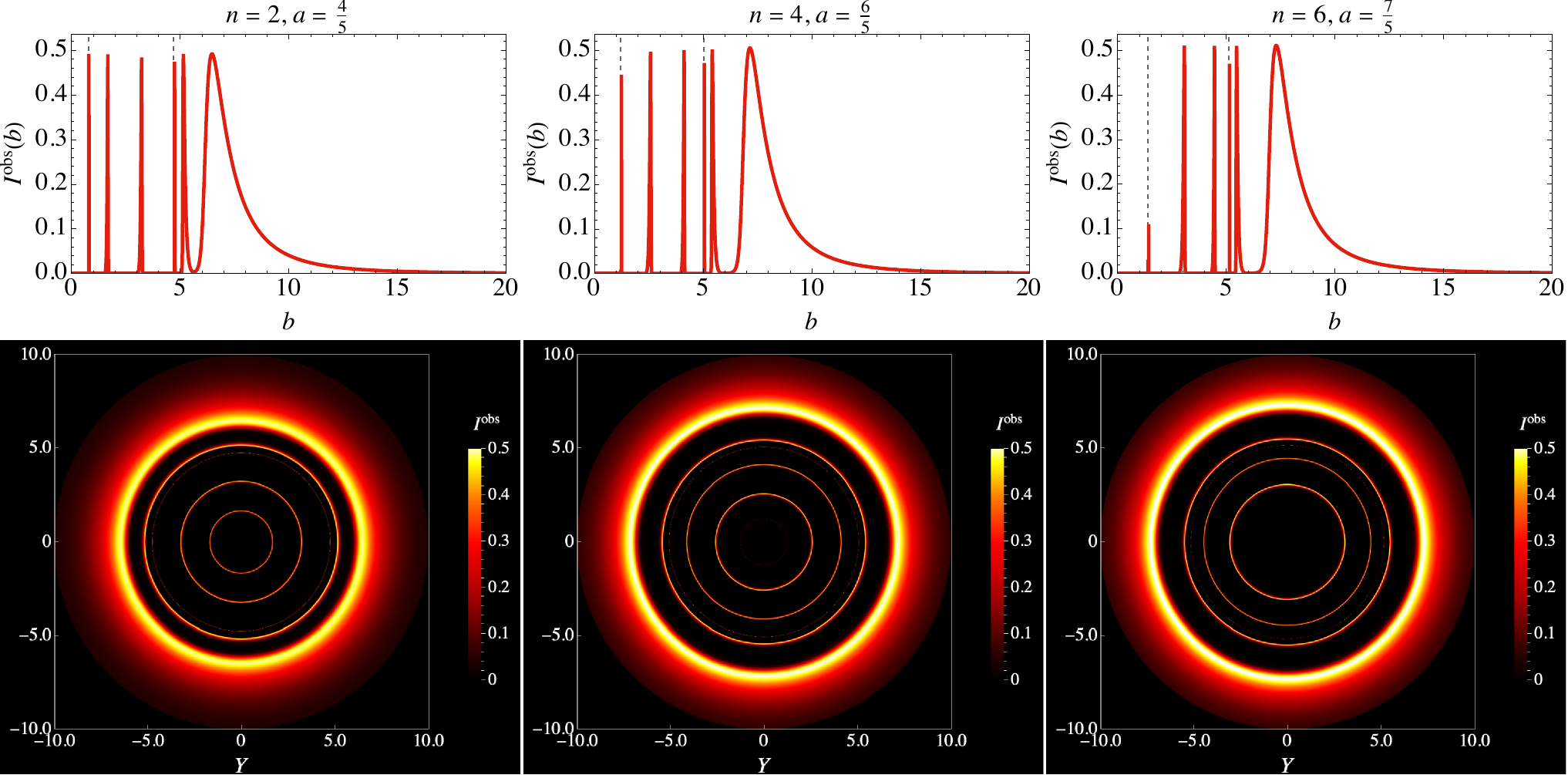}
    \caption{Observed intensity profiles and corresponding two-dimensional optical images for the GBL wormhole configurations with $M=1$. From left to right, the columns correspond to $(n,a)=(2,4/5)$, $(4,6/5)$, and $(6,7/5)$. These parameter choices place each configuration in the horizonless wormhole regime while ensuring the simultaneous existence of an outer unstable circular photon orbit and an ISCO. The first, second, and third pairs of rows correspond, respectively, to emission profiles associated with the throat, the outer unstable photon sphere, and the ISCO. Within each pair, the upper panels show the total observed intensity $I^{\rm obs}$, while the lower panels display the corresponding face-on optical images. The vertical dashed lines indicate the critical impact parameters.}
    \label{fig:Obs_GBL_WH_Multiple}
\end{figure*}

A considerably richer optical structure is found for the GBL wormholes with multiple light rings, shown in Fig.~\ref{fig:Obs_GBL_WH_Multiple}. In these configurations, the unstable photon sphere at the throat and the outer one introduce two distinct critical scales, while the stable one located between them modifies the propagation of photons in the intermediate region. As a result, the lensed and first photon ring contributions are divided into several disconnected or rapidly varying branches, producing multiple narrow peaks in $I^{\rm obs}(b)$ and a sequence of nested rings in the corresponding images.

When the emission profile is associated with the throat, the optical appearance is dominated by a compact and bright inner annulus close to the critical curve generated by the throat photon sphere. The structures associated with the outer critical scale are still present, but they are considerably fainter. By contrast, when the emission is concentrated near the outer unstable circular photon orbit, several annular components acquire comparable visibility, reflecting the simultaneous influence of the two unstable photon orbits and of the stable orbit between them. For the ISCO profile, the broadest and brightest component is displaced outward, while the higher-order images remain visible as a sequence of thin rings inside the main emission annulus. The number, separation, and relative brightness of these rings vary across the selected configurations. These differences should be understood as the combined effect of changing both $n$ and $a$, since the parameter pairs were chosen to preserve the multiple light ring wormhole regime rather than to isolate the dependence on $n$ alone. Nevertheless, the comparison clearly shows that the presence of a stable circular photon orbit and two unstable critical scales can produce a much more structured optical appearance than in the black hole branch.

\begin{figure*}[!htb]
    \includegraphics[width=.9\linewidth]{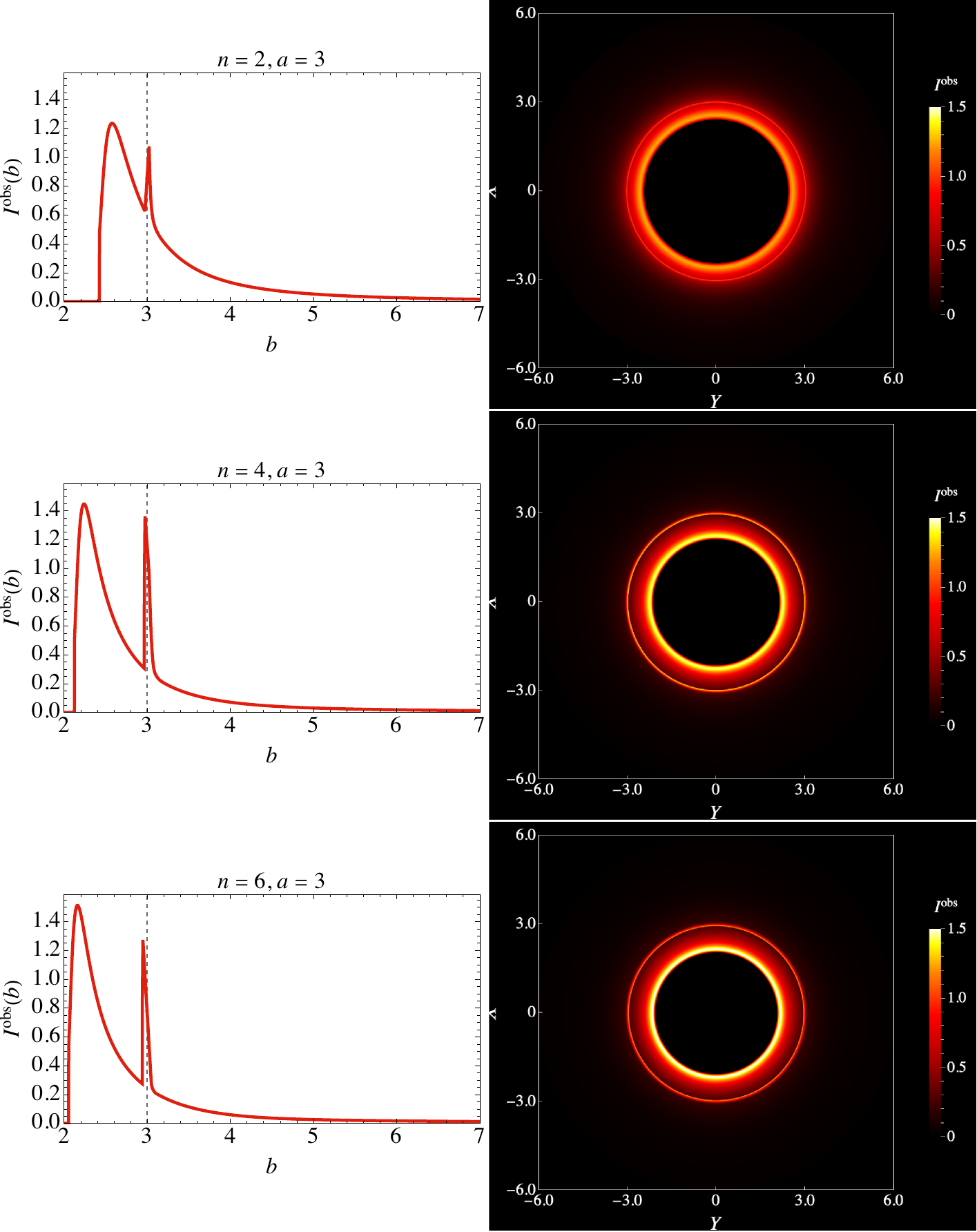}
    \caption{Observed intensity profiles and corresponding optical images
for the GBL wormhole configurations with $M=1$ and $a=3$. From top to
bottom, the rows correspond to $n=2$, $4$, and $6$, respectively. The
left panels show the total observed intensity $I^{\rm obs}$ obtained
by summing the direct, lensed, and first photon-ring contributions,
while the right panels display the corresponding face-on optical
images. All configurations employ the same SU emission profile associated with the throat, with $\mu=0$,
$\sigma=1/4$, and $\gamma=-2$. Only disk intersections in the region
$x\geq0$ are retained. The vertical dashed line indicates the unique
critical impact parameter $b_c=3$, associated with the circular photon orbit at
the throat. The central dark region results from the absence of
emission from the opposite asymptotic region.}
    \label{fig:Obs_GBL_WH}
\end{figure*}

The configurations shown in Fig.~\ref{fig:Obs_GBL_WH} provide a direct massive counterpart of the GEB wormholes discussed previously. In both cases, the areal function is characterized by the same values $a=3$ and $n=2,4,6$, the only unstable circular photon orbit is located at the throat, and the corresponding critical impact parameter is $b_c=3$. Moreover, the same throat emission profile and the same restriction to the region $x\geq0$ are employed. Consequently, the characteristic radii of the main annular structures and the origin of the central dark region are closely related in the two families. The introduction of the mass parameter does not change the throat radius or the position of the critical curve in this particular comparison, but it modifies both the geodesic mapping and the intensity received from each disk intersection. Unlike the massless GEB geometry, the GBL contributions are weighted by the gravitational redshift factor. The combined effect of this redshift and of the modified transfer functions changes the relative balance between the broad direct component and the narrow feature near $b_c$. In particular, the critical peak is less prominent relative to the broad inner maximum in the massive GBL configurations than in their massless GEB counterparts, especially for $n=4$ and $n=6$. The corresponding two-dimensional images therefore preserve a similar overall ring scale while exhibiting different radial brightness distributions. This comparison illustrates that the critical impact parameter alone is insufficient to distinguish the massless and massive wormhole geometries: the mass deformation may leave the characteristic radius unchanged while producing observable differences in the relative intensity and width of the direct and higher-order image components.

Taken together, these results emphasize the distinction between a critical curve, a central dark region, and a genuine black hole shadow. The critical curves are determined by unstable photon orbits, whereas the complete brightness distribution depends on the disk emissivity, the transfer functions, and the gravitational redshift. 
\section{Conclusions and final remarks}\label{sec:con}

In this work, we presented a general framework to construct massive black bounce geometries starting from massless wormhole seed space-times. By introducing a position-dependent mass function $M(x)$ into the temporal and radial metric components without altering the intrinsic areal radius profile $r(x)$, we successfully supplied an ADM mass to the geometry while preserving its underlying throat structure. By promoting massless wormhole geometries to black bounce space-times, it becomes possible to study and compare these geometries against the latest astrophysical observational data on compact objects.

We applied this framework to a class of wormholes that has recently gained significant attention in the literature: the GEB wormhole. By applying our framework to this geometry, we promoted it to what we call GBL black bounce. This structure smoothly interpolates between a regular black hole and a traversable wormhole, retaining a throat structure analogous to that of the GEB wormhole. With this geometry in hand, we conducted a detailed study of the trajectories of neutral massive particles in this space-time. We concluded that, contrary to the GEB case, where there is only a single unstable orbit located at the throat, the GBL black bounce possesses a highly rich structure of massive particle orbits. Depending on the parameters, it allows for the presence of two unstable particle orbits or even ISCOs, as shown in the diagram presented in Fig.~\ref{fig:massivebb} for the horizonless case.

To analyze the optical properties of these geometries, we performed a detailed study of photon trajectories. For the GEB case, due to the simplified form of the effective potential, photon trajectories exhibit a behavior quite similar to that of massive particles, yielding only a single identifiable unstable photon orbit. On the other hand, for different values of the deformation parameters $(a,m)$, the photon orbits in the GBL black bounce present a rather exotic structure, allowing for the existence of multiple light rings.

Continuing our analysis, we calculated the theoretical shadow radius of the GBL black bounce and compared our results with the observational data obtained by the EHT for Sgr A*. The results in Fig.~\ref{fig:shadow_radius}, which displays the shadow radius as a function of the throat radius, both normalized by the ADM mass, shows that when the parameters allow for the formation of a horizon, the theoretical shadow radius remains within the $1\sigma$ confidence band. Furthermore, the shadow radius remains approximately constant for larger values of $n$. This behavior can be understood by examining the effective potential for photons, shown in Fig.~\ref{fig:pot_massless_BD}, where a flattening of the potential maxima characterizing the photon sphere is observed; that is, as $n$ increases, the maxima increasingly resemble a broad potential barrier. For the horizonless case, considering the photon orbit located at the throat, we are able to constrain the normalized throat radius to $4.55 \lesssim a/m \lesssim 5.22$ within $1\sigma$; within $2\sigma$, we obtain $4.21 \lesssim a/m \lesssim 5.56$.

Furthermore, we investigated the optical appearance of these geometries when illuminated by a geometrically and optically thin equatorial accretion disk modeled via the phenomenological SU emission profile. By integrating null geodesics backwards from a face-on observer, we evaluated the direct, lensed, and first photon ring contributions to the total observed intensity. As a key result, we showed how the metric parameters dictate the luminous ring structure of the optical images. For instance, in the GEB case, where gravitational redshift is absent, varying $n$ modifies the brightness contrast between image orders without altering the critical impact parameter $b_c$. In contrast, for GBL wormholes featuring multiple circular photon orbits (two unstable and one intermediate stable), the lensed and photon ring transfer functions exhibit a non-monotonic structure, producing a rich sequence of discrete, nested rings in the observed intensity profile. The inclusion of the gravitational redshift factor in the GBL geometry further alters the relative power between direct and higher-order contributions, demonstrating that the full brightness profile, rather than the critical impact parameter alone, is essential for distinguishing between these compact object configurations.

This rich ring morphology, illustrated in Fig.~\ref{fig:Obs_GBL_WH_Multiple}, represents a highly exotic structure that markedly differs from the classic ring signatures produced by standard space-times such as Schwarzschild or Kerr black holes. Similar features were reported in Ref.~\cite{Guerrero:2022qkh}, which corresponds to a particular realization of the GBL geometry for $n=2$. For $n>2$, we observe that the spatial separation between the multiple observed light rings becomes progressively smaller, a behavior that is also clearly visible in the ray-tracing integration shown in Fig.~\ref{raytracingWH}.

\begin{figure*}[!htb]
    \includegraphics[width=.785\linewidth]{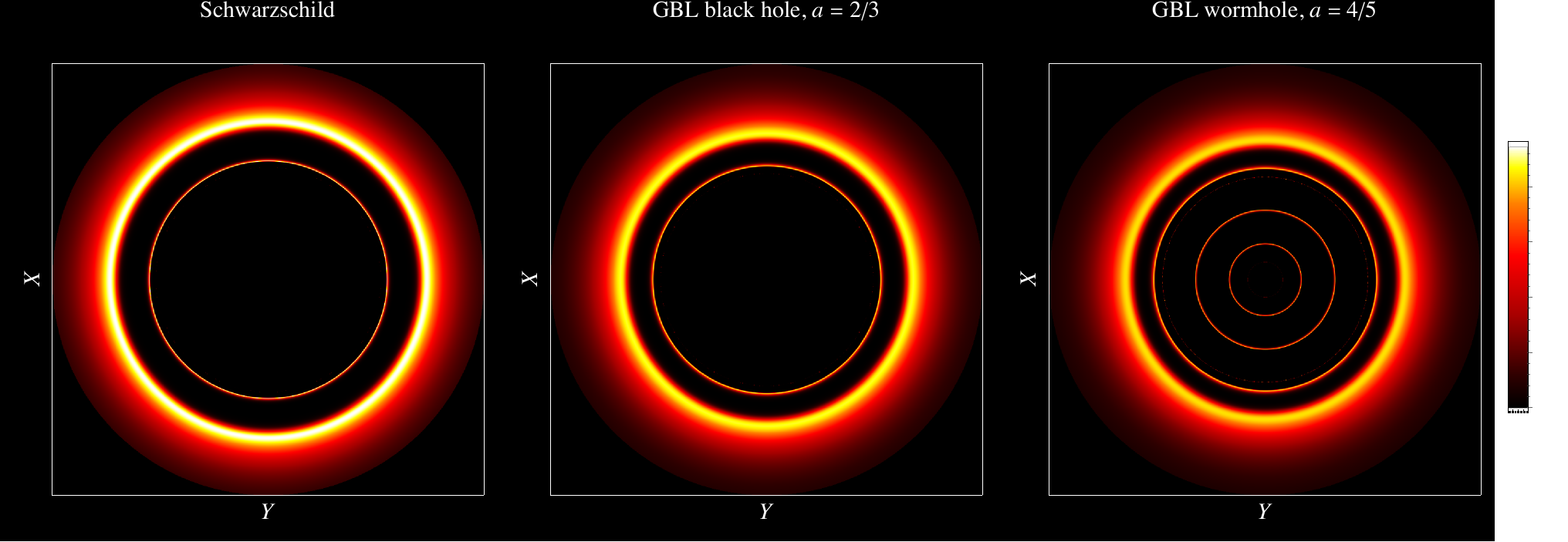}
    \includegraphics[width=.785\linewidth]{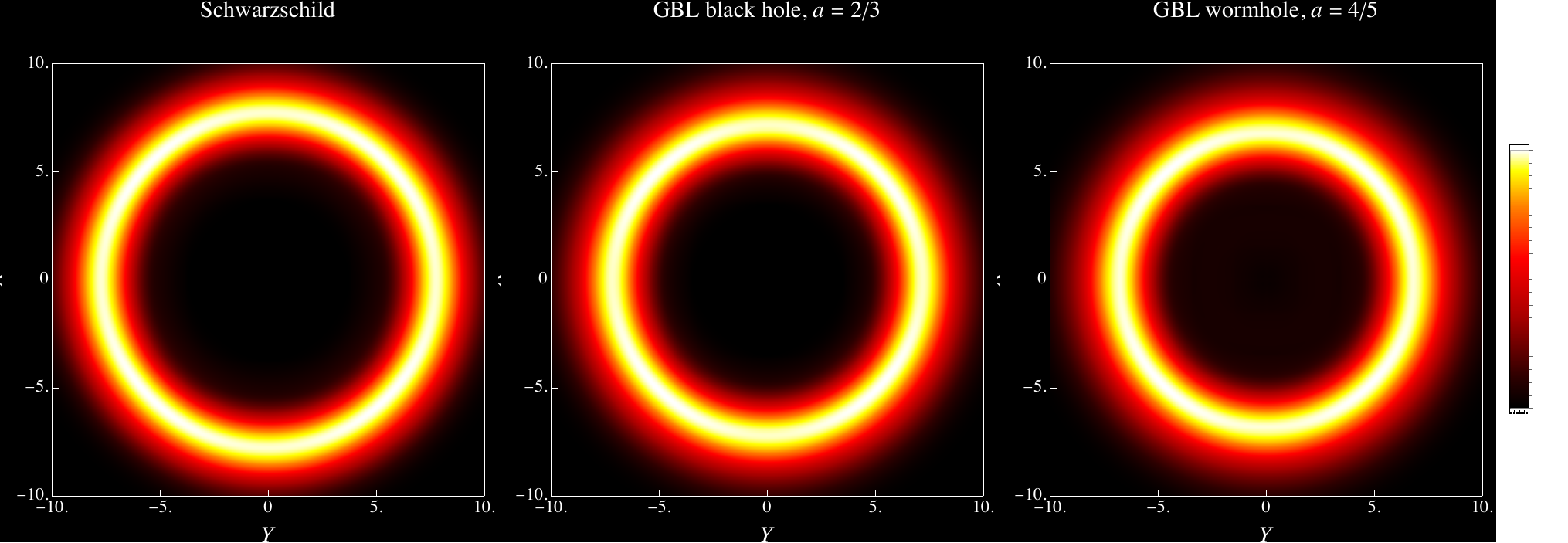}
    
    \caption{In the top panel, we display the optical images corresponding to a Schwarzschild black hole (left), a GBL black hole with $n=2$ and $a=2/3$ (middle), and a GBL wormhole with $n=2$ and $a=4/5$ (right). In the bottom panel, we show the corresponding images after applying a two-dimensional Gaussian filter designed to mimic the nominal observational resolution of the EHT.}
    \label{fig:Obs_GBL_WH_Multiplefilter}
\end{figure*}

To assess the actual observational feasibility of detecting such features, one must account for the finite resolution of current telescope arrays. To model this effect, we convolve our high resolution ray-traced images with a two-dimensional Gaussian filter matching the nominal beam width (Full Width at Half Maximum, FWHM) of EHT~\cite{EventHorizonTelescope:2019dse,Gralla:2019hhd}. The standard deviation $\sigma$ of the Gaussian filter is chosen to match the nominal EHT resolution, with the physical angular scale converted to pixels using the standard relation $\sigma = \frac{\text{FWHM}}{2\sqrt{2\ln 2}}$. The resulting blurred images, presented in Fig.~\ref{fig:Obs_GBL_WH_Multiplefilter}, show that the intricate sub-structures, such as the discrete multiple light rings and narrow photon rings generated near the throat, are entirely washed out by the observational smearing, merging into a single, smooth, and continuous ring surrounding a central dark region~\cite{Gralla:2019hhd,Johnson:2019ljv}.

It is important to emphasize that this procedure constitutes a simplified theoretical approximation. In practice, the EHT does not operate as a traditional optical telescope, but rather as a VLBI array, which incompletely samples spatial frequencies in the Fourier domain (the $u$-$v$ plane). A fully rigorous observational simulation would require generating synthetic interferometric visibilities, introducing atmospheric and instrumental noise, and applying non-linear image reconstruction algorithms.

Nevertheless, for the scope of the present work, such detailed modeling is not strictly required. Convolution with an equivalent Gaussian filter is a standard approach widely adopted in the theoretical literature and is well-suited to illustrate our main point: demonstrating how complex sub-structures, such as discrete multiple light rings and narrow photon rings generated by the throat dynamics, are completely washed out by spatial resolution limits. After filtering, these sharp features merge into a single, smooth, continuous emission ring surrounding a central dark region.

This beam smearing reveals a significant observational degeneracy: under current EHT resolution, the optical appearance of exotic compact objects like GBL black bounces or GEB wormholes becomes practically indistinguishable from that of a standard Schwarzschild or Kerr black hole of equivalent mass. The fine throat dynamics and light ring structures remain hidden behind the observational blur. Consequently, disentangling these exotic space-times from standard GR black holes will strictly require next-generation VLBI or space-based interferometers capable of reaching the extreme angular resolution necessary to resolve these sharp, high-order photon rings~\cite{Johnson:2019ljv,Ayzenberg:2023uhx}. In the meantime, further observational signatures of this generalized class of black bounces can be systematically investigated, such as studying their quasinormal modes and evaluating their correspondence with the photon ring/shadow parameters, a mapping already explored for the simpler $n=2$ case (see Ref.~\cite{Duran-Cabaces:2025sly}). Therefore, extending this analysis to the general $n$ regime, as well as considering new classes of wormholes with even richer ring structures, remain important directions for future work.

\section*{Acknowledgments}
K.B. acknowledges partial funding by the Ministry of Science and Higher Education of the Russian Federation, Project FSWU-2026-0010. M.S. thanks the Conselho Nacional de Desenvolvimento Científico e Tecnológico (CNPq), Brazil, for financial support through CNPq/PDE Grant No. 200218/2025-5. G.A. acknowledges support from CNPq, Brazil, through a PQ-1B Research Productivity Fellowship (Grant No. 308182/2025-1).  This study was financed in part by the Coordenação de Aperfeiçoamento de Pessoal de Nível Superior- Brasil (CAPES)- Finance Code 001. This work is also supported by the Spanish National Grant PID2024-157196NB-I00 funded by MICIU/AEI/10.13039/501100011033; the Department of Education, Junta de Castilla y Le\'on and FEDER Funds, Ref. CLU-2023-1-05.


\bibliography{Ref-a.bib}

@article{Bronnikov:2001tv,
    author = "Bronnikov, Kirill A.",
    title = "{Spherically symmetric false vacuum: No go theorems and global structure}",
    eprint = "gr-qc/0104092",
    archivePrefix = "arXiv",
    doi = "10.1103/PhysRevD.64.064013",
    journal = "Phys. Rev. D",
    volume = "64",
    pages = "064013",
    year = "2001"
}

@book{Bronnikov:2012ws,
    author = "Bronnikov, Kirill A. and Rubin, Sergey G.",
    title = "{Black Holes, Cosmology and Extra Dimensions}",
    doi = "10.1142/12186",
    isbn = "978-981-4374-20-0, 978-981-4440-02-8",
    publisher = "WSP",
    year = "2012"
}

@article{Bronnikov:2017kvq,
    author = "Bronnikov, K. A. and Baleevskikh, K. A. and Skvortsova, M. V.",
    title = "{Wormholes with fluid sources: A no-go theorem and new examples}",
    eprint = "1708.02324",
    archivePrefix = "arXiv",
    primaryClass = "gr-qc",
    doi = "10.1103/PhysRevD.96.124039",
    journal = "Phys. Rev. D",
    volume = "96",
    number = "12",
    pages = "124039",
    year = "2017"
}

@article{KordZangeneh:2025qel,
    author = "Kord Zangeneh, Mahdi and Lobo, Francisco S. N.",
    title = "{Evolving Wormholes in a Cosmological Background}",
    eprint = "2507.14750",
    archivePrefix = "arXiv",
    primaryClass = "gr-qc",
    doi = "10.3390/universe11070236",
    journal = "Universe",
    volume = "11",
    number = "7",
    pages = "236",
    year = "2025"
}

@article{Bambi:2021qfo,
    author = "Bambi, Cosimo and Stojkovic, Dejan",
    title = "{Astrophysical Wormholes}",
    eprint = "2105.00881",
    archivePrefix = "arXiv",
    primaryClass = "gr-qc",
    doi = "10.3390/universe7050136",
    journal = "Universe",
    volume = "7",
    number = "5",
    pages = "136",
    year = "2021"
}

@article{Hochberg:1998ha,
    author = "Hochberg, David and Visser, Matt",
    title = "{Dynamic wormholes, anti-trapped surfaces, and energy conditions}",
    eprint = "gr-qc/9802046",
    archivePrefix = "arXiv",
    reportNumber = "LAEFF-98-01",
    doi = "10.1103/PhysRevD.58.044021",
    journal = "Phys. Rev. D",
    volume = "58",
    pages = "044021",
    year = "1998"
}

@article{Bronnikov:2010hu,
    author = "Bronnikov, Kirill A. and Sushkov, Sergey V.",
    title = "{Trapped ghosts: a new class of wormholes}",
    eprint = "1001.3511",
    archivePrefix = "arXiv",
    primaryClass = "gr-qc",
    doi = "10.1088/0264-9381/27/9/095022",
    journal = "Class. Quant. Grav.",
    volume = "27",
    pages = "095022",
    year = "2010"
}

@article{Rodrigues:2025plw,
    author = "Rodrigues, Manuel E. and Silva, Marcos V. de S.",
    title = "{Spherically symmetric and static black bounces with multiple horizons, throats, and anti-throats in four dimensions}",
    eprint = "2502.00502",
    archivePrefix = "arXiv",
    primaryClass = "gr-qc",
    doi = "10.1088/1361-6382/adaf01",
    journal = "Class. Quant. Grav.",
    volume = "42",
    number = "5",
    pages = "055005",
    year = "2025"
}

@article{Bronnikov:2016xvj,
    author = "Bronnikov, K. A. and Galiakhmetov, A. M.",
    title = "{Wormholes and black universes without phantom fields in Einstein-Cartan theory}",
    eprint = "1607.07791",
    archivePrefix = "arXiv",
    primaryClass = "gr-qc",
    doi = "10.1103/PhysRevD.94.124006",
    journal = "Phys. Rev. D",
    volume = "94",
    number = "12",
    pages = "124006",
    year = "2016"
}

@article{Misner:1964je,
    author = "Misner, Charles W. and Sharp, David H.",
    title = "{Relativistic equations for adiabatic, spherically symmetric gravitational collapse}",
    doi = "10.1103/PhysRev.136.B571",
    journal = "Phys. Rev.",
    volume = "136",
    pages = "B571--B576",
    year = "1964"
}

@article{Bugaev:2024dte,
    author = "Bugaev, Mikhail and Novikov, Igor and Repin, Serge and Samorodskaya, Polina and I. D. Novikov, jr.",
    title = "{Observing of Background Electromagnetic Radiation of the Real Sky through the Throat of a Wormhole}",
    eprint = "2412.06872",
    archivePrefix = "arXiv",
    primaryClass = "gr-qc",
    doi = "10.1134/S1063772925701549",
    journal = "Astron. Rep.",
    volume = "69",
    number = "2",
    pages = "67--76",
    year = "2025"
}

@article{EventHorizonTelescope:2019dse,
    author = "Akiyama, Kazunori and others",
    collaboration = "Event Horizon Telescope",
    title = "{First M87 Event Horizon Telescope Results. I. The Shadow of the Supermassive Black Hole}",
    eprint = "1906.11238",
    archivePrefix = "arXiv",
    primaryClass = "astro-ph.GA",
    doi = "10.3847/2041-8213/ab0ec7",
    journal = "Astrophys. J. Lett.",
    volume = "875",
    pages = "L1",
    year = "2019"
}

@article{Xavier:2024iwr,
    author = "Xavier, S\'ergio V. M. C. B. and Herdeiro, Carlos A. R. and Crispino, Lu\'\i{}s C. B.",
    title = "{Traversable wormholes and light rings}",
    eprint = "2404.02208",
    archivePrefix = "arXiv",
    primaryClass = "gr-qc",
    doi = "10.1103/PhysRevD.109.124065",
    journal = "Phys. Rev. D",
    volume = "109",
    number = "12",
    pages = "124065",
    year = "2024"
}

@article{Ellis:1973yv,
    author = "Ellis, H. G.",
    title = "{Ether flow through a drainhole - a particle model in general relativity}",
    doi = "10.1063/1.1666161",
    journal = "J. Math. Phys.",
    volume = "14",
    pages = "104--118",
    year = "1973"
}

@article{Bronnikov:1973fh,
    author = "Bronnikov, K. A.",
    title = "{Scalar-tensor theory and scalar charge}",
    journal = "Acta Phys. Polon. B",
    volume = "4",
    pages = "251--266",
    year = "1973"
}

@article{Morris1988WormholesIS,
  title={Wormholes in spacetime and their use for interstellar travel: A tool for teaching general relativity},
  author={Michael S. Morris and Kip S. Thorne},
  journal={American Journal of Physics},
  year={1988},
  volume={56},
  pages={395-412}
}

@article{Nilton:2022hrp,
    author = "Nilton, M. and Furtado, J. and Alencar, G. and Landim, R. R.",
    title = "{Generalized Ellis\textendash{}Bronnikov wormholes in asymptotically safe gravity}",
    eprint = "2203.08860",
    archivePrefix = "arXiv",
    primaryClass = "gr-qc",
    doi = "10.1016/j.aop.2022.169195",
    journal = "Annals Phys.",
    volume = "448",
    pages = "169195",
    year = "2023"
}

@article{Bronnikov:2000vy,
    author = "Bronnikov, Kirill A.",
    title = "{Regular magnetic black holes and monopoles from nonlinear electrodynamics}",
    eprint = "gr-qc/0006014",
    archivePrefix = "arXiv",
    doi = "10.1103/PhysRevD.63.044005",
    journal = "Phys. Rev. D",
    volume = "63",
    pages = "044005",
    year = "2001"
}

@article{Bronnikov:2022bud,
    author = "Bronnikov, K. A.",
    title = "{Black bounces, wormholes, and partly phantom scalar fields}",
    eprint = "2206.09227",
    archivePrefix = "arXiv",
    primaryClass = "gr-qc",
    doi = "10.1103/PhysRevD.106.064029",
    journal = "Phys. Rev. D",
    volume = "106",
    number = "6",
    pages = "064029",
    year = "2022"
}

@article{Simpson:2018tsi,
    author = "Simpson, Alex and Visser, Matt",
    title = "{Black-bounce to traversable wormhole}",
    eprint = "1812.07114",
    archivePrefix = "arXiv",
    primaryClass = "gr-qc",
    doi = "10.1088/1475-7516/2019/02/042",
    journal = "JCAP",
    volume = "02",
    pages = "042",
    year = "2019"
}

@article{Alencar:2024nxi,
    author = "Alencar, G. and Nilton, M. and Rodrigues, Manuel E. and de S. Silva, Marcos V.",
    title = "{Field sources for f(R,R{\ensuremath{\mu}}{\ensuremath{\nu}}) black-bounce solutions: The case of K-gravity}",
    eprint = "2409.12101",
    archivePrefix = "arXiv",
    primaryClass = "gr-qc",
    doi = "10.1016/j.dark.2025.102060",
    journal = "Phys. Dark Univ.",
    volume = "49",
    pages = "102060",
    year = "2025"
}

@article{Lobo:2020ffi,
    author = "Lobo, Francisco S. N. and Rodrigues, Manuel E. and de Sousa Silva, Marcos V. and Simpson, Alex and Visser, Matt",
    title = "{Novel black-bounce spacetimes: wormholes, regularity, energy conditions, and causal structure}",
    eprint = "2009.12057",
    archivePrefix = "arXiv",
    primaryClass = "gr-qc",
    doi = "10.1103/PhysRevD.103.084052",
    journal = "Phys. Rev. D",
    volume = "103",
    number = "8",
    pages = "084052",
    year = "2021"
}

@article{Crispim:2024yjz,
    author = "Crispim, Tiago M. and Estrada, Milko and Muniz, C. R. and Alencar, G.",
    title = "{Braneworld black bounce to transversable wormhole}",
    eprint = "2405.08048",
    archivePrefix = "arXiv",
    primaryClass = "hep-th",
    doi = "10.1088/1475-7516/2024/10/063",
    journal = "JCAP",
    volume = "10",
    pages = "063",
    year = "2024"
}

@article{Pereira:2024rtv,
    author = "Pereira, Carlos F. S. and C. Rodrigues, Denis and Silva, Marcos V. de S. and Fabris, J{\'u}lio C. and Rodrigues, Manuel E. and Belich, H.",
    title = "{Magnetically charged black-bounce solution via nonlinear electrodynamics in a k-essence theory}",
    eprint = "2409.09182",
    archivePrefix = "arXiv",
    primaryClass = "gr-qc",
    doi = "10.1103/PhysRevD.111.084025",
    journal = "Phys. Rev. D",
    volume = "111",
    number = "8",
    pages = "084025",
    year = "2025"
}

@article{Javed:2022dfn,
    author = "Javed, Faisal and Fatima, G. and Mustafa, G. and Ovgun, Ali",
    title = "{Effects of variable equations of state on the stability of nonlinear electrodynamics thin-shell wormholes}",
    eprint = "2212.09607",
    archivePrefix = "arXiv",
    primaryClass = "gr-qc",
    doi = "10.1142/S021988782350010X",
    journal = "Int. J. Geom. Meth. Mod. Phys.",
    volume = "20",
    number = "01",
    pages = "2350010",
    year = "2023"
}

@article{Canate:2022dzb,
    author = "Ca\~nate, Pedro and Maldonado-Villamizar, F. H.",
    title = "{Novel traversable wormhole in general relativity and Einstein-Scalar-Gauss-Bonnet theory supported by nonlinear electrodynamics}",
    eprint = "2202.12463",
    archivePrefix = "arXiv",
    primaryClass = "gr-qc",
    doi = "10.1103/PhysRevD.106.044063",
    journal = "Phys. Rev. D",
    volume = "106",
    number = "4",
    pages = "044063",
    year = "2022"
}

@article{Kamal:2018oxw,
    author = "Kamal, Vivek and Kumar, Sanjeev and Kulshreshtha, Usha and Kulshreshtha, Daya Shankar",
    title = "{Wormholes in a Theory of Scalar Phantom Field}",
    doi = "10.1007/s00601-018-1392-9",
    journal = "Few Body Syst.",
    volume = "59",
    number = "4",
    pages = "70",
    year = "2018"
}

@article{Crispim:2024dgd,
    author = "Crispim, T. M. and Alencar, G. and Muniz, C. R.",
    title = "{Field sources for generalized Ellis{\textendash}Bronnikov wormhole}",
    eprint = "2410.11147",
    archivePrefix = "arXiv",
    primaryClass = "gr-qc",
    doi = "10.1088/1361-6382/ae74ac",
    journal = "Class. Quant. Grav.",
    volume = "43",
    number = "11",
    pages = "115013",
    year = "2026"
}

@book{Visser:1995cc,
    author = "Visser, Matt",
    title = "{Lorentzian wormholes: From Einstein to Hawking}",
    isbn = "978-1-56396-653-8",
    year = "1995"
}

@article{Bronnikov:2024izh,
    author = "Bronnikov, Kirill A.",
    title = "{Regular black holes as an alternative to black bounce}",
    eprint = "2404.14816",
    archivePrefix = "arXiv",
    primaryClass = "gr-qc",
    doi = "10.1103/PhysRevD.110.024021",
    journal = "Phys. Rev. D",
    volume = "110",
    number = "2",
    pages = "024021",
    year = "2024"
}

@article{Bronnikov:2021uta,
    author = "Bronnikov, Kirill A. and Walia, Rahul Kumar",
    title = "{Field sources for Simpson-Visser spacetimes}",
    eprint = "2112.13198",
    archivePrefix = "arXiv",
    primaryClass = "gr-qc",
    doi = "10.1103/PhysRevD.105.044039",
    journal = "Phys. Rev. D",
    volume = "105",
    number = "4",
    pages = "044039",
    year = "2022"
}

@article{Alencar:2024yvh,
    author = "Alencar, G. and Bronnikov, Kirill A. and Rodrigues, Manuel E. and S\'aez-Chill\'on G\'omez, Diego and de S. Silva, Marcos V.",
    title = "{On black bounce space-times in non-linear electrodynamics}",
    eprint = "2403.12897",
    archivePrefix = "arXiv",
    primaryClass = "gr-qc",
    doi = "10.1140/epjc/s10052-024-13119-4",
    journal = "Eur. Phys. J. C",
    volume = "84",
    number = "7",
    pages = "745",
    year = "2024"
}

@article{Sharma:2022dbx,
    author = "Sharma, Vivek and Ghosh, Suman",
    title = "{Geodesic congruences in 5D warped Ellis\textendash{}Bronnikov spacetimes}",
    eprint = "2205.08865",
    archivePrefix = "arXiv",
    primaryClass = "gr-qc",
    doi = "10.1140/epjp/s13360-022-03086-8",
    journal = "Eur. Phys. J. Plus",
    volume = "137",
    number = "8",
    pages = "881",
    year = "2022"
}

@article{Alencar:2025jvl,
    author = "Alencar, G. and Duran-Cabac\'es, Albert and Rubiera-Garcia, Diego and S\'aez-Chill\'on G\'omez, Diego",
    title = "{General spherically symmetric black bounces within nonlinear electrodynamics}",
    eprint = "2501.03909",
    archivePrefix = "arXiv",
    primaryClass = "gr-qc",
    doi = "10.1103/PhysRevD.111.104020",
    journal = "Phys. Rev. D",
    volume = "111",
    number = "10",
    pages = "104020",
    year = "2025"
}

@article{Sharma:2022tiv,
    author = "Sharma, Vivek and Ghosh, Suman",
    title = "{Geodesics in generalised Ellis\textendash{}Bronnikov spacetime embedded in warped 5D background}",
    eprint = "2205.05973",
    archivePrefix = "arXiv",
    primaryClass = "gr-qc",
    doi = "10.1140/epjc/s10052-022-10682-6",
    journal = "Eur. Phys. J. C",
    volume = "82",
    number = "8",
    pages = "702",
    year = "2022"
}

@article{Sharma:2021kqb,
    author = "Sharma, Vivek and Ghosh, Suman",
    title = "{Generalised Ellis\textendash{}Bronnikov wormholes embedded in warped braneworld background and energy conditions}",
    eprint = "2111.07329",
    archivePrefix = "arXiv",
    primaryClass = "gr-qc",
    doi = "10.1140/epjc/s10052-021-09789-z",
    journal = "Eur. Phys. J. C",
    volume = "81",
    number = "11",
    pages = "1004",
    year = "2021"
}

@article{Godani:2023jhq,
    author = "Godani, Nisha and Kala, Shubham",
    title = "{A new class of generalized Ellis\textendash{}Bronnikov wormhole in asymptotically safe gravity}",
    doi = "10.1142/S0218271823500670",
    journal = "Int. J. Mod. Phys. D",
    volume = "32",
    number = "10",
    pages = "2350067",
    year = "2023"
}

@article{Muniz:2022aal,
    author = "Muniz, R. C. and Maluf, R. V.",
    title = "{Generalized Ellis-Bronikov traversable wormholes in $f(R)$ gravity with anisotropic dark matter}",
    eprint = "2203.09263",
    archivePrefix = "arXiv",
    primaryClass = "gr-qc",
    month = "3",
    journal="",
    year = "2022"
}

@article{Kar:1995jz,
    author = "Kar, Sayan and Minwalla, Shiraz and Mishra, D. and Sahdev, D.",
    title = "{Resonances in the transmission of massless scalar waves in a class of wormholes}",
    doi = "10.1103/PhysRevD.51.1632",
    journal = "Phys. Rev. D",
    volume = "51",
    pages = "1632--1638",
    year = "1995"
}

@article{deSouza:2022ioq,
    author = "de Souza, T. F. and Ramos, A. C. A. and Costa Filho, R. N. and Furtado, J.",
    title = "{Generalized Ellis-Bronnikov bilayer graphene wormholelike surface}",
    doi = "10.1103/PhysRevB.106.165426",
    journal = "Phys. Rev. B",
    volume = "106",
    number = "16",
    pages = "165426",
    year = "2022"
}

@article{Rueda:2023val,
    author = "Rueda, A. and Contreras, E.",
    title = "{Geodesic analysis and steady accretion on a traversable wormhole}",
    eprint = "2311.08344",
    archivePrefix = "arXiv",
    primaryClass = "gr-qc",
    doi = "10.1016/j.aop.2023.169540",
    journal = "Annals Phys.",
    volume = "459",
    pages = "169540",
    year = "2023"
}

@article{Bhattacharya:2025ljc,
    author = "Bhattacharya, Soumya and Ghosh, Suman",
    title = "{Displacement memory and B-memory in generalised Ellis-Bronnikov wormholes}",
    eprint = "2502.03007",
    archivePrefix = "arXiv",
    primaryClass = "gr-qc",
    month = "2",
    year = "2025"
}

@article{Jana:2024pdh,
    author = "Jana, Soumya and Sharma, Vivek and Ghosh, Suman",
    title = "{Gravitational Lensing and Deflection Angle by generalised Ellis-Bronnikov wormhole Embedded in Warped Braneworld Background}",
    eprint = "2411.10804",
    archivePrefix = "arXiv",
    primaryClass = "gr-qc",
    month = "11",
    year = "2024"
}

@article{Mitra:2023yjf,
    author = "Mitra, Antariksha and Ghosh, Suman",
    title = "{Signature quasinormal modes of Ellis-Bronnikov wormhole embedded in warped braneworld background}",
    eprint = "2310.17479",
    archivePrefix = "arXiv",
    primaryClass = "gr-qc",
    doi = "10.1103/PhysRevD.109.064005",
    journal = "Phys. Rev. D",
    volume = "109",
    number = "6",
    pages = "064005",
    year = "2024"
}

@article{Crispim:2024lzf,
    author = "Crispim, T. M. and de S. Silva, Marcos V. and Alencar, G. and Muniz, Celio R. and S\'aez-Chill\'on G\'omez, Diego",
    title = "{Field sources for wormholes with multiple throats/anti-throats}",
    eprint = "2412.05236",
    archivePrefix = "arXiv",
    primaryClass = "gr-qc",
    doi = "10.1088/1361-6382/adc654",
    journal = "Class. Quant. Grav.",
    volume = "42",
    number = "8",
    pages = "085005",
    year = "2025"
}

@article{Ovalle2017,
  author         = {Ovalle, Jorge},
  title          = {Decoupling gravitational sources in general relativity: The MGD approach},
  journal        = {Phys. Rev. D},
  volume         = {95},
  pages          = {104019},
  year           = {2017},
  doi            = {10.1103/PhysRevD.95.104019},
  eprint         = {1704.05899},
  archivePrefix  = {arXiv},
  primaryClass   = {gr-qc}
}

@article{Ovalle2019,
  author         = {Ovalle, Jorge},
  title          = {Decoupling gravitational sources in general relativity: An extended review},
  journal        = {Phys. Lett. B},
  volume         = {788},
  pages          = {213--219},
  year           = {2019},
  doi            = {10.1016/j.physletb.2018.11.029},
  eprint         = {1812.03000},
  archivePrefix  = {arXiv},
  primaryClass   = {gr-qc}
}

@article{Misyura:2024fho,
    author = "Misyura, Maxim and Rincon, Angel and Vertogradov, Vitalii",
    title = "{Non-singular black hole by gravitational decoupling and some thermodynamic properties}",
    eprint = "2405.05370",
    archivePrefix = "arXiv",
    primaryClass = "gr-qc",
    doi = "10.1016/j.dark.2024.101717",
    journal = "Phys. Dark Univ.",
    volume = "46",
    pages = "101717",
    year = "2024"
}

@article{Do:2019txf,
    author = "Do, Tuan and others",
    title = "{Relativistic redshift of the star S0-2 orbiting the Galactic center supermassive black hole}",
    eprint = "1907.10731",
    archivePrefix = "arXiv",
    primaryClass = "astro-ph.GA",
    doi = "10.1126/science.aav8137",
    journal = "Science",
    volume = "365",
    number = "6454",
    pages = "664--668",
    year = "2019"
}

@article{GRAVITY:2020gka,
    author = "Abuter, R. and others",
    collaboration = "GRAVITY",
    title = "{Detection of the Schwarzschild precession in the orbit of the star S2 near the Galactic centre massive black hole}",
    eprint = "2004.07187",
    archivePrefix = "arXiv",
    primaryClass = "astro-ph.GA",
    doi = "10.1051/0004-6361/202037813",
    journal = "Astron. Astrophys.",
    volume = "636",
    pages = "L5",
    year = "2020"
}

@article{Vagnozzi:2022moj,
    author = "Vagnozzi, Sunny and others",
    title = "{Horizon-scale tests of gravity theories and fundamental physics from the Event Horizon Telescope image of Sagittarius A}",
    eprint = "2205.07787",
    archivePrefix = "arXiv",
    primaryClass = "gr-qc",
    reportNumber = "UCI-HEP-TR-2022-07",
    doi = "10.1088/1361-6382/acd97b",
    journal = "Class. Quant. Grav.",
    volume = "40",
    number = "16",
    pages = "165007",
    year = "2023"
}

@article{EventHorizonTelescope:2022xqj,
    author = "Akiyama, Kazunori and others",
    collaboration = "Event Horizon Telescope",
    title = "{First Sagittarius A* Event Horizon Telescope Results. VI. Testing the Black Hole Metric}",
    eprint = "2311.09484",
    archivePrefix = "arXiv",
    primaryClass = "astro-ph.HE",
    reportNumber = "FERMILAB-PUB-22-422-PPD",
    doi = "10.3847/2041-8213/ac6756",
    journal = "Astrophys. J. Lett.",
    volume = "930",
    number = "2",
    pages = "L17",
    year = "2022"
}

@article{Perlick:2021aok,
    author = "Perlick, Volker and Tsupko, Oleg Yu.",
    title = "{Calculating black hole shadows: Review of analytical studies}",
    eprint = "2105.07101",
    archivePrefix = "arXiv",
    primaryClass = "gr-qc",
    doi = "10.1016/j.physrep.2021.10.004",
    journal = "Phys. Rept.",
    volume = "947",
    pages = "1--39",
    year = "2022"
}

@article{Bronnikov:2021liv,
    author = "Bronnikov, Kirill A. and Konoplya, Roman A. and Pappas, Thomas D.",
    title = "{General parametrization of wormhole spacetimes and its application to shadows and quasinormal modes}",
    eprint = "2102.10679",
    archivePrefix = "arXiv",
    primaryClass = "gr-qc",
    doi = "10.1103/PhysRevD.103.124062",
    journal = "Phys. Rev. D",
    volume = "103",
    number = "12",
    pages = "124062",
    year = "2021"
}

@article{Johnson:2019ljv,
    author = "Johnson, Michael D. and others",
    title = "{Universal interferometric signatures of a black hole{\textquoteright}s photon ring}",
    eprint = "1907.04329",
    archivePrefix = "arXiv",
    primaryClass = "astro-ph.IM",
    doi = "10.1126/sciadv.aaz1310",
    journal = "Sci. Adv.",
    volume = "6",
    number = "12",
    pages = "eaaz1310",
    year = "2020"
}

@article{Gralla:2020srx,
    author = "Gralla, Samuel E. and Lupsasca, Alexandru and Marrone, Daniel P.",
    title = "{The shape of the black hole photon ring: A precise test of strong-field general relativity}",
    eprint = "2008.03879",
    archivePrefix = "arXiv",
    primaryClass = "gr-qc",
    doi = "10.1103/PhysRevD.102.124004",
    journal = "Phys. Rev. D",
    volume = "102",
    number = "12",
    pages = "124004",
    year = "2020"
}

@article{EventHorizonTelescope:2019pgp,
    author = "Akiyama, Kazunori and others",
    collaboration = "Event Horizon Telescope",
    title = "{First M87 Event Horizon Telescope Results. V. Physical Origin of the Asymmetric Ring}",
    eprint = "1906.11242",
    archivePrefix = "arXiv",
    primaryClass = "astro-ph.GA",
    doi = "10.3847/2041-8213/ab0f43",
    journal = "Astrophys. J. Lett.",
    volume = "875",
    number = "1",
    pages = "L5",
    year = "2019"
}

@article{Olmo:2025ctf,
    author = "Olmo, Gonzalo J. and Rosa, Jo{\~a}o Lu{\'\i}s and Rubiera-Garcia, Diego and Rueda, Alejandro and S{\'a}ez-Chill{\'o}n G{\'o}mez, Diego",
    title = "{Shadows from thin accretion disks of parametrized black hole solutions}",
    eprint = "2507.16580",
    archivePrefix = "arXiv",
    primaryClass = "gr-qc",
    doi = "10.1103/s968-npmt",
    journal = "Phys. Rev. D",
    volume = "112",
    number = "8",
    pages = "084059",
    year = "2025"
}

@article{Cardoso:2008bp,
    author = "Cardoso, Vitor and Miranda, Alex S. and Berti, Emanuele and Witek, Helvi and Zanchin, Vilson T.",
    title = "{Geodesic stability, Lyapunov exponents and quasinormal modes}",
    eprint = "0812.1806",
    archivePrefix = "arXiv",
    primaryClass = "hep-th",
    doi = "10.1103/PhysRevD.79.064016",
    journal = "Phys. Rev. D",
    volume = "79",
    number = "6",
    pages = "064016",
    year = "2009"
}

@article{Gralla:2019yty,
    author = "Gralla, Samuel E. and Holz, Daniel E. and Wald, Robert M.",
    title = "{Black Hole Shadows, Photon Rings, and Lensing Rings}",
    eprint = "1906.00873",
    archivePrefix = "arXiv",
    primaryClass = "astro-ph.HE",
    doi = "10.1103/PhysRevD.100.024018",
    journal = "Phys. Rev. D",
    volume = "100",
    number = "2",
    pages = "024018",
    year = "2019"
}

@article{Guerrero:2021ues,
    author = "Guerrero, M. and Olmo, G. J. and Rubiera-Garcia, D. and S\'aez-Chill\'on G\'omez, D.",
    title = "{Shadows and optical appearance of black bounces illuminated by an accretion disk}",
    eprint = "2105.15073",
    archivePrefix = "arXiv",
    primaryClass = "gr-qc",
    doi = "10.1088/1475-7516/2021/08/036",
    journal = "JCAP",
    volume = "08",
    pages = "036",
    year = "2021"
}

@article{deSSilva:2025ncg,
    author = "de S. Silva, Marcos V. and Pereira, Carlos F. S. and Bragato, Bruna and Rodrigues, Manuel E. and Fabris, J{\'u}lio C. and Belich, H.",
    title = "{Sources of matter for wormholes in a k-essence theory}",
    eprint = "2512.11018",
    archivePrefix = "arXiv",
    primaryClass = "gr-qc",
    doi = "10.1088/1361-6382/ae85a8",
    journal = "Class. Quant. Grav.",
    volume = "43",
    number = "14",
    pages = "145004",
    year = "2026"
}

@article{Guerrero:2022qkh,
    author = "Guerrero, Merce and Olmo, Gonzalo J. and Rubiera-Garcia, Diego and  S{\'a}ez-Chill{\'o}n G{\'o}mez, Diego",
    title = "{Light ring images of double photon spheres in black hole and wormhole spacetimes}",
    eprint = "2202.03809",
    archivePrefix = "arXiv",
    primaryClass = "gr-qc",
    doi = "10.1103/PhysRevD.105.084057",
    journal = "Phys. Rev. D",
    volume = "105",
    number = "8",
    pages = "084057",
    year = "2022"
}

@article{Gralla:2019hhd,
    author = "Gralla, Samuel E. and Lupsasca, Alexandru and Marrone, Daniel P.",
    title = "{Physics of the Photon Ring of a Black Hole}",
    eprint = "1902.04511",
    archivePrefix = "arXiv",
    primaryClass = "astro-ph.HE",
    doi = "10.1103/PhysRevD.99.124016",
    journal = "Phys. Rev. D",
    volume = "99",
    number = "12",
    pages = "124016",
    year = "2019"
}

@article{Ayzenberg:2023uhx,
    author = "Ayzenberg, D. and others",
    title = "{Fundamental physics opportunities with future ground-based mm/sub-mm VLBI arrays}",
    eprint = "2312.02130",
    archivePrefix = "arXiv",
    primaryClass = "astro-ph.HE",
    doi = "10.1007/s41114-025-00057-0",
    journal = "Living Rev. Rel.",
    volume = "28",
    number = "1",
    pages = "4",
    year = "2025",
    note = "[Erratum: Living Rev.Rel. 28, 7 (2025)]"
}

@article{Duran-Cabaces:2025sly,
    author = "Duran-Cabac{\'e}s, Albert and Rubiera-Garcia, Diego and S{\'a}ez-Chill{\'o}n G{\'o}mez, Diego",
    title = "{Quasinormal modes and echoes of generalized black hole bounces and their correspondence with shadows}",
    eprint = "2506.10814",
    archivePrefix = "arXiv",
    primaryClass = "gr-qc",
    doi = "10.1103/3v89-z8rf",
    journal = "Phys. Rev. D",
    volume = "112",
    number = "4",
    pages = "044016",
    year = "2025"
}

@article{Damour:2007ap,
    author = "Damour, Thibault and Solodukhin, Sergey N.",
    title = "{Wormholes as black hole foils}",
    eprint = "0704.2667",
    archivePrefix = "arXiv",
    primaryClass = "gr-qc",
    reportNumber = "IHES-P-07-19",
    doi = "10.1103/PhysRevD.76.024016",
    journal = "Phys. Rev. D",
    volume = "76",
    pages = "024016",
    year = "2007"
}

@article{Konoplya:2025mvj,
    author = "Konoplya, Roman A. and Khrabustovskyi, Andrii and K{\v{r}}{\'\i}{\v{z}}, Jan and Zhidenko, Alexander",
    title = "{Quasinormal ringing and shadows of black holes and wormholes in dark matter-inspired Weyl gravity}",
    eprint = "2501.16134",
    archivePrefix = "arXiv",
    primaryClass = "gr-qc",
    doi = "10.1088/1475-7516/2025/04/062",
    journal = "JCAP",
    volume = "04",
    pages = "062",
    year = "2025"
}

@article{Bronnikov:2013coa,
    author = "Bronnikov, K. A. and Lipatova, L. N. and Novikov, I. D. and Shatskiy, A. A.",
    title = "{Example of a stable wormhole in general relativity}",
    eprint = "1312.6929",
    archivePrefix = "arXiv",
    primaryClass = "gr-qc",
    doi = "10.1134/S0202289313040038",
    journal = "Grav. Cosmol.",
    volume = "19",
    pages = "269--274",
    year = "2013"
}

@article{Alencar:2026qeb,
    author = "Alencar, G. and Crispim, T. M. and S{\'a}ez-Chill{\'o}n G{\'o}mez, Diego and de S. Silva, V., Marcos",
    title = "{Black bounce as a quantum correction from string T-duality: Thermodynamics, energy conditions, and observational imprints from EHT}",
    eprint = "2603.05543",
    archivePrefix = "arXiv",
    primaryClass = "gr-qc",
    doi = "10.1016/j.dark.2026.102427",
    journal = "Phys. Dark Univ.",
    volume = "53",
    pages = "102427",
    year = "2026"
}

@article{Lobo:2005us,
    author = "Lobo, Francisco S. N.",
    title = "{Phantom energy traversable wormholes}",
    eprint = "gr-qc/0502099",
    archivePrefix = "arXiv",
    doi = "10.1103/PhysRevD.71.084011",
    journal = "Phys. Rev. D",
    volume = "71",
    pages = "084011",
    year = "2005"
}

@article{Lobo:2005yv,
    author = "Lobo, Francisco S. N.",
    title = "{Stability of phantom wormholes}",
    eprint = "gr-qc/0506001",
    archivePrefix = "arXiv",
    doi = "10.1103/PhysRevD.71.124022",
    journal = "Phys. Rev. D",
    volume = "71",
    pages = "124022",
    year = "2005"
}

@article{Hernandez:1966zf,
    author = "Hernandez, W. C. and Misner, C. W.",
    title = "{Observer Time as a Coordinate in Relativistic Spherical Hydrodynamics}",
    doi = "10.1086/148525",
    journal = "Astrophys. J.",
    volume = "143",
    pages = "452",
    year = "1966"
}

@inproceedings{Bardeen:1968pbd,
    author = "Bardeen, James M.",
    title = "{Non-singular general-relativistic gravitational collapse}",
    booktitle = "{Proceedings of the 5th International Conference on Gravitation and the Theory of Relativity (GR5)}",
    address = "Tbilisi, USSR",
    pages = "174",
    year = "1968"
}

@article{Ayon-Beato:2000mjt,
    author = "Ayon-Beato, Eloy and Garcia, Alberto",
    title = "{The Bardeen model as a nonlinear magnetic monopole}",
    eprint = "gr-qc/0009077",
    archivePrefix = "arXiv",
    doi = "10.1016/S0370-2693(00)01125-4",
    journal = "Phys. Lett. B",
    volume = "493",
    pages = "149--152",
    year = "2000"
}

@article{Penrose1965,
  title   = {Gravitational Collapse and Space-Time Singularities},
  author  = {Penrose, Roger},
  journal = {Physical Review Letters},
  volume  = {14},
  number  = {3},
  pages   = {57--59},
  year    = {1965},
  publisher = {American Physical Society},
  doi     = {10.1103/PhysRevLett.14.57}
}

@article{Hawking:1967ju,
    author = "Hawking, Stephen",
    title = "{The occurrence of singularities in cosmology. III. Causality and singularities}",
    doi = "10.1098/rspa.1967.0164",
    journal = "Proc. Roy. Soc. Lond. A",
    volume = "300",
    pages = "187--201",
    year = "1967"
}

@article{Hawking:1970zqf,
    author = "Hawking, S. W. and Penrose, R.",
    title = "{The Singularities of gravitational collapse and cosmology}",
    doi = "10.1098/rspa.1970.0021",
    journal = "Proc. Roy. Soc. Lond. A",
    volume = "314",
    pages = "529--548",
    year = "1970"
}

@article{Einstein:1935tc,
    author = "Einstein, Albert and Rosen, N.",
    title = "{The Particle Problem in the General Theory of Relativity}",
    doi = "10.1103/PhysRev.48.73",
    journal = "Phys. Rev.",
    volume = "48",
    pages = "73--77",
    year = "1935"
}

@article{Herdeiro:2018ldf,
    author = "Herdeiro, Carlos A. R. and Lemos, Jos{\'e} P. S.",
    title = "{The black hole fifty years after: Genesis of the name}",
    eprint = "1811.06587",
    archivePrefix = "arXiv",
    primaryClass = "physics.hist-ph",
    month = "11",
    journal="",
    year = "2018"
}

@article{Oppenheimer:1939ue,
    author = "Oppenheimer, J. R. and Snyder, H.",
    title = "{On Continued gravitational contraction}",
    doi = "10.1103/PhysRev.56.455",
    journal = "Phys. Rev.",
    volume = "56",
    pages = "455--459",
    year = "1939"
}

@article{Bambi:2008jg,
    author = "Bambi, Cosimo and Freese, Katherine",
    title = "{Apparent shape of super-spinning black holes}",
    eprint = "0812.1328",
    archivePrefix = "arXiv",
    primaryClass = "astro-ph",
    reportNumber = "IPMU08-0095",
    doi = "10.1103/PhysRevD.79.043002",
    journal = "Phys. Rev. D",
    volume = "79",
    pages = "043002",
    year = "2009"
}

@article{Ortiz:2015rma,
    author = "Ortiz, N{\'e}stor and Sarbach, Olivier and Zannias, Thomas",
    title = "{Shadow of a naked singularity}",
    eprint = "1505.07017",
    archivePrefix = "arXiv",
    primaryClass = "gr-qc",
    doi = "10.1103/PhysRevD.92.044035",
    journal = "Phys. Rev. D",
    volume = "92",
    number = "4",
    pages = "044035",
    year = "2015"
}

@article{Bambi:2013nla,
    author = "Bambi, Cosimo",
    title = "{Can the supermassive objects at the centers of galaxies be traversable wormholes? The first test of strong gravity for mm/sub-mm very long baseline interferometry facilities}",
    eprint = "1304.5691",
    archivePrefix = "arXiv",
    primaryClass = "gr-qc",
    doi = "10.1103/PhysRevD.87.107501",
    journal = "Phys. Rev. D",
    volume = "87",
    pages = "107501",
    year = "2013"
}

@article{Shaikh:2018kfv,
    author = "Shaikh, Rajibul",
    title = "{Shadows of rotating wormholes}",
    eprint = "1803.11422",
    archivePrefix = "arXiv",
    primaryClass = "gr-qc",
    doi = "10.1103/PhysRevD.98.024044",
    journal = "Phys. Rev. D",
    volume = "98",
    number = "2",
    pages = "024044",
    year = "2018"
}

@article{
Novikov:2025elo,
    author = "Novikov, I. D. and Repin, S. V. and Paksivatova, D. A.",
    title = "{Observing an accretion disk inside a wormhole shadow}",
    eprint = "2508.02752",
    archivePrefix = "arXiv",
    primaryClass = "gr-qc",
    month = "8",
    journal="",
    year = "2025"
}

@article{Bugaev:2023mlc,
    author = "Bugaev, Mikhail A. and Samorodskaya, Polina S. and Novikov, Igor D. and Repin, Serge V.",
    title = "{Uniform sky glow observed through the throat of a wormhole}",
    eprint = "2305.18041",
    archivePrefix = "arXiv",
    primaryClass = "gr-qc",
    doi = "10.1103/PhysRevD.108.124059",
    journal = "Phys. Rev. D",
    volume = "108",
    number = "12",
    pages = "124059",
    year = "2023"
}

@article{Bronnikov:2012ch,
    author = "Bronnikov, K. A. and Konoplya, R. A. and Zhidenko, A.",
    title = "{Instabilities of wormholes and regular black holes supported by a phantom scalar field}",
    eprint = "1205.2224",
    archivePrefix = "arXiv",
    primaryClass = "gr-qc",
    doi = "10.1103/PhysRevD.86.024028",
    journal = "Phys. Rev. D",
    volume = "86",
    pages = "024028",
    year = "2012"
}

@article{Bronnikov:2006fu,
    author = "Bronnikov, K. A. and Melnikov, V. N. and Dehnen, Heinz",
    title = "{Regular black holes and black universes}",
    eprint = "gr-qc/0611022",
    archivePrefix = "arXiv",
    doi = "10.1007/s10714-007-0430-6",
    journal = "Gen. Rel. Grav.",
    volume = "39",
    pages = "973--987",
    year = "2007"
}

@article{Bronnikov:2019sbx,
    author = "Bronnikov, Kirill A. and Konoplya, Roman A.",
    title = "{Echoes in brane worlds: ringing at a black hole--wormhole transition}",
    eprint = "1912.05315",
    archivePrefix = "arXiv",
    primaryClass = "gr-qc",
    doi = "10.1103/PhysRevD.101.064004",
    journal = "Phys. Rev. D",
    volume = "101",
    number = "6",
    pages = "064004",
    year = "2020"
}

@article{Bolokhov:2012kn,
    author = "Bolokhov, S. V. and Bronnikov, K. A. and Skvortsova, M. V.",
    title = "{Magnetic black universes and wormholes with a phantom scalar}",
    eprint = "1208.4619",
    archivePrefix = "arXiv",
    primaryClass = "gr-qc",
    doi = "10.1088/0264-9381/29/24/245006",
    journal = "Class. Quant. Grav.",
    volume = "29",
    pages = "245006",
    year = "2012"
}

@article{Alencar:2026ypy,
    author = "Alencar, G. and Duran-Cabac{\'e}s, Albert and Lima, A. and Rinc{\'o}n, {\'A}ngel and S{\'a}ez-Chill{\'o}n G{\'o}mez, Diego",
    title = "{Analyzing Quasi-normal modes spectrum in asymptotically de Sitter black bounces and their optical appearance}",
    eprint = "2608.20081",
    archivePrefix = "arXiv",
    primaryClass = "gr-qc",
    month = "8",
    journal="",
    year = "2026"
}

@article{Alencar:2026quc,
    author = "Alencar, G. and Crispim, T. M. and Muniz, C. R.",
    title = "{Geometrically Regular Black Object Solutions in Lower-Dimensional Gauss-Bonnet Gravity and Its Unimodular Extension}",
    eprint = "2606.13635",
    archivePrefix = "arXiv",
    primaryClass = "gr-qc",
    month = "6",
    journal="",
    year = "2026"
}

@article{Crispim:2025cql,
    author = "Crispim, T. M. and de S. Silva, Marcos V. and Alencar, G. and S{\'a}ez-Chill{\'o}n G{\'o}mez, Diego",
    title = "{Tidal stretching and compression in black bounce backgrounds}",
    eprint = "2507.00311",
    archivePrefix = "arXiv",
    primaryClass = "gr-qc",
    doi = "10.1140/epjc/s10052-025-14837-z",
    journal = "Eur. Phys. J. C",
    volume = "85",
    number = "10",
    pages = "1186",
    year = "2025",
    note = "[Erratum: Eur.Phys.J.C 85, 1248 (2025)]"
}

@article{Muniz:2024wiv,
    author = "Muniz, C. R. and Alencar, G. and Cunha, M. S. and Olmo, Gonzalo J.",
    title = "{Static and stationary black bounces inspired by loop quantum gravity}",
    eprint = "2408.08542",
    archivePrefix = "arXiv",
    primaryClass = "gr-qc",
    doi = "10.1103/h7rn-4ht6",
    journal = "Phys. Rev. D",
    volume = "112",
    number = "2",
    pages = "024018",
    year = "2025"
}

@article{deSousaSilva:2026kvp,
    author = "de Sousa Silva, Marcos V. and Rodrigues, Manuel E. and Pereira, C. F. S.",
    title = "{Three dimensional black bounces in f(R) gravity}",
    eprint = "2601.17848",
    archivePrefix = "arXiv",
    primaryClass = "gr-qc",
    doi = "10.1016/j.dark.2026.102293",
    journal = "Phys. Dark Univ.",
    volume = "52",
    pages = "102293",
    year = "2026"
}

@article{Silva:2026mlo,
    author = "Silva, Marcos V. de S.",
    title = "{Roche limit and stellar disruption in the Simpson{\textendash}Visser spacetime}",
    eprint = "2601.16082",
    archivePrefix = "arXiv",
    primaryClass = "gr-qc",
    doi = "10.1016/j.dark.2026.102326",
    month = "1",
    journal="Phys. Dark Univ.",
    year = "2026"
}

@article{Pereira:2026ffn,
    author = "Pereira, C. F. S. and Belich, H. and Soares, A. R. and Silva, Marcos V. de S. and Vit{\'o}ria, R. L. L. and Ara{\'u}jo Filho, A. A.",
    title = "{Light propagation and quasinormal modes of a topologically charged Schwarzschild-Klinkhamer wormhole}",
    eprint = "2601.16305",
    archivePrefix = "arXiv",
    primaryClass = "gr-qc",
    doi = "10.1016/j.dark.2026.102365",
    journal = "Phys. Dark Univ.",
    volume = "53",
    pages = "102365",
    year = "2026"
}

@article{Pereira:2025fvg,
    author = "Pereira, C. F. S. and Soares, A. R. and Silva, M. V. de S. and Vit{\'o}ria, R. L. L. and Belich, H.",
    title = "{Light deflection and gravitational lensing effects in acoustic black-bounce spacetime}",
    eprint = "2505.12577",
    archivePrefix = "arXiv",
    primaryClass = "gr-qc",
    doi = "10.1103/fqvd-8nl7",
    journal = "Phys. Rev. D",
    volume = "112",
    number = "6",
    pages = "064012",
    year = "2025"
}

@article{Silva:2025fqj,
    author = "Silva, Marcos V. de S. and Crispim, T. M. and Alencar, G. and Landim, R. R. and Rodrigues, Manuel E.",
    title = "{Generalized black-bounces solutions in f(R) gravity and their field sources}",
    eprint = "2502.19186",
    archivePrefix = "arXiv",
    primaryClass = "gr-qc",
    doi = "10.1088/1361-6382/ae2730",
    journal = "Class. Quant. Grav.",
    volume = "43",
    number = "1",
    pages = "015005",
    year = "2026"
}

@article{Cardoso:2016rao,
    author = "Cardoso, Vitor and Franzin, Edgardo and Pani, Paolo",
    title = "{Is the gravitational-wave ringdown a probe of the event horizon?}",
    eprint = "1602.07309",
    archivePrefix = "arXiv",
    primaryClass = "gr-qc",
    doi = "10.1103/PhysRevLett.116.171101",
    journal = "Phys. Rev. Lett.",
    volume = "116",
    number = "17",
    pages = "171101",
    year = "2016",
    note = "[Erratum: Phys.Rev.Lett. 117, 089902 (2016)]"
}

@article{Konoplya:2016hmd,
    author = "Konoplya, R. A. and Zhidenko, A.",
    title = "{Wormholes versus black holes: quasinormal ringing at early and late times}",
    eprint = "1606.00517",
    archivePrefix = "arXiv",
    primaryClass = "gr-qc",
    doi = "10.1088/1475-7516/2016/12/043",
    journal = "JCAP",
    volume = "12",
    pages = "043",
    year = "2016"
}
\end{document}